\documentclass[aps,prd, notitlepage, onecolumn,superscriptaddress,floatfix,letterpaper,nofootinbib, longbibliography]{revtex4-1}

\usepackage{amsmath, amsthm, amssymb,slashed,mathtools,tabu}
\usepackage{pifont}

\usepackage[usenames, dvipsnames]{color}
\usepackage[svgnames]{xcolor}
\usepackage[colorlinks,citecolor=RoyalBlue, urlcolor=RoyalBlue, linkcolor=RoyalBlue ]{hyperref} %BlueViolet  NavyBlue RoyalBlue MidnightBlue

\usepackage{multirow}

\newcommand{\cmark}{\ding{51}}
\newcommand{\xmark}{\ding{55}}

\usepackage[normalem]{ulem}

\newcommand{\cred}[1]{\textcolor{black}{#1}}
\newcommand{\cblue}[1]{\textcolor{black}{#1}}

\newcommand{\cpurple}[1]{\textcolor{black}{#1}}

\usepackage{booktabs}

\newcommand{\ccred}[1]{\textcolor{black}{#1}}

\definecolor{mygray}{gray}{0.6}

\usepackage{upgreek}
\usepackage{bbm}

\newenvironment{myfont}[2][]{\csname#2\endcsname[#1]}{}

\usepackage{slashed}
\usepackage[makeroom]{cancel}
\usepackage[normalem]{ulem}
\usepackage{soul}
\newcommand{\stkout}[1]{\ifmmode\text{\sout{\ensuremath{#1}}}\else\sout{#1}\fi}

\usepackage{sseq}
\usepackage[all,cmtip]{xy}
\usepackage{tikz-cd}
\usepackage{tikz}
\usetikzlibrary{matrix}
\usetikzlibrary{decorations.markings}
\usetikzlibrary{tikzmark,decorations.pathreplacing,positioning}
\usepackage{amsfonts}

\newcommand{\bea}{\begin{eqnarray}}
\newcommand{\eea}{\end{eqnarray}}
\def\be{\begin{equation}}
\def\ee{\end{equation}}

\newcommand{\e}{\hspace{1pt}\mathrm{e}}
\newcommand{\ii}{\hspace{1pt}\mathrm{i}\hspace{1pt}}

\def\CP{{\mathbb{CP}}}

\definecolor{red}{rgb}{1,0,0}
\definecolor{blue}{rgb}{0,0,1}
\definecolor{dblue}{rgb}{0,0,0.4}
\definecolor{green}{rgb}{0,1,0}
\definecolor{black}{rgb}{0,0,0}
\definecolor{white}{rgb}{1,1,1}

\definecolor{brn}{rgb}{.8,.4,.0}
\definecolor{redo}{rgb}{1,.5,.0}
\definecolor{ddgrn}{rgb}{0,0.4,0}
\definecolor{dgrn}{rgb}{0,0.55,0}
\definecolor{dbl}{rgb}{0,0,0.5}

\usepackage[bbgreekl]{mathbbol}
\usepackage{amscd}

\newcommand{\Z}{\mathbb{Z}}
\newcommand{\C}{\mathbb{C}}
\newcommand{\R}{\mathbb{R}}

\newcommand{\dd}{\hspace{.5pt}\mathrm{d}}
\newcommand{\<}{\langle} 
\renewcommand{\>}{\rangle} 

\newcommand{\Refe}[1]{Ref.~[\onlinecite{#1}]}
\newcommand{\Eq}[1]{Eq.~(\ref{#1})} 
\newcommand{\eq}[1]{(\ref{#1})}

\newcommand{\Eqn}[1]{Eqn.~(\ref{#1})} 

\newcommand{\Tr}{{\rm Tr}}

\newcommand{\ch}{{\rm ch}} 
 
\newcommand{\prt}{\partial} 
\newcommand{\pt}{\partial} 
\newcommand{\sgn}{{\rm sgn}}

\newcommand{\bpm}{\begin{pmatrix}}
\newcommand{\epm}{\end{pmatrix}}
\newcommand{\bmm}{\begin{matrix}}
\newcommand{\emm}{\end{matrix}}

\newcommand{\cA}{ {\cal A} }

\newcommand{\cD}{ {\cal D} } 
\newcommand{\cE}{ {\cal E} } 
\newcommand{\cG}{ {\cal G} }

\newcommand{\cL}{ {\cal L} }

\newcommand{\al}{\alpha} 
\newcommand{\bt}{\beta}

\newcommand{\ga}{\gamma}

\def\CD{{\cal D}}

\def\CO{{\cal O}}
\def\CS{{\rm CS}}

\def\CX{{\cal X}}
\def\CY{{\cal Y}}
\def\CZ{{\cal Z}}

\def\Z{{\mathbb{Z}}}
\def\Q{{\mathbb{Q}}}
\def\R{{\mathbb{R}}}
\def\C{{\mathbb{C}}}

\def\pt{\mathrm{pt}}
\def\Tr{{\mathrm{Tr}}}

\def \Hom{\operatorname{Hom}}

\def \H{\operatorname{H}}

\def \Z{\mathbb{Z}}

\def \CP{\mathbb{CP}}

\newcommand {\emptycomment}[1]{}

\def\TP{\mathrm{TP}}

\def\B{\mathrm{B}}

\newcommand{\SO}{{\rm SO}}
\newcommand{\Spin}{{\rm Spin}}
\newcommand{\U}{{\rm U}}
\newcommand{\SU}{{\rm SU}}
\newcommand{\PSU}{{\rm PSU}}
\newcommand{\Pin}{{\rm Pin}}

\def\bZ{{\mathbf{Z}}}
\usepackage{centernot}

\newcommand{\Sec}[1]{Sec.~\ref{#1}} 

\usepackage{enumitem} %{enumerate}
\usepackage{mathtools,amssymb,varwidth}
\usepackage[utf8]{inputenc}

\newcommand{\Table}[1]{Table \ref{#1}}

\usepackage{datetime}

\newcommand{\rT}{{\rm T}}

\newcommand{\rF}{{\rm F}}

\newcommand{\GL}{{\rm GL}}

\newcommand{\PD}{{\rm PD}}

\def\ra{\mathrm{a}}
\def\rb{\mathrm{b}}
\def\rc{\mathrm{c}}
\def\rd{\mathrm{d}}
\def\re{\mathrm{e}}

\newcommand{\SM}{{\rm SM}}

\newcommand{\rA}{{\rm A}}
\newcommand{\q}{{\rm q}}

\newcommand{\bT}{{\mathbf T}}

\def\GCS{\mathrm{GCS}}
\def\ch{\mathrm{ch}}

\begin{document}

%\begin{titlepage}

\title{Categorical Symmetry of the Standard Model
from Gravitational Anomaly}
%{B+L-and-Higher-Symmetry Anomaly}

%%%%%%
\author{Pavel Putrov}
\email[]{putrov@ictp.it}
\affiliation{The Abdus Salam International Centre for Theoretical Physics, %Strada Costiera 11, 
Trieste 34151, Italy}
 
%%%%%%
\author{Juven Wang}
\email[]{jw@cmsa.fas.harvard.edu}
%\href{http://sns.ias.edu/~juven/}{http://sns.ias.edu/$\sim$juven/} 
%\homepage{http://sns.ias.edu/~juven/}
\affiliation{Center of Mathematical Sciences and Applications, Harvard University, MA 02138, USA}

\begin{abstract}

In the Standard Model, %the $\bf B - L$ 
some combination of 
the baryon $\bf B$ and lepton $\bf L$ number symmetry is free of mixed anomalies with strong and electroweak $su(3) \times su(2) \times u(1)_{\tilde Y}$ gauge forces. However, it
can still suffer from a mixed gravitational anomaly, 
hypothetically pertinent to leptogenesis in the very early universe. 
This happens when the \cred{total} ``sterile right-handed'' 
neutrino number $n_{\nu_R}$ is not equal to the family number $N_f$. 
Thus the invertible $\bf B - L$ symmetry current conservation 
can be violated quantum mechanically by gravitational backgrounds such as gravitational instantons. 
In specific, we show that a 
noninvertible categorical %counterpart of the $\bf B - L$ generalized symmetry 
$\bf B - L$ generalized symmetry
still survives 
in gravitational backgrounds.
In general, we propose a construction of 
noninvertible symmetry charge operators as topological defects derived from 
invertible anomalous symmetries that suffer from mixed gravitational anomalies. 
Examples include the perturbative local and nonperturbative global anomalies classified by $\mathbb{Z}$ and $\mathbb{Z}_{16}$ respectively.
For this construction, we utilize the anomaly inflow bulk-boundary correspondence,
\cred{the 4d Pontryagin class and the gravitational Chern-Simons 3-form,}
the 3d Witten-Reshetikhin-Turaev-type topological quantum field theory \cred{with a framing anomaly} 
corresponding to a 2d rational conformal field theory
with an appropriate \cred{rational} chiral central charge,
and the 4d \cred{$\mathbb{Z}_4^{\rm TF}$}-time-reversal symmetric topological superconductor with 
3d boundary topological order.

\end{abstract}

%\pacs{}

\maketitle

\tableofcontents

\section{Introduction and Summary}

\subsection{Introduction and the Plan}

The Standard Model (SM) \cite{Glashow1961trPartialSymmetriesofWeakInteractions, Salam1964ryElectromagneticWeakInteractions, Salam1968, Weinberg1967tqSMAModelofLeptons} 
has a specific combination of 
the baryon {\bf B} and lepton {\bf L} number symmetry, known as a continuous $\U(1)_{\bf B - L}$, preserved within the SM gauge interactions, 
thanks to the SM lagrangian interaction structure
and thanks to the $\U(1)_{\bf B - L}$ being mixed gauge anomaly-free 
with strong and electroweak gauge forces of Lie algebra 
$\mathcal{G}_{\rm SM} \equiv su(3) \times su(2) \times u(1)_{\tilde Y}$ \cite{Weinberg1996Vol2}.
In the past, the $\U(1)_{\bf B - L}$ symmetry current conservation is  
checked quantum mechanically by perturbative local anomalies, 
captured by Feynman graphs (see references in \cite{Weinberg1996Vol2}).
The $\U(1)_{\bf B - L}$ symmetry preservation
is a remarkable fact  because of the following reason.
In 4d spacetime, the U(1) symmetry current of a single Weyl fermion number alone
is known to suffer from Adler-Bell-Jackiw (ABJ)
perturbative local anomaly \cite{Adler1969gkABJ, Bell1969tsABJ}
via triangle Feynman diagram calculations
with three vertices of $\U(1)^3$, $\U(1)$-$G^2$ and $\U(1)$-gravity$^2$,
through
abelian $\U(1)$ or 
nonabelian $G$ instantons \cite{BelavinBPST1975, tHooft1976ripPRL, JackiwRebbi:1976pf, 
CallanDashenGross:1976je}
and gravitational instantons \cite{Eguchi:1976db, AlvarezGaume1983igWitten1984},
characterized respectively by Chern class \cite{chern1946characteristicAOM} and Pontryagin class \cite{Pontryagin1947,milnor1974characteristic}.
Recently,
thanks to the development of
cobordism classifications of bulk topological phases and their boundary anomalies
(\cite{Kapustin2014tfa1403.1467, 1405.7689, Kapustin1406.7329, 
2016arXiv160406527F, WanWang2018bns1812.11967, Witten2019bou1909.08775} and references therein) 
via the classic anomaly inflow idea \cite{1984saCallanHarvey},
both perturbative local anomalies and nonperturbative global anomalies
in the SM have been checked systematically, 
via the cobordism group supplemented with quantum field theory (QFT) calculations
\cite{Freed0607134, GarciaEtxebarriaMontero2018ajm1808.00009, WangWen2018cai1809.11171, 
DavighiGripaiosLohitsiri2019rcd1910.11277, WW2019fxh1910.14668, 
JW2006.16996, JW2008.06499, JW2012.15860, WangWanYou2112.14765, WangWanYou2204.08393}.
However, the $\bf B - L$ symmetry 
suffers from a mixed gravitational anomaly,
when the total ``sterile right-handed'' 
neutrino number $n_{\nu_R}$ is not equal to the family number $N_f$.
Namely, the invertible $\bf B - L$ symmetry current or charge conservation 
can be violated by gravitational backgrounds under curved spacetime geometries
or by gravitational instantons. 
{Phenomenological applications of this mixed $\bf B - L$-gravitational anomaly
include the gravitational leptogenesis \cite{Davidson0802.2962:2008bu,PhysRevLett.96.081301, Adshead1711.04800:2017znw} and beyond the Standard Model (BSM) new exotic sectors 
\cite{JW2006.16996, JW2008.06499, JW2012.15860, WangWanYou2112.14765, WangWanYou2204.08393}
obtained from canceling this gravitational anomaly}.                                                                                                                                                                                                                                                                                                                                                                                                                                                   

Although physicists had confirmed at least $N_f=3$ families of quarks and leptons by experiments
\cite{Harari:1977kv,ParticleDataGroupPDG2022},
we do not yet identify the detailed properties of sterile neutrinos, nor know how many $n_{\nu_R}$ there are in nature \cite{ParticleDataGroupPDG2022}.
Following the set-up advocated in 
\cite{JW2006.16996, JW2008.06499, JW2012.15860, WangWanYou2112.14765, WangWanYou2204.08393},
the index $-N_f+n_{\nu_{R}}$ counting the difference between the family and the total right-hand neutrino number 
 will become important.
As we will review in \Sec{sec:SM-Polynomial}
that a nonzero $-N_f+n_{\nu_{R}}$ implies nontrivial \emph{perturbative local anomalies}, classified by $\Z^2$ and captured by small gauge-diffeomorphism transformations
via the ABJ %Adler-Bell-Jackiw
triangle Feynman diagram \cite{Adler1969gkABJ, Bell1969tsABJ}
with three vertices of $\U(1)_{\bf B - L}^3$ and $\U(1)_{\bf B - L}$-gravity$^2$ types.  
When the continuous ${\bf B - L}$ symmetry 
is combined with the $\tilde Y$ electroweak hypercharge gauge symmetry and then
restricted to a discrete $\Z_{4,X}$ subgroup, 
where $X \equiv 5({ \mathbf{B}-  \mathbf{L}})-\frac{2}{3} {\tilde Y}$ 
with properly integer quantized hypercharge $\tilde Y$ \cite{Wilczek1979hcZee, WilczekZeePLB1979},
the aforementioned perturbative local anomaly classified by $\Z^2$ 
becomes a nonperturbative global anomaly classified by $\Z_{16}$
\cite{GarciaEtxebarriaMontero2018ajm1808.00009}.
All the quarks and leptons have a unit charge 1 under $\Z_{4,X}$ 
(see \Table{table:SMfermion} in Appendix \ref{app:rep}).
Thus, the index $-N_f+n_{\nu_{R}}$ mod 16 also implies a nontrivial 
\emph{nonperturbative global anomaly} classified by $\Z_{16}$ 
\cite{JW2006.16996, JW2008.06499, JW2012.15860, WangWanYou2112.14765, WangWanYou2204.08393} and captured by the large gauge-diffeomorphism transformations.

In this work, 
specifically,
we reinterpret the %familiar 
nonconservation of the invertible $\bf B - L$ symmetry current due 
to a gravitational background
as the replacement of the Noether charge operators by their noninvertible analogs. 
\emph{Noninvertible categorical symmetry} 
\cite{Simons-Collaboration-Categorical}
is a concept growing out of the recent development on 
the generalized global symmetry \cite{Gaiotto2014kfa1412.5148} 
(see reviews \cite{McGreevy2204.03045,Cordova2205.09545}).
\cred{For the terminology on the measurement of any global symmetry,
there is a \emph{charge operator} that measures,
while there is also a \emph{charged object} that is being measured.}
Ref.~\cite{Gaiotto2014kfa1412.5148} emphasize anew that the symmetry charge operator $U$
is a topological defect. 
A topology-preserving deformation of $U$
around a relatively charged object $\CO$ would not affect the measurement of the symmetry charge.
While an ordinary global symmetry with a group $G$ 
implies that the fusion rules of the symmetry charge operators (a.k.a. topological defects)
is described by the corresponding group law,
there are also symmetries with charge operators that obey fusion rules described by a fusion category that goes beyond an ordinary group.
These symmetries are called \emph{noninvertible symmetries} 
(since some of the charge operators do not have inverse operators)
or \emph{categorical symmetries} 
(since charge operators form a category).
While fusion categories of topological defects in 2d conformal field theories (CFT) 
were appreciated long ago 
\cite{Verlinde:1988sn,Petkova:2000ip,Frohlich:2004ef, Frohlich:2006ch, Frohlich:2009gb},
Ref.~\cite{Bhardwaj:2017xup, Tachikawa2017gyf1712.09542, Chang:2018iay, ThorngrenWang1912.02817, KongLanWenZhangZheng2005.14178, 
KomargodskiOhmoriRoumpedakisSeifnashri2008.07567, ThorngrenWang2106.12577} 
advocate their noninvertible symmetry nature.  
Only recently, noninvertible symmetries have been explored more systematically
in higher spacetime dimensions,
see selective references \cite{ChoiCordovaHsinLamShao2111.01139, KaidiOhmoriZheng2111.01141, WangYou2111.10369GEQC, Choi2205.05086:2022jqy, Cordova2205.06243:2022ieu, 
Cordova2211.07639:2022fhg, CordovaKoren2212.13193:2022qtz} relevant for the 4d SM or BSM
context, 
see selective mathematically grandiose encyclopedic references
\cite{MonteroRudelius2104.07036, 
Bhardwaj2204.06564:2022yxj, Bhardwaj2208.05973:2022lsg, Bartsch2208.05993:2022mpm,
Freed2209.07471:2022qnc, Bhardwaj2212.06159:2022kot, Bhardwaj2212.06842:2022maz, Bartsch2212.07393:2022ytj,
Heckman2212.09743:2022xgu},
and references therein. 
Importantly Ref.~\cite{Choi2205.05086:2022jqy, Cordova2205.06243:2022ieu} shows that
although the invertible $\U(1)$ symmetry can be broken by the dynamical $\U(1)'$ gauge theory
via the $\U(1)$-$\U(1)'^2$ ABJ perturbative local anomaly \cite{Adler1969gkABJ, Bell1969tsABJ}
\cred{(such as the axial $\U(1)=\U(1)_{\rm A}$ symmetry of the \cpurple{vector-gauged} 
$\U(1)'=\U(1)_{\rm V}$ quantum electrodynamics (QED))}, 
a subgroup of the broken invertible $\U(1)$ can be revived as a noninvertible symmetry. Namely, it is the subgroup of elements $\e^{\ii \alpha}\in \U(1)$ such that 
\begin{equation}
    \alpha = 2\pi p/N  \in 2\pi\cdot (\mathbb{Q}/\mathbb{Z}) \subset 
2\pi\cdot (\mathbb{R}/\mathbb{Z})  \cong \U(1)
\end{equation}
 for some integers $p$ and $N$ (one can always assume that $\alpha\in [0,2\pi)$, so that $N>p\geq 0$, and that $p$ and $N$ are coprime). That is
the rational $\mathbb{Q}/\mathbb{Z}$ 
part of the original $\mathbb{R}/\mathbb{Z}\cong \U(1)$ invertible symmetry
is revived as a noninvertible symmetry,
meaning that the modified symmetry charge operators beget noninvertible fusion rules.

For the invertible U(1) symmetry,
there is a one-to-one correspondence between the elements 
$\alpha\in 2\pi\cdot (\R/\Z) \cong \R/(2 \pi \Z) \cong \U(1)$ \cred{and} the invertible symmetry charge operators $U_\alpha$, with the fusion corresponding to the group binary operation $\alpha_1+\alpha_2 \in 2\pi\cdot (\R/\Z)\cong \U(1)$:   
\begin{equation}
U_{\alpha_1} \,  U_{\alpha_2} =  U_{\alpha_1 + \alpha_2}.
\end{equation}

For the full (i.e. closed under fusion) noninvertible symmetry, however,
there is no longer a one-to-one correspondence between $\Q/\Z$ group elements and the topological operators. The operators however can be labelled by elements of a certain commutative monoid $\mathfrak{M}$, such that the noninvertible fusion rules correspond to the monoid's binary operation and there is surjective homomorphism of monoids $\mathfrak{M}\rightarrow \Q/\Z$ \cite{Putrov2208.12071:2022pua}.

We will encounter an analogous structure in our setup with gravitational anomalies. The plan of this article goes as follows:

In \Sec{subsec:Summary}, we outline and summarize our strategy and interpretations
in a friendly and nontechnical way.

In \Sec{sec:SM-Polynomial}, we recall and setup 
the 4d SM, its anomaly associated with the quark number $\U(1)_{{\bf Q}}$
and lepton number $\U(1)_{{\bf L}}$ symmetry (whose combination gives the ${\bf B - L}$),
and the anomaly associated with the spacetime diffeomorphisms or, equivalently, gravity.
We will write down the 4d anomaly in terms of 
a 5d invertible topological field theory (iTFT)\footnote{We shall call an 
invertible topological quantum field theory simply as an invertible topological field theory (iTFT),
because the iTFT can be written in terms of a partition function of 
the classical non-dynamical background field.}, 
or 6d anomaly polynomial.
We put the emphasis on the two integers, belonging to the $\Z^2$ group that classifies local anomalies 
(the $\U(1)_{\bf B - L}^3$ pure gauge anomaly and $\U(1)_{\bf B - L}$-gravity$^2$ mixed gauge-gravity anomaly), and also $\nu\in\Z_{16}$ that classifies global anomalies 
(the mixed $\Z_{4,X}$-gravity anomaly).  
This story will turn out to match exactly the cobordism results previously obtained in 
\cite{WangWanYou2112.14765, WangWanYou2204.08393}.

In \Sec{sec:U1-grav}, we discuss the construction of 
the noninvertible categorical symmetry topological defects from 
the mixed $\U(1)$-gravitational anomaly and the pure $\U(1)$ anomaly.

In \Sec{sec:Z4-grav}, we discuss the construction of 
the noninvertible categorical symmetry topological defect from 
the mixed $\Z_{4}$-gravitational anomaly classified by $\Z_{16}$.

In \Sec{sec:Conclusion}, we conclude with final remarks. We enlist future directions pertinent to the leptogenesis, baryogenesis, and
possible BSM implications of the theoretical proposals on
replacing the right-handed neutrinos with interacting topological quantum field theory
(TQFT) or CFT sectors together \cite{JW2006.16996, JW2008.06499, JW2012.15860}.

In Appendix \ref{app:rep}, for the reader's convenience, 
we gather the representations of Weyl fermions
in various gauge or global symmetries,
the SM's $su(3) \times su(2) \times u(1)_{\tilde Y}$,
\cred{the vector} $\U(1)_{{\bf Q}-N_c {\bf L}}$ 
(the precise form of $\U(1)_{\bf B - L}$ with properly quantized charges, with the color number $N_c=3$), 
\cred{the vector} $\Z_{2N_cN_f, {{\bf Q} + {N_c} {\bf L}} } \subset \U(1)_{{\bf Q} + {N_c} {\bf L}}$ 
(the precise form of $\Z_{2 N_f,\bf B + L} \subset \U(1)_{\bf B + L}$ 
with properly quantized charges), 
\cred{the chiral} $\Z_{4,X}$ symmetry, and others.

\cred{In Appendix \ref{app:Notations-and-Conventions},
we review the notations and conventions \cpurple{about characteristic classes and their differential form representatives}.
}

In Appendix \ref{app:anomaly-polynomial-quantization} we review the classification of anomalies for 
$\Spin \times \U(1)$, 
$\frac{\Spin\times \U(1)}{{\Z_2^\rF}} \equiv \Spin\times_{\Z_2^\rF}\U(1) \equiv \Spin^c$, 
and $\SO \times \U(1)$
symmetries in 4d \cite{WanWang2018bns1812.11967}
in terms of degree 6 anomaly polynomial and its relation to the classification of anomalies for 
$\frac{\Spin\times \Z_4}{{\Z_2^\rF}} \equiv 
\Spin\times_{\Z_2^\rF} \Z_4$ symmetry.

\subsection{Summary}

\label{subsec:Summary}

In this work, 
we show that a noninvertible counterpart 
of certain kinds of mixed-gravitational anomalous
symmetry still survives 
in gravitational backgrounds.
Below are some strategies, steps, and interpretations that we will take to achieve that goal.

\begin{enumerate}

\item We will {flourish} further the general idea about the trade-off between anomalies and noninvertability of symmetries. A presence of anomaly of a global $p$-form symmetry in a $d$-dimensional QFT implies that the na\"ive (i.e. classically defined) extended charge operators of dimension $d-p-1$ are no longer topological\footnote{In general, to capture the full anomaly one may need to consider networks of the charge operators and the deformations involving the moves of the network. This, in particular, will be relevant for the construction in Sec. \ref{sec:Z4-grav}.} or require some additional noncanonical choice to be unambiguously defined. However, one can consider modifying the charge operator by introducing a $(d-p-1)$-dimensional 
topological quantum field theory
(TQFT, typically we mean noninvertible TQFT) 
supported on its worldvolume and coupled in a nontrivial way to bulk fields. If the TQFT itself has an anomaly, its partition function may also change when the charge operator is deformed, or it may also require some noncanonical choice to be made. It then can happen that such a pathology of the TQFT cancels the pathology of the na\"ive charge operator and together they will form a well-defined topological defect. However, for a TQFT to have an anomaly, it must be noninvertible. Therefore the new topological effects will be noninvertible as well. A version of such construction was in particular initiated in \cite{Choi2205.05086:2022jqy,Cordova2205.06243:2022ieu}.

\item There are two types of anomalies involving $\U(1)$ global symmetry in 4d:
\begin{enumerate}
    \item Pure $\U(1)$ anomaly, with the  \cpurple{following} corresponding \cred{term in the degree 6 anomaly polynomial}
    \begin{equation}
        I_6=\kappa_1 \,\frac{c_1^3}{3!}+\ldots\equiv \kappa_1\, \frac{F^3}{3!\,(2\pi)^3}+\ldots
    \end{equation}
    where $F=\dd A$ is the field strength \cred{2-form} of the corresponding \cred{1-form} $\U(1)$ gauge field $A$ and $c_1\equiv F/(2\pi)$ is the Chern-Weil representative of the first Chern class of the $\U(1)$ principle bundle. 
    The wedge product $\wedge$ \cpurple{is implicit}. 
    This anomaly implies that the $\U(1)$ symmetry current 1-form $j$ is not conserved in a general background gauge field configuration:
    \begin{equation}
        \dd \star j=\kappa_1\frac{F^2}{8\pi^2}+\ldots 
    \end{equation}
    In particular, in the presence of such anomaly, the $\U(1)$ symmetry cannot be \cred{dynamically} gauged 
    \cpurple{(within just the original 4-dimensional spacetime)}, 
    i.e. the corresponding gauge field cannot be made dynamical. On the other hand, 
    \cred{unlike ABJ mixed anomaly \cite{Adler1969gkABJ, Bell1969tsABJ}},
    such an anomaly is typically \emph{not} considered to be breaking $\U(1)$ as a global symmetry, as the current is still conserved in the trivial background. 
    
    In the nontrivial background however, the nonconservation of the current implies that the corresponding 3-dimensional charge operators are no longer topological. As we will consider in more detail later, their topological but noninvertible counterparts can be constructed essentially in the same way as it was done in \cite{Choi2205.05086:2022jqy,Cordova2205.06243:2022ieu} in the case of ABJ mixed anomaly between a global $\U(1)$ and a different gauged $\U(1)'$. The difference is that in our case the second $\U(1)$ is not gauged and identified with the first $\U(1)$. According to the general prescription outlined above, the new operators are constructed by introducing a 3d TQFT supported on the worldvolume of the original charge operators and coupled to the bulk background $\U(1)$ gauge field. This construction works for the charge operators corresponding to the torsion elements of $\U(1)$, i.e. the operators realizing rotations by fractions of the full $\U(1)$ rotation.

    Let us also note that if $\Z_2^\rF\subset \U(1)$ (which is the case of $\mathbf{B-L}$ symmetry considered in this work), the 4d theory can be considered on a nonspin spacetime manifold $M$. 
    In that case, the background $\U(1)$ gauge field is necessarily nontrivial (in particular, for the corresponding first Chern class, we must have $2 c_1=w_2(TM)\neq 0 \mod 2$,
    where $w_j(TM)$ is the $j$-th Stiefel-Whitney class of tangent bundle $TM$, see App. 
    \ref{app:anomaly-polynomial-quantization} for review). 

    \item Mixed gravitational anomaly,  with the \cpurple{following} corresponding \cred{term in the degree 6 anomaly polynomial}:
    \begin{equation}
        I_6=\kappa_2 \,{c_1p_1}+\ldots\equiv -\kappa_2\, \frac{1}{\,2(2\pi)^3}F\cred{\wedge}\Tr[R\wedge R] +\ldots
    \end{equation}
    where $R=\dd \omega +\omega\wedge \omega$ is the curvature 2-form of the Levi-Civita spin-connection 1-form $\omega$ and $p_1=-\Tr[ R\wedge R]/(8\pi^2)$ is the standard representative of the first Pontryagin class. \cred{Note that here and in the rest of the article, we consider Euclidean spacetime, so that $\omega$ is a ${so}(4)$ valued connection 1-form. However, we will comment later on how the constructed topological defects should be modified in the case of Lorentzian spacetime.}  The anomaly implies the following nonconservation of the $\U(1)$ current:
    \begin{equation}
        \dd \star j=-\frac{\kappa_2}{8\pi^2}\Tr[ R\wedge R]+\ldots 
    \end{equation}
    Note that in principle one can get rid of the term in the right-hand side by introducing a local counterterm \cite{AlvarezGaume1983igWitten1984}. However, it will break the general covariance of the theory. Since we expect that Standard Model can be coupled to gravity in a consistent way, we will assume the absence of such a counterterm. 
    
    This type of anomaly ordinarily is also not considered to be breaking the $\U(1)$ symmetry. The current is still conserved on a flat spacetime. Unlike in the previous case, however, we do expect the spacetime to be curved in a physical theory due to gravitational effects. Such an anomaly in particular plays a crucial role in gravitational leptogenesis \cite{PhysRevLett.96.081301}.

    On a curved spacetime, the presence of such anomaly implies that the na\"ive $\U(1)$ charge operators will not be topological anymore and therefore the corresponding charge will not be conserved. In particular, the change of the total charge in some time interval is given by
    \begin{equation}
        \Delta Q= -\frac{\kappa_2}{8\pi^2} \int_{\Delta M^4}{\Tr[ R\wedge R]}
    \end{equation}
    where $\Delta M^4$ is the spacetime between the initial and final time slices.

    As we will show later, the charge operators can be again modified to be topological, at the cost of losing invertibility. According to the general prescription outlined above this is done by introducing a 3d TQFT which is supported on the worldvolume of the defect and coupled to the bulk gravity via the framing anomaly. Existence of such extended topological operators can be interpreted as a certain modified charge conservation law.

When the spacetime topology and metric are dynamical \cred{in a UV-complete theory}, in principle one expects no global symmetries at all in quantum gravity \cite{KalloshKLLS9502069,Banks2010zn1011.5120,HarlowOoguri1810.05338, McNamara1909.10355:2019rup}, including noninvertible ones \cite{MonteroRudelius2104.07036}. Our construction then shows that if a $\U(1)$ symmetry has a mixed gravitational anomaly, it does not become completely broken in quantum gravity just by this anomaly. Rather, it should be either (1) broken by some other method or (2) dynamically gauged in the UV-complete theory. 
    
\end{enumerate}

When the anomalies of both types are present, the construction of the noninvertible counterparts to the na\"ive charge operators can be combined by stacking together the two anomalous 3d TQFTs used the individual cases. 

\end{enumerate}

\section{Standard Model: 4d Anomaly, 5d Invertible Phase, %Field Theory 
 and 6d Polynomial}
 \label{sec:SM-Polynomial}

Standard Model (SM) \cite{Glashow1961trPartialSymmetriesofWeakInteractions, Salam1964ryElectromagneticWeakInteractions, Salam1968, Weinberg1967tqSMAModelofLeptons}
is a 4d chiral gauge theory with Yang-Mills spin-1 gauge fields of
the  Lie algebra 
\bea \label{eq:SMLieAlgebra}
\cG_{\rm SM} \equiv su(3) \times  su(2) \times u(1)_{\tilde Y}
\eea
coupling to $N_f=3$ families of 15 or 16 Weyl fermions (spin-$\frac{1}{2}$ Weyl spinor 
is in the ${\bf 2}_L$ representation \cpurple{of} the spacetime symmetry Spin(1,3),
written as a left-handed 15- or 16-plet $\psi_L$)
in the following $\cG_{\rm SM}$ representation
\begin{multline}
    \label{eq:SMrep}
({\psi_L})_{\rm I} =
( \bar{d}_R \oplus {l}_L  \oplus q_L  \oplus \bar{u}_R \oplus   \bar{e}_R  
)_{\rm I}
\oplus
n_{\nu_{{\rm I},R}} {\bar{\nu}_{{\rm I},R}}
\\
\sim 
\big((\overline{\bf 3},{\bf 1})_{2} \oplus ({\bf 1},{\bf 2})_{-3}  
\oplus
({\bf 3},{\bf 2})_{1} \oplus (\overline{\bf 3},{\bf 1})_{-4} \oplus ({\bf 1},{\bf 1})_{6} \big)_{\rm I}
\oplus n_{\nu_{{\rm I},R}} {({\bf 1},{\bf 1})_{0}}
\end{multline} 
for each family. Both of the left-handed particles, 
$q_L$ and $l_L$, are the weak force SU(2) doublets, for quarks and leptons respectively.
The right-handed anti-particles,
up quark $\bar{u}_R$, down quark $\bar{d}_R$, neutrino ${\bar{\nu}_{R}}$, and electron $\bar{e}_R$ are the  weak force SU(2) singlets.
There is also a spin-0 Higgs scalar $\phi$ in $({\bf 1},{\bf 2})_{3}$. 
Hereafter the family index is denoted by \cpurple{symbols in roman font} ${\rm I},{\rm J}=1,2,3$; 
with ${\psi_L}_1$ for $u,d,e$ type,
${\psi_L}_2$ for $c,s,\mu$ type,
and 
${\psi_L}_3$ for $t,b,\tau$ type of quarks and leptons.
We use ${\rm I}=1,2,3$ for $n_{\nu_{e,R}}, n_{\nu_{\mu,R}}, n_{\nu_{\tau,R}} \in \{ 0, 1\}$
to label either the absence or presence of electron $e$, muon $\mu$, or tauon $\tau$ types of sterile neutrinos (i.e., ``right-handed'' neutrinos sterile to $\cG_{\rm SM}$ gauge forces).
Below we consider $N_f$ families 
(typically $N_f=3$) of fermions (including quarks and leptons),
and sterile neutrinos of the total number $n_{\nu_{R}} \equiv \sum_{\rm I} n_{\nu_{{\rm I},R}}$    
{which can be equal, smaller, or larger than 3 (here ${\rm I}=1,2,3,\dots$ for $e,\mu,\tau,\dots$ type of neutrinos)}.
Following the set-up in \cite{WangWanYou2112.14765, WangWanYou2204.08393},
the index counting the difference between the family number and the right-hand neutrino number 
is important:
\bea
-N_f+n_{\nu_{R}} \equiv 
-N_f+
\sum_{\rm I} 
 n_{\nu_{{\rm I},R}}
= -3+n_{\nu_{e,R}} +  n_{\nu_{\mu,R}} + n_{\nu_{\tau,R}} + \dots.
\eea

The SM action \cred{as a real scalar on a curved spacetime 
(pseudo-)Riemannian 4-manifold $M^4$ with a metric ${\rm g}_{\mu \nu}$ and its determinant ${\rm g}$
} schematically reads 
\be
 \label{eq:SM-action}
S_{\rm SM} \equiv \int_{M^4} \big(\sum_{{ I}=1,2,3} \frac{-1}{g_I^2}\Tr[F_I \wedge \star F_I] + 
\dd^4 x \cred{\sqrt{|{\rm g}|}}  ({\psi}^\dagger_L  (\ii \bar  \sigma^\mu {D}_{\mu, A} ) \psi_L
 -( {\psi}^\dagger_L \phi \psi_R +{\rm h.c.})
+ | {D}_{\mu, A}\phi |^2 
-{\rm U}(\phi)) +\dots \big)
\equiv \int_{M^4} \cred{\sqrt{|{\rm g}|}} \dd^4 x \, \hat{\cL}_{\rm SM}, 
\ee
\cred{see Appendix \ref{app:Notations-and-Conventions} for notation conventions.}
The SM lagrangian \cred{scalar} $\cred{\hat{\cL}}_{\rm SM}$ contains
the Yang-Mills spin-1 gauge field term 
$-\frac{1}{4 {g_I^2}} F_{{I},\mu\nu}^\ra F_{{I}}^{\ra \mu\nu}$
with the field strength $F_I$ (with gauge sector indices in \emph{italic} ${I}=1,2,3$ for $u(1),su(2),su(3)$), 
the Weyl spin-$\frac{1}{2}$ fermions coupled to the Yang-Mills gauge fields,
Yukawa-Higgs term, and 
the electroweak Higgs kinetic-potential term of spin-0 Higgs scalar $\phi$.
The ``$\dots$'' includes a possible theta term for $su(3)$ with nearly zero theta-angle. 
\cred{The $\bar\sigma^\mu \equiv \bar{\hat \sigma}^a e^\mu{}_a$ 
\cpurple{is the generalized sigma matrix
in the curved spacetime, with the vielbein $e^\mu{}_a$,
relating it to the standard generators of the algebra of 2-by-2 matrices} 
$\bar{\hat \sigma}^a = (\hat \sigma^0, - \hat \sigma^1, - \hat \sigma^2,- \hat \sigma^3)$.
The ${D}_{\mu, A}$ contains a covariant derivative $\nabla_\mu$ involving
\cpurple{Levi-Civita spin-connection when acting on  a spinor field} $\psi$.}
The $\cred{\hat{\cL}}_{\rm YH}={\psi}^\dagger_L \phi \psi_R+{\rm h.c.}$ is a shorthand of
$\cred{\hat{\cL}}_{\rm YH}^d
+\cred{\hat{\cL}}_{\rm YH}^u
+\cred{\hat{\cL}}_{\rm YH}^e=\lambda^{d}_{\rm IJ} {{q}^{\rm I \dagger}_L} \phi d_R^{\rm J} 
  +\lambda^{u}_{\rm IJ} \epsilon^{ab} {{q}^{\rm I \dagger}_{L a}}\phi_b^* u_R^{\rm J}
  +\lambda^{e}_{\rm IJ} {{l}^{\rm I \dagger}_L} \phi e_R^{\rm J}+{\rm h.c.}$
  with $a,b$ labeling the component of  $su(2)$ fundamental representation, and the ``h.c.'' standing for the hermitian conjugate.
Diagonalization of Yukawa-Higgs term of the quark sector implies {that} the $W^{\pm}$ boson induces a flavor-changing current mixing  different families,
thus we only have a $\U(1)_{\bf Q}$ quark symmetry for all quarks (instead of {an} individual $\U(1)$ for each quark family), at least \emph{semiclassically}.
The diagonalization of Yukawa-Higgs term of the lepton sector without neutrino mass term
$\cred{\hat{\cL}}_{\rm YH}^{\nu}=\lambda^{\nu}_{\rm IJ} \epsilon^{ab} {{l}^{\rm I \dagger}_{L a}}\phi_b^* \nu_R^{\rm J}+{\rm h.c.}$
implies that $\cred{\hat{\cL}}_{\rm SM}$ has individual lepton $\U(1)_{{\bf L}_e}$, $\U(1)_{{\bf L}_\mu}$, $\U(1)_{{\bf L}_\tau}$ symmetries for each lepton family.
However, \cblue{established experiments show that each lepton $\U(1)$ symmetry is violated such as by neutrino oscillations 
\cite{ParticleDataGroupPDG2022, Super-Kamiokande:1998kpq9807003, SNO:2002tuh0204008, KamLAND:2002uet0212021}, 
only the total lepton number $\U(1)_{\bf L}$ should be considered, at least \emph{semiclassically}.
}

Thus, we can focus on $\U(1)_{\bf Q}$ and $\U(1)_{\bf L}$
transformations:
\bea \label{eq:U1QL}
{\U(1)_{\bf Q}}: ({\psi_L})_{\rm I}  &\mapsto&
((\e^{-\ii {\alpha_{\bf Q}}} \mathbb{I}_3 \cdot \bar{d}_R) \oplus
 {l}_L  \oplus 
(\e^{\ii {\alpha_{\bf Q}}} \mathbb{I}_6  \cdot q_L)  \oplus 
(\e^{-\ii {\alpha_{\bf Q}}} \mathbb{I}_3  \cdot \bar{u}_R) \oplus   \bar{e}_R 
)_{\rm I}
\oplus
n_{\nu_{{\rm I},R}} {\bar{\nu}_{{\rm I},R}},\cr
\U(1)_{\bf L}: 
({\psi_L})_{\rm I}  &\mapsto&
(\bar{d}_R \oplus  (\e^{\ii {\alpha_{\bf L}}} \mathbb{I}_2 \cdot {l}_L)  \oplus q_L  \oplus \bar{u}_R \oplus   (\e^{-\ii {\alpha_{\bf L}}}\bar{e}_R)
)_{\rm I}
\oplus
 (\e^{-\ii {\alpha_{\bf L}}} n_{\nu_{{\rm I},R}} {\bar{\nu}_{{\rm I},R}}).
\eea

The quark's $\U(1)_{{\bf Q}}$ is related to baryon's $\U(1)_{\bf B}$ via ${\alpha_{{\bf Q}}} ={\alpha_{\bf B}}/N_c={\alpha_{\bf B}}/3 \in [0, 2 \pi)$.
Here $\mathbb{I}_{\rm N}$ means a rank-N  identity matrix that can act on an N-plet.
To have properly quantized charges, we shall consider the linear combination of $\U(1)_{{\bf Q}}$ and $\U(1)_{{\bf L}}$, see \Table{table:SMfermion}.
So what one may informally call the $\U(1)_{{\bf B-L}}$ or $\U(1)_{{\bf B+L}}$ symmetry
mathematically really means the
$\U(1)_{{\bf Q}-N_c {\bf L}}$ or $\U(1)_{{\bf Q}+N_c {\bf L}}$ symmetry
that has properly quantized integer charges.
\Refe{KorenProtonStability2204.01741} and \cite{WangWanYou2204.08393} recap that
although $\U(1)_{{\bf Q}-N_c {\bf L}}$  
stays free from mixed gauge anomalies with SM gauge forces \emph{quantum mechanically},
the classical $\U(1)_{{\bf Q}+N_c {\bf L}}$ symmetry is broken \emph{quantum mechanically}
down to a discrete $\Z_{2N_cN_f, {{\bf Q} + {N_c} {\bf L}}}$ subgroup
(which is a finite abelian elementary group of order $2N_cN_f$ embedded inside $\U(1)_{{\bf Q}+N_c {\bf L}}$).\footnote{In Ref.~\cite{WangWanYou2204.08393},
the $\U(1)_{{\bf Q}-N_c {\bf L}}$ 
and $\Z_{2N_cN_f, {{\bf Q} + {N_c} {\bf L}}}$
are loosely speaking written as $\U(1)_{{\bf B}- {\bf L}}$,
and $\Z_{2N_f, {\bf B}- {\bf L}}$
respectively. Hereafter we have to be precise to 
have a proper charge quantization for any $\U(1)$ or $\Z_N$ symmetry.}

Let us write down the full \emph{invertible} spacetime-internal symmetry structure 
of the SM \cite{WangWanYou2112.14765, WangWanYou2204.08393}.
To specify the spacetime-internal symmetries of a theory, we follow Freed-\cpurple{Hopkins'} notation \cite{2016arXiv160406527F} \cred{$\frac{G_1 \times G_2}{N} \equiv G_1 \times_N G_2$}
to write
\bea
G\equiv ({\frac{{G_{\text{spacetime} }} \ltimes  {{G}_{\text{internal}} } }{{N_{\text{shared}}}}}) \equiv {{G_{\text{spacetime} }} \ltimes_{{N_{\text{shared}}}}  {{G}_{\text{internal}} } }.
\eea
The semi-direct product $\ltimes$ specifies a group extension.
The ${N_{\text{shared}}}$ is the shared common normal subgroup symmetry between ${G_{\text{spacetime} }}$ and ${{G}_{\text{internal}} }$, 
e.g. ${N_{\text{shared}}}$ can be the fermion parity symmetry $\Z_2^\rF$, which acts on fermions by $\psi \mapsto - \psi$.
The Lie algebra of the internal symmetry of SM is $\cG_{\rm SM}$,
but the global structure of Lie group $G_{\text{SM}_\q}$ has four possible versions \cite{AharonyASY2013hdaSeiberg1305.0318, Tong2017oea1705.01853, Wan2019sooWWZHAHSII1912.13504,  AnberPoppitz2110.02981}
all compatible with the SM matter field representation \eq{eq:SMrep}:
\bea
G_{\SM_\q} \equiv \frac{\SU(3) \times   \SU(2) \times \U(1)_{\tilde Y}}{\Z_\q},  \quad \text{ with } \q=1,2,3,6.
\eea

Following \cite{WangWanYou2112.14765, WangWanYou2204.08393},
if we treat the $G_{\SM_\q}$ as an internal global symmetry,
we shall consider the  spacetime-internal symmetry of SM as
\bea \label{eq:GSM-0form}
G={\Spin \times_{\Z_2^\rF} \U(1)_{{{\bf Q}} - {N_c}{\bf L}}  \times_{\Z_2^{\rF}} \Z_{2N_cN_f, {{\bf Q} + {N_c} {\bf L}} }  \times  G_{\SM_\q}}.
\eea
However, $G_{\SM_\q}$ is an SM dynamical gauge group, such that dynamically gauging it induces a generalized global symmetry \cite{Gaiotto2014kfa1412.5148}, 
including a 1-form electric symmetry and a 1-form magnetic symmetry, as
\cite{Wan2019sooWWZHAHSII1912.13504, AnberPoppitz2110.02981, WangYou2111.10369GEQC,WangWanYou2112.14765}
\bea \label{eq:GSM-1form}
G={\Spin \times_{\Z_2^\rF} \U(1)_{{{\bf Q}} - {N_c}{\bf L}}  \times_{\Z_2^{\rF}} \Z_{2N_cN_f, {{\bf Q} + {N_c} {\bf L}} }  \times \Z_{6/\q,[1]}^e \times \U(1)_{[1]}^m}.
\eea
Ref.~\cite{WangWanYou2112.14765, WangWanYou2204.08393} looks into the 4d SM's anomaly
via the 5d cobordism group TP$_5$ calculation.
Here instead we start from deriving the 6d anomaly polynomial.

As was described in \cite{AlvarezGaume1983igWitten1984,Alvarez-Gaume:1984zlq}, the anomaly polynomial of Weyl fermions 
can be computed using Atiyah-Singer index theorem.
The contribution of a single Weyl fermion in 4d is the degree 6 part of
$\hat{\rA} \, \ch(\mathcal{E})$ where $\hat{\rA}$ is the A-roof genus of the spacetime tangent bundle 
\cpurple{$TM$ over the base spacetime manifold $M$, expressed in terms of $j$-th Pontryagin classes $p_j$, while the ch is the total Chern character expressed in terms of $j$-th Chern classes $c_j$}, and
$\mathcal{E}$ is the complex vector bundle associated with the representation of the fermion. The explicit expression in terms of Pontryagin and Chern characteristic classes 
 \cite{chern1946characteristicAOM, Pontryagin1947,milnor1974characteristic}
 can be obtained using the expansions of $\hat{\rA}$
and $\ch(\mathcal{E})$: 
\bea
\label{eq:hatA}
\hat{\rA} &=&1-\frac{p_1}{24}+ \frac{7 p_1^2 - 4 p_2}{5760}+ \ldots,
\\
\label{eq:ch}
\ch(\mathcal{E})&=&\mathrm{rank}\,\mathcal{E}+c_1(\mathcal{E})+
    \frac{1}{2}\left(c_1^2(\mathcal{E})-2c_2(\mathcal{E})\right)+
    \frac{1}{6}\left(
(c_1^3(\mathcal{E})-3c_1(\mathcal{E})c_2(\mathcal{E})+3c_3(\mathcal{E})
    \right)+\ldots
\eea
and also using the properties $\ch(\mathcal{E}_1\oplus \mathcal{E}_2)=\ch(\mathcal{E}_1)+\ch( \mathcal{E}_2)$, $\ch(\mathcal{E}_1\otimes \mathcal{E}_2)=\ch(\mathcal{E}_1)\,\ch (\mathcal{E}_2)$. The explicit anomaly polynomial for the gauge, global, and diffeomorphism symmetries of the 4d SM, with the matter representation given in \eq{eq:SMrep}, reads\footnote{To obtain the polynomial coefficients correctly,
here we use the convention in Table \ref{table:SMfermion}
such that
every fermion is written as a left-handed Weyl spinor 
(left-handed particle $\psi_L$ or
right-handed anti-particle $\ii \sigma_2 \psi_R^*$). 
Every particle contributes $+1$ (e.g., $\psi_L$)
and every anti-particle contributes $-1$ (e.g., $\ii \sigma_2 \psi_R^*$), 
to the quark $\mathbf{Q}$ or lepton $\mathbf{L}$ number,
namely the integer charge representation of $\U(1)_\mathbf{Q}$ or $\U(1)_\mathbf{L}$.
}:
%$\cG_{\SM}$, $\U(1)_\mathbf{Q}$ and $\U(1)_\mathbf{L}$,
%and by also including also the Spin group diffeomorphism couplings to spacetime geometry and gravity.}
%This gives rise to a partition function
%$\exp(\ii  \theta \int_{M^6} I_6 )$ with
\begin{multline}
   \label{SM-6d-I6} 
I_6 \equiv
\left(N_c   c_1(\U(1)_\mathbf{Q})
+  c_1(\U(1)_\mathbf{L})
\right) N_f \left(\cred{-} 18 \,\frac{ c_1(\U(1)_{\tilde{Y}})^2}{2} \cred{-} c_2(\SU(2)) \right)  \\
+\cred{(N_f - n_{\nu_R})} \,\left(\frac{c_1(\U(1)_\mathbf{L})^3}{6}-\frac{c_1(\U(1)_\mathbf{L}) p_1(TM)}{24}\right), 
\; \;  
\end{multline}
%
%
%  \cred{JW: Are the coefficients of \eq{SM-6d-I6} off by $\pm$ signs previously}
%
where \cpurple{we abbreviate as $c_j(\mathcal{E}_G) \equiv c_j(G)$ the $j$-th Chern class of 
the vector bundle $\mathcal{E}_G$ associated with the defining representation of $G$,\footnote{\cred{\cpurple{More precisely,
the vector bundle $\mathcal{E}_G \equiv P \times_{\rho} V$ over $M$ is said to be associated with a principle $G$ bundle $P$ over $M$ and a representation $(V, \rho)$ of $G$, consisting of}
a vector space $V$ and a homomorphism $\rho:  G \to \GL(V,\rF)$ for a field $\rF$. Here $\rF=\C$ is complex.}} 
and $p_j(TM)$ is the $j$-th Pontryagin class of the spacetime tangent bundle $TM$. 
In the form more familiar to physicists, we have
$p_1 \coloneqq- \frac{1}{8 \pi^2}   \Tr[ R \wedge R]$ 
and $- c_2 +\frac{1}{2} c_1^2 \coloneqq \frac{1}{8\pi^2} \Tr({F}\wedge {F})$}.

When $M^6$ is a closed 6-manifold, then $\int_{M^6} I_6 \in \Z$,
\cred{\cpurple{and there is a 6d iTFT with the partition function $\exp(\ii \int  \theta I_6)$ where 
$\theta \in [0, 2 \pi)$}}.
When $M^6$ has a boundary $\prt M^6 = M^5$, 
 we can consider this $M^5$ as a 5d interface between two 6d bulks with the lagrangian density
$\theta  I_6$ such that $\theta=0$ 
on one 6d side and $\theta=2 \pi$ on the other 6d side.
On the $M^5$ interface, we have an \emph{invertible} topological field theory (iTFT)
with the action
$S_5= 2 \pi \int_{ M^5} I_5 \in 2 \pi \R$ from $I_6 = \dd I_5$. 
\cred{The 5d iTFT partition function is $\exp(\ii   S_5) \in \U(1)$}. 
\cred{The $S_5$} value modulo $2 \pi $ is independent of the choice of $M^6$.
The explicit 5d iTFT related in this way to the anomaly polynomial (\ref{SM-6d-I6}) reads
\bea\label{SM-5d-S5-iTFT} 
&&S_5 \equiv \int_{M^5}
   (N_c A_{\mathbf{Q}} + A_{\mathbf{L}} ) N_f \left( \cred{-} 18 \,\frac{ c_1(\U(1)_{\tilde{Y}})^2}{2} \cred{-} c_2(\SU(2))\right)
     +\cred{(N_f - n_{\nu_R})}\,A_{\mathbf{L}} \,\left(\frac{c_1(\U(1)_\mathbf{L})^2}{6}-\frac{p_1(TM)}{24}\right).       
\eea
Here $A_\mathbf{Q}$ and $A_\mathbf{L}$ are background gauge fields for $\U(1)_\mathbf{Q}$ and $\U(1)_\mathbf{L}$ symmetries respectively.
This 5d TQFT encodes the anomaly of the 4d SM by the standard anomaly inflow setup. 
Note that in principle, there is an ambiguity of adding a total derivative: $I_5\rightarrow I_5+\dd I_4$. Such a change corresponds to the addition of a counterterm $I_4$ to the action of the 4d theory. In 
\Eq{SM-5d-S5-iTFT}, we have made the choice that preserves gauge invariance for the 4d dynamical gauge fields and general covariance. 
\cred{See more discussions in Appendix \ref{app:Notations-and-Conventions}.}

Here are some comments on the symmetries and anomalies in 4d read from the 5d iTFT \eq{SM-5d-S5-iTFT}:
\begin{enumerate}

\item The coefficients of each term in \eq{SM-6d-I6} matches with the corresponding triangle Feynman diagram, based on the data of \Table{table:SMfermion} in Appendix \ref{app:rep}:\\
$\bullet$ $\U(1)_{{\bf Q}}$-$\U(1)_{\tilde{Y}}^2$ triangle diagram shows $(2 \cdot 1^2 - 2^2 - (-4)^2)N_c N_f=-18 N_c N_f$ for the $c_1(\U(1)_\mathbf{Q}) \frac{ c_1(\U(1)_{\tilde{Y}})^2}{2}$ coefficient.\\
$\bullet$ $\U(1)_{{\bf L}}$-$\U(1)_{\tilde{Y}}^2$ triangle diagram shows $(2 \cdot (-3)^2-6^2) N_f=-18 N_f$ for the $c_1(\U(1)_\mathbf{L}) \frac{ c_1(\U(1)_{\tilde{Y}})^2}{2}$ coefficient.\\
$\bullet$ $\U(1)_{{\bf Q}}$-$\SU(2)^2$ triangle diagram shows 1 multiplied by $N_c N_f$ for the $c_1(\U(1)_\mathbf{Q}) c_2(\SU(2))$ coefficient.\\
$\bullet$ $\U(1)_{{\bf L}}$-$\SU(2)^2$ triangle diagram shows 1 multiplied by $N_f$ for the $c_1(\U(1)_\mathbf{L}) c_2(\SU(2))$ coefficient.\\

\item Two particular linear combinations of $\U(1)_\mathbf{Q}$ and $\U(1)_\mathbf{L}$,
written as $\U(1)_{{{\bf Q}} - {N_c}{\bf L}}$ and $\U(1)_{{{\bf Q}} + {N_c}{\bf L}}$
are particularly convenient. 
Because both $\U(1)_{{{\bf Q}} - {N_c}{\bf L}}$ and $\U(1)_{{{\bf Q}} + {N_c}{\bf L}}$
contain the fermion parity $\Z_2^\rF$ normal subgroup, we have two types of 
$\Spin^c\equiv \Spin \times_{\Z_2^\rF} \U(1)$ structures from both ${{\bf Q}} - {N_c}{\bf L}$ and ${{\bf Q}} + {N_c}{\bf L}$, agreeing with \eq{eq:GSM-0form} and \eq{eq:GSM-1form}.

\item \cred{\bf $\U(1)_{{{\bf Q}} - {N_c}{\bf L}}$ symmetry is \cpurple{free of mixed anomaly} with $G_{\SM_\q}$}: 
The invertible $\U(1)_{{{\bf Q}} - {N_c}{\bf L}}$ ordinary 0-form symmetry
couples to the 1-form background gauge fields satisfying the constraint $N_c A_{\mathbf{Q}} + A_{\mathbf{L}}=0$
 or simply $N_c A_{\mathbf{Q}} = - A_{\mathbf{L}} = N_c A_{{\bf Q} - {N_c}{\bf L}}$. 
The vanishing of the first term in \eq{SM-5d-S5-iTFT}
tells that the ABJ-type anomalies of the form $\U(1)_{{{\bf Q}} - {N_c}{\bf L}}$-$\U(1)_{\tilde{Y}}^2$
and $\U(1)_{{{\bf Q}} - {N_c}{\bf L}}$-$\SU(2)^2$ are absent.  

\item \cred{\bf $\Z_{2 N_cN_f, {{\bf Q} + {N_c} {\bf L}} }$ symmetry is \cpurple{free of mixed anomaly} with $G_{\SM_\q}$}:
The invertible $\U(1)_{{{\bf Q}} + {N_c}{\bf L}}$ ordinary 0-form symmetry
couples to the 1-form background gauge fields satisfying the constraint 
$N_c A_{\mathbf{Q}} - A_{\mathbf{L}}=0$ or 
simply $N_c A_{\mathbf{Q}} = A_{\mathbf{L}} = N_c A_{{\bf Q} + {N_c}{\bf L}}$. 
The nonvanishing of the first term in \eq{SM-5d-S5-iTFT}
with the coefficient 
$\cred{-} 2 N_f N_c  A_{{\bf Q} + {N_c}{\bf L}}(18 \frac{ c_1(\U(1)_{\tilde{Y}})^2}{2} +c_2(\SU(2)))$
tells that: (1) The ABJ-type $\U(1)_{{{\bf Q}}+ {N_c}{\bf L}}$-$\U(1)_{\tilde{Y}}^2$ anomaly breaks
$\U(1)_{{{\bf Q}}+ {N_c}{\bf L}}$ down to  $\Z_{36N_cN_f, {{\bf Q} + {N_c} {\bf L}} }$
via the U(1) instanton number $n^{(1)}= \int \frac{ c_1(\U(1)_{\tilde{Y}})^2}{2} \in \Z$ on spin manifolds.
(2) Meanwhile,
the ABJ-type $\U(1)_{{{\bf Q}}+ {N_c}{\bf L}}$-$\SU(2)^2$ anomaly breaks
$\U(1)_{{{\bf Q}}+ {N_c}{\bf L}}$ down to  $\Z_{2 N_cN_f, {{\bf Q} + {N_c} {\bf L}} }$
via the SU(2) instanton number $n^{(2)}= - \int c_2(\SU(2)) \in \Z$ on arbitrary 4-manifolds.

\item {\bf No noninvertible symmetry for the ${{\bf Q}} + {N_c}{\bf L}$} (or ${\bf B + L}$ symmetry):
The SM is compatible with four global structure versions of the Lie gauge group $G_{\SM_\q}$.
When $q=1$ or $3$, the SM admits the $\SU(2) \times \U(1)_{\tilde{Y}}$ instantons.
When $q=2$ or $6$, the SM admits the $\U(2)_{\tilde{Y}}$ instantons.
The $q=1,3$ and $q=2,6$ are related by gauging the 1-form electric symmetry in \eq{eq:GSM-1form}.
Let us compare $\SU(2) \times \U(1)_{\tilde{Y}}$ instanton versus 
$\U(2)_{\tilde{Y}}\equiv\frac{\SU(2) \times \U(1)_{\tilde{Y}}}{\Z_2}$ instanton.

$\bullet$ Because of the 
$\Spin \times_{\Z_2^\rF} \U(1)_{{{\bf Q}} - {N_c}{\bf L}}=\Spin^c$
structure in \eq{eq:GSM-0form} and \eq{eq:GSM-1form},
we can allow different instanton number quantizations 
on the spin manifolds versus nonspin manifolds.

$\bullet$ Regardless of the $\SU(2) \times \U(1)_{\tilde{Y}}$ instanton numbers from
$n^{(2)}= - \int c_2(\SU(2))$ and $n^{(1)}= \int \frac{ c_1(\U(1)_{\tilde{Y}})^2}{2}$,
or the $\U(2)$ instantons from $n^{(2)}=  \int (- c_2(\U(2)) +\frac{ c_1(\U(2)_{\tilde{Y}})^2}{2})$,
we are concerned only with the quantization of the second Chern class and the first \cred{Chern class} squared.

$\bullet$ The Chern numbers $\int c_1(\U(1)_{\tilde Y}) \in \Z$, 
$\int c_1(\U(2)_{\tilde Y}) \in \Z$, and
$\int c_2(\U(2)_{\tilde Y}) \in \Z$
 are all integer-valued
for both spin and nonspin manifolds.
But the $\U(1)_{\tilde Y}$ instanton number
$\int \frac{1}{2}c_1^2(\U(1)_{\tilde Y}) \in \Z$ on spin manifolds,
while $\int \frac{1}{2}c_1^2(\U(1)_{\tilde Y}) \in \frac{\Z}{2}$ becomes
half-integer valued on nonspin manifolds.
However the fractional $\frac{1}{2}$ $\U(1)_{\tilde Y}$ instanton
only means to break
$\U(1)_{{{\bf Q}}+ {N_c}{\bf L}}$ down to  $\Z_{18 N_cN_f, {{\bf Q} + {N_c} {\bf L}} }$,
which would not be enough to affect the symmetry already broken down to
$\Z_{2 N_cN_f, {{\bf Q} + {N_c} {\bf L}} }$ by SU(2) instantons.
Namely, there is \emph{no}  noninvertible symmetry to be constructed
out of the invertible $\Z_{2 N_cN_f, {{\bf Q} + {N_c} {\bf L}} }$
symmetry because this $\Z_{2 N_cN_f, {{\bf Q} + {N_c} {\bf L}} }$ 
remains preserved and anomaly-free under any instantons
on spin and nonspin manifolds compatible with the SM structure
\eq{eq:GSM-0form} and \eq{eq:GSM-1form}.

\item \cred{\bf{No 2-group structure in the SM within \eq{SM-5d-S5-iTFT}}}:
Eq.~\eq{SM-5d-S5-iTFT} also shows that the anomaly cancellation 
for triangle Feynman diagrams with three vertices $\U(1)_{\bf Q}^2$-$\U(1)_{\tilde{Y}}$,
$\U(1)_{\bf L}^2$-$\U(1)_{\tilde{Y}}$,
$\U(1)_{\bf Q}^2$-$\SU(2)$,
and $\U(1)_{\bf L}^2$-$\SU(2)$ always holds because their coefficients are always zero in the SM (based on \Table{table:SMfermion}'s data  in Appendix \ref{app:rep}):\\
$\bullet$ $\U(1)_{\bf Q}^2$-$\U(1)_{\tilde{Y}}$ triangle diagram shows the coefficient $(+ 1 \cdot (- 1)^2 \cdot 2 + 2 \cdot 1^2 \cdot 1  + 1 \cdot (- 1)^2 \cdot (-4)) N_c N_f=0$.\\
$\bullet$ $\U(1)_{\bf L}^2$-$\U(1)_{\tilde{Y}}$ triangle diagram shows the coefficient $(2 \cdot 1^2 \cdot (- 3) +  (-1)^2 \cdot 6 ) N_f=0$.\\
$\bullet$ $\U(1)_{\bf Q}^2$-$\SU(2)$ triangle diagram shows the coefficient 0. \\
$\bullet$ $\U(1)_{\bf L}^2$-$\SU(2)$ triangle diagram shows the coefficient 0. \\
So we do not obtain a 2-group-like structure \cite{Cordova1802.04790:2018cvg} 
in the SM within \eq{SM-5d-S5-iTFT}.

\item {\bf From $\Spin^c$ to ${\Spin \times_{\Z_2^\rF} {\Z_{4,X}}}$-structure manifold}:
When we replace the continuous $\U(1)_{{{\bf Q}} - {N_c}{\bf L}}$
symmetry by a discrete $\Z_{4,X}$
with $X \equiv 5({ \mathbf{B}-  \mathbf{L}})-\frac{2}{3} {\tilde Y}
=\frac{5}{N_c}({ \mathbf{Q}-  N_c \mathbf{L}})-\frac{2}{3} {\tilde Y}$,
we also map the 5d iTFT in \eq{SM-5d-S5-iTFT} classified by $\Z^2$ to the 5d iTFT classified by 
$\Z_{16}$ evaluated on a 5d $M^5$:
\bea  \label{SM-Z16-iTFT} 
S_5 \equiv 
     (-N_f+n_{\nu_R})\,
     \frac{2\pi }{16}  \eta_{4{\dd}}(\text{PD}( A_{{\Z_{2,X}}} )) \big\vert_{M^5}.
\eea

$\bullet$ Because all the quarks and leptons have charge 1 under ${\Z_{4,X}}$ (see \Table{table:SMfermion} in Appendix \ref{app:rep}),
there is no $N_c$ factor in this formula.

$\bullet$ The background gauge field $A_{{\Z_{2,X}}}\in \H^1(M^5,\Z_2)$ is 
obtained by the quotient map down to $\Z_{2,X}\equiv {\Z_{4,X}}/{\Z_2^\rF}$ from 
the ${\Spin \times_{\Z_2^\rF} {\Z_{4,X}}}$-structure on the 5d spacetime manifold $M^5$.
%the generator of the cohomology group $\H^1(\B{\Z_{2,X}}, \Z_2)$ with ${\Z_{2,X}} \equiv {\Z_{4,X}}/{\Z_2^\rF}$
%of ${\Spin \times_{\Z_2^\rF} {\Z_{4,X}}}$-structure manifold $M$.
%The classifying space $\B^{n}\rA  \equiv K(\rA,n)$ is the Eilenberg-MacLane space for an abelian group $\rA$.
%Hereafter we use the standard convention, 
%all cohomology classes are pulled back to the manifold $M$ along the maps given in the definition of cobordism groups,
%thus $\H^1(\B{\Z_{2,X}}, \Z_2)$ can be pulled back to $\H^1(M, \Z_2)$.

$\bullet$ Here the 5d  Atiyah-Patodi-Singer (APS \cite{Atiyah1975jfAPS}) eta-invariant 
 $\eta_{5{\dd}} = \eta_{4{\dd}}(\text{PD}( A_{{\Z_{2,X}}}))$ is valued in 
 {$\Z_{16} \equiv \Z/(16\Z)$} and
is written as the 4d eta invariant ${\eta}_{4{\dd}} \in \Z_{16}$ \footnote{
The normalization is different from the normalisation used in  \cite{Witten2016cio1605.02391}: $\eta_{4{\dd}}^\text{Here}=4\eta_{4{\dd}}^\text{There}$.}
on the  4d ${\Pin^+}$ submanifold representing Poincar\'e dual (PD)  to $A_{{\Z_{2,X}}}$. The ${\Pin^+}$ structure is obtained from the 5d bulk ${\Spin \times_{\Z_2^\rF} {\Z_{4,X}}}$-structure by Smith isomorphism:
$\Omega_5^{\Spin \times_{\Z_2} \Z_4} \cong \Omega_4^{\Pin^+} \cong \Z_{16}$
\cite{2018arXiv180502772T, Hsieh2018ifc1808.02881, GuoJW1812.11959, Hason1910.14039}.
The  eta invariant ${\eta}_{4{\dd}}\in \Z_{16}$ is the effective topological action of the interacting fermionic time-reversal symmetric topological superconductor
of condensed matter in three spatial dimensions 
with an \emph{anti-unitary} time-reversal symmetry $\Z_4^{\rm TF}$ such that 
the time-reversal symmetry generator $T$ squares to the fermion parity operator, namely $\rT^2=(-1)^\rF$.
The symmetry can be defined by the nontrivial group extension
$1 \to \Z_2^{\rF} \to  \Z_4^{\rm TF} \to \Z_2^{\rT} \to 1$
(see a review \cite{Senthil1405.4015, 1711.11587GPW}).
In contrast, in the SM, we have the \emph{unitary} ${\mathbf{B}-  \mathbf{L}}$-like symmetry ${\Z_{4,X}}$ whose generator $X$ squares to $X^2=(-1)^\rF$. The symmetry again
can be defined by the nontrivial group extension
$1 \to \Z_2^{\rF} \to  \Z_{4,X} \to \Z_{2,X} \to 1$.

%Here we derive the anomaly polynomial $I_6$ for the 4d SM,
%by including the dynamical or background gauge field couplings to
%$\cG_{\SM}$, $\U(1)_\mathbf{Q}$ and $\U(1)_\mathbf{L}$,
%and by also including also the Spin group diffeomorphism couplings to spacetime geometry and gravity.

\item {\bf The 4d SM and 5d iTFT coupled path integral}: 
We can write down a fully
gauge-diffeomorphism-invariant path integral
by coupling a 4d SM action \eq{eq:SM-action}
on $M^4$
with a 5d iTFT action \eq{SM-6d-I6} on $M^5$ with $M^4 {=\prt M^5}$:
\bea \label{eq:4d-5d-Z}
\bZ[M^4, M^5; A_{\mathbf{Q}}, A_{\mathbf{L}}]
=
\big(\int [\cD \psi_L][\cD \psi_L^\dagger]
[\cD A_I][\cD \phi]
\e^{\ii S_{\rm SM}[M^4; A_{\mathbf{Q}}, A_{\mathbf{L}}]}
\big) \cdot
\e^{\ii S_5[M^5; A_{\mathbf{Q}}, A_{\mathbf{L}}]}.
\eea
Here we have included the dynamical gauge fields 
(namely $A_{I=1,2,3}$ for $\cG_{\SM} \equiv su(3) \times  su(2) \times u(1)_{\tilde Y}$), background gauge fields 
(namely $A_\mathbf{Q}$ for $\U(1)_\mathbf{Q}$, 
 and $A_\mathbf{L}$ for $\U(1)_\mathbf{L}$),
and the background gravity fields. Importantly, the dynamical gauge fields $A_{I}$ are restricted to the 4d manifold $M^4$,
while both background gauge fields, $A_\mathbf{Q}$ and $A_\mathbf{L}$, 
and background gravity can couple to and propagate between the 4d SM theory and the 5d bulk.

Two certain convenient combinations of $A_\mathbf{Q}$ and $A_\mathbf{L}$ gauge fields 
contain indeed two kinds of $\Spin^c$ gauge fields: 
 $A_{{\bf Q}-N_c {\bf L}}$ and $A_{{\bf Q}+N_c {\bf L}}$. 
 So we can probe the anomalies associated with $\Spin^c$ structures.

The continuous $\U(1)_{\bf Q}$ and $\U(1)_{\bf L}$ symmetry transformations 
give $\alpha$ phase variations on the Weyl fermions as in \eq{eq:U1QL}, 
or \cred{effectively induce} the background gauge field transformation 
$A_{\mathbf{Q}} \mapsto A_{\mathbf{Q}}+ \dd \alpha_{\mathbf{Q}}$ and 
$A_{\mathbf{L}} \mapsto A_{\mathbf{L}}+ \dd \alpha_{\mathbf{L}}$
in terms like $A \wedge \star j_{\rm 4d}$ and $A \wedge \star J_{\rm 5d}$.
Following Noether's theorem with quantum anomaly, we can derive the anomalous current nonconservation in the 4d SM's path integral
via integration by parts:
\footnote{Here we use $A \wedge \star j \mapsto
(A+\dd \alpha )\wedge \star j 
=A \wedge \star j +\dd (\alpha   \star j) -\alpha (\dd \star j)$ to keep the 4d term
$-\alpha (\dd \star j)$ on $M^4$, but we drop $\dd (\alpha   \star j)$ when $\prt{M^4}=0$ has no 3d boundary;
while we use $A \wedge \star J \mapsto
(A+\dd \alpha) \wedge \star J 
=A \wedge \star J +\dd (\alpha   \star J) -\alpha (\dd \star J)$ to keep the 4d term
$\int_{M^4}  \alpha   \star j
=\int_{M^5} \dd (\alpha   \star j)$
by \cpurple{Stokes} theorem when ${M^4}=\prt{M^5}$.
The anomaly inflow \eq{eq:instanton-breaking-diff} 
shows $\dd \star j= \star J$ on the 4d $M^4$, while $\dd \star J= 0$ thanks to $\dd^2=0$ inside the 5d $M^5$ bulk.}
\begin{multline}
\label{eq:instanton-breaking-diff}
\int [\cD \psi_L][\cD \psi_L^\dagger][\cD A_I][\cD \phi]
\e^{\ii \big( 
\int_{M^4}\big( (\dd^4x \cred{\sqrt{|{\rm g}|}} \, \cred{\hat{\cL}}_{\rm SM}[A_{\mathbf{Q}}, A_{\mathbf{L}}])
\cred{-}\alpha_{\bf Q}(
\dd \star j_{\bf Q}
)
\cred{-}\alpha_{\bf L}(
\dd \star  j_{\bf L}
)}\\
{}^{+
(N_c \alpha_{\mathbf{Q}} + \alpha_{\mathbf{L}} ) 
N_f \big(\cred{-}18 \frac{ c_1(\U(1)_{\tilde{Y}})^2}{2} \cred{-} c_2(\SU(2))\big)
\cred{+}(N_f-n_{\nu_R})\alpha_{\bf L}\big(\frac{c_1(\U(1)_\mathbf{L})^2}{6}-\frac{p_1(TM)}{24}\big)
\big) 
\big)}.
\end{multline}

Here 
\bea
\text{
$j_{\bf Q} =j_{\bf Q \mu} \dd  x^\mu$
%$= \bar{\Psi}_{\bf Q}\gamma_\mu  {\Psi}_{\bf Q} \dd  x^\mu$ 
$= q_{\bf Q}(\psi_{L \bf Q}^\dagger \bar\sigma_\mu\psi_{L \bf Q}) \dd  x^\mu$
and 
$j_{\bf L} = j_{\bf L \mu} \dd  x^\mu$
%$=\bar{\Psi}_{\bf L}\gamma_\mu  {\Psi}_{\bf L} \dd  x^\mu$
$= q_{\bf L}(\psi_{L \bf L}^\dagger \bar\sigma_\mu\psi_{L \bf L}) \dd  x^\mu$,}
\eea
where $\psi_{L \bf Q}$ and $\psi_{L \bf L}$ respectively contain 
the quark and lepton sectors of the Weyl fermion multiplet $\psi_{L}$ in \eq{eq:SMrep}.
The quark number $q_{\bf Q}$ is $+1$ for left-handed quarks
and $-1$ for right-handed anti-quarks.
The lepton number $q_{\bf L}$ is $+1$ for left-handed leptons
and $-1$ for right-handed anti-leptons.
The divergence of the currents are given by
\ccred{$\dd \star  j_{\bf Q}= (-1)^s\prt_\mu ({\sqrt{|g|}} j_{\bf Q}^\mu) \dd^4 x$
and $\dd \star  j_{\bf L}= (-1)^s\prt_\mu ({\sqrt{|g|}} j_{\bf L}^\mu) \dd^4 x$}.\footnote{
Here and below, when gravity and curved spacetime is involved,
$(\star j)_{\mu_1 \mu_2 \mu_3}= g^{\nu \nu'} \epsilon_{\nu' \mu_1 \mu_2 \mu_3} j_\nu= 
{\sqrt{|g|}} \tilde\epsilon_{\nu' \mu_1 \mu_2 \mu_3} j^{\nu'}$, while
$(\dd \star  j_{})_{\mu \mu_1 \mu_2 \mu_3}=4 \prt_{[\mu} (\star j)_{\mu_1 \mu_2 \mu_3]}$,
and $\tilde\epsilon_{\nu' \mu_1 \mu_2 \mu_3}  \tilde \epsilon^{\mu \mu_1 \mu_2 \mu_3}
=(-1)^s 3!\delta_{\nu'}^\mu$
with $s$ as the number of negative eigenvalues in the metric, or
$(-1)^s$ as the sign of the metric determinant.
So we have
$(\dd \star  j_{}) =
\frac{1}{4!}(\dd \star  j_{})_{\mu \mu_1 \mu_2 \mu_3} (\dd x^\mu \wedge \dd x^\mu_1 \wedge \dd x^\mu_2 \wedge \dd x^\mu_3)
=\frac{1}{4!} (\dd \star  j_{})_{\mu \mu_1 \mu_2 \mu_3}  \tilde \epsilon^{\mu \mu_1 \mu_2 \mu_3} \dd^4 x
=(-1)^s \prt_\mu (\sqrt{|{\rm g}|} { j^{ \mu}}) \dd^4 x
=(-1)^s  (\nabla_\mu  { j^{ \mu}}) {\sqrt{|g|}} \dd^4 x$ with the spacetime metric ${\rm g}_{\mu \nu}$ involved.
See Appendix \ref{app:Notations-and-Conventions}.
}
The violation of the quark ${\bf Q}$ and lepton ${\bf L}$ currents 
by the mixed gauge anomalies 
or mixed gravitational anomalies on the  quantum level reads:
\bea \label{eq:dJQL}
\begin{array}{ccccl}
\dd \star j_{\bf Q} &=& - N_c N_f (18 \frac{ c_1(\U(1)_{\tilde{Y}})^2}{2} +c_2(\SU(2))). & &   \cr
\dd \star j_{\bf L} &=&  - N_f (18 \frac{ c_1(\U(1)_{\tilde{Y}})^2}{2} +c_2(\SU(2)))  &+ & 
(N_f - n_{\nu_R}) \,(\frac{c_1(\U(1)_\mathbf{L})^2}{6}-\frac{p_1(TM)}{24}).\cr
\dd \star j_{{\bf Q}-N_c {\bf L}} &=& &+& 
 (-N_f+n_{\nu_R}) \,(N_c^3 \frac{c_1(\U(1)_{{\bf Q}-N_c {\bf L}})^2}{6}-N_c\frac{p_1(TM)}{24}).
\cr
\dd \star j_{{\bf Q}+N_c {\bf L}} &=& -2  N_c N_f (18 \frac{ c_1(\U(1)_{\tilde{Y}})^2}{2} +c_2(\SU(2))) 
&+& 
(N_f - n_{\nu_R}) \,(N_c^3 \frac{c_1(\U(1)_{{\bf Q}+N_c {\bf L}})^2}{6}-N_c\frac{p_1(TM)}{24}).
\end{array}
\eea
\Eq{eq:instanton-breaking-diff} shows the 4d SM perspective. 
But from the anomaly inflow perspective, those are the boundary currents in 4d
that inflow to the bulk currents in 5d. 
The 5d bulk currents (denoted by $J_{\bf Q}$ and $J_{\bf L}$) 
can be introduced by adding an extra term 
$ \int (A_{\bf Q} \wedge \star J_{\bf Q} + A_{\bf L} \wedge \star J_{\bf L})$ to the original 5d bulk action $S_5[M^5; A_{\mathbf{Q}}, A_{\mathbf{L}}]$.
We then obtain the equalities following \eq{eq:dJQL}
as the boundary-bulk current inflow
relations $\dd \star j_{\bf Q}=\star J_{\bf Q}$
and $\dd \star j_{\bf L}=\star J_{\bf L}$,
similarly for $\dd  \star j_{{\bf Q}-N_c {\bf L}} =  \star J_{{\bf Q}-N_c {\bf L}}$
and $\dd  \star j_{{\bf Q} + N_c {\bf L}} = \star J_{{\bf Q} + N_c {\bf L}}$.

When the continuous $\U(1)_{{{\bf Q}} - {N_c}{\bf L}}$
 is replaced with a discrete $\Z_{4,X}$ symmetry,
one can make an analogous analysis by adjusting the 5d iTFT action $S_5$ of
\eq{eq:4d-5d-Z} to the 5d iTFT in \eq{SM-Z16-iTFT}. Schematically, we have a 4d-5d coupled path integral
\bea \label{eq:4d-5d-Z-Z4X}
\bZ[M^4, M^5;A_{\Z_{4,X}}]
=
\big(\int [\cD \psi_L][\cD \psi_L^\dagger]
[\cD A_I][\cD \phi]
\e^{\ii S_{\rm SM}[M^4; A_{\Z_{4,X}}]}
\big) \cdot
\e^{\ii S_5[M^5; A_{\Z_{4,X}}]},
\eea
where $A_{\Z_{4,X}}$ is precisely a $\Spin\times_{\Z_2^\rF}\Z_{4,X}$ gauge field
that couples to and communicates between the 4d SM theory and the 5d bulk.

\end{enumerate}

\section{Categorical Symmetry from Mixed $\U(1)$-Gravitational Anomaly}
\label{sec:U1-grav}

In \Sec{sec:SM-Polynomial}, we reviewed the violation of the continuous ${\bf B-L}$
symmetry (more precisely, $\U(1)_{{\bf Q}-N_c {\bf L}}$) 
and thus the nonconservation of its current
$j_{{\bf Q}-N_c {\bf L}}$ of the SM,
due to the pure $\U(1)^3$ anomaly and the mixed $\U(1)$-gravity$^2$ anomaly in \eq{eq:dJQL}:
$$
\dd \star j_{{\bf Q}-N_c {\bf L}} = 
(-N_f+n_{\nu_R}) \,\left(N_c^3 \frac{c_1(\U(1)_{\cred{{\bf Q}-N_c {\bf L}}})^2}{6}-N_c\frac{p_1(TM)}{24}\right).
$$
In this section, we simply denote $j_{{\bf Q}-N_c {\bf L}}$ as $j$,
and consider a mathematically motivated general expression (with the reason to be explained),
\bea
\dd \star j
=\kappa_1\,\frac{c_1^2}{3!}+\kappa_2\,p_1
\eea
that corresponds to a general degree 6 anomaly polynomial for the $\U(1)$ gauge theory:
\begin{equation}
    I_6=\kappa_1\,\frac{c_1^3}{3!}+\kappa_2\,c_1p_1
    \label{I6-lk} 
\end{equation}
where, as in Section \ref{subsec:Summary}, $c_1=F/{(2\pi)}$, with $F=\dd A$ \cred{and $A \equiv A_{{\bf Q}-N_c {\bf L}}$}, is the Chern-Weil representative 2-form of the first Chern class of the $\U(1)$ bundle with connection 1-form $A$, and $p_1=-{\Tr[R\wedge R]}/{(8\pi^2)}$, with $R=\dd \omega+\omega\wedge\omega$, is the \cred{curvature} 2-form representative of 
the first Pontryagin class of the spacetime tangent bundle $TM^4$ with Levi-Civita spin-connection 1-form $\omega$. 
Note that in general the coefficients $\kappa_{1}$ and $\kappa_{2}$ cannot be arbitrary numbers. Their possible values are determined by Atiyah-Singer index theorem \cite{Alvarez-Gaume:1984zlq} (see Appendix \ref{app:anomaly-polynomial-quantization} for a review). Assuming $\Z_2^\rF\subset \U(1)$, as in the case of ${\bf B -L}$ symmetry in Standard Model
\cred{which endorses a $\Spin \times_{\Z_2^\rF} \U(1) \equiv \Spin^c$ structure}, they must satisfy the conditions
\begin{equation}
    \kappa_1=24\ell+k,\qquad \kappa_2=-\frac{k}{24},
\end{equation}
for some $k,\ell\in\Z$ (if $\Z_2^\rF\not\subset {\U(1)}$, 
\cred{which endorses a $\Spin \times \U(1)$ structure}, 
we have instead $\ell\in \frac{1}{4}\Z$). Moreover, any values of $k$ and $\ell$ can be realized by considering all possible 4d QFTs. 

{Note that, generically, when $k \neq 0\mod 24$, the presence of mixed $\U(1)$-gravitational anomaly ($\kappa_{2} \neq 0$) implies the presence of the pure $\U(1)$ anomaly ($\kappa_{1} \neq 0$).} 
When $k=0\mod 24$, it is possible to have the mixed $\U(1)$-gravitational anomaly ($\kappa_{2} \neq 0$) but without any pure $\U(1)$ anomaly ($\kappa_{1} = 0$).

For the Standard Model setup considered in Section \ref{sec:SM-Polynomial}: 
\begin{equation}
\label{eq:kandell}
    k=(-N_f+n_{\nu_R})\,N_c,\qquad \ell=(-N_f+n_{\nu_R})\,\frac{N_c^3-N_c}{24}.
\end{equation}
To consider first the effect of the mixed $\U(1)$-gravitational anomaly only, we assume that we are on a connected spacetime spin 4-manifold $M^4$ and in the trivial background $\U(1)$ gauge field. The current $j$ of the global $\U(1)$ symmetry is not conserved, but satisfies:
\begin{equation}
    \label{j-p1-divergence}
    \dd \star j=-\frac{k}{24}p_1
    =\frac{k}{24}\frac{1}{8 \pi^2}   \Tr[ R \wedge R] .
\end{equation}
\cred{This current $j$ nonconservation means that fermions
can be created or annihilated locally
out of the vacuum by distorting the curved spacetime, whenever $k \neq 0$ and $\Tr[ R \wedge R] \neq 0$, hypothetically pertinent to gravitational leptogenesis 
\cite{PhysRevLett.96.081301, Adshead1711.04800:2017znw} in the very early universe.}
%\begin{enumerate}[leftmargin=-0mm]

%\item  
Consider the na\"ive Noether charge operator, corresponding to the rotation by the angle 
$\alpha\in 2\pi\cdot (\R/\Z) \cong \U(1)$ and supported on an \emph{oriented connected} dimension 3 submanifold\footnote{In general, it is not required to be a submanifold, just a 3-cycle with $\U(1)$ coefficients. Such a cycle corresponds to a network of charge operators. The discussion in principle can be generalized to this more general case. However, all the ingredients of the construction of noninvertible defects will have locality property, therefore the final result will give a local definition of the noninvertible defect.} $\CY\subset M^4$:
\begin{equation}
    U_\alpha(\CY)=\e^{\ii \alpha \int_{\CY} \star j}.
\end{equation}
By slightly abusing the notations, we will use the same symbol for a chosen lift of $\alpha$ to $\R$ (we can always choose a representative 
$\alpha$ to be in the interval $[0,2\pi)$).

\begin{figure}[h]
    \centering
\includegraphics[scale=0.5]{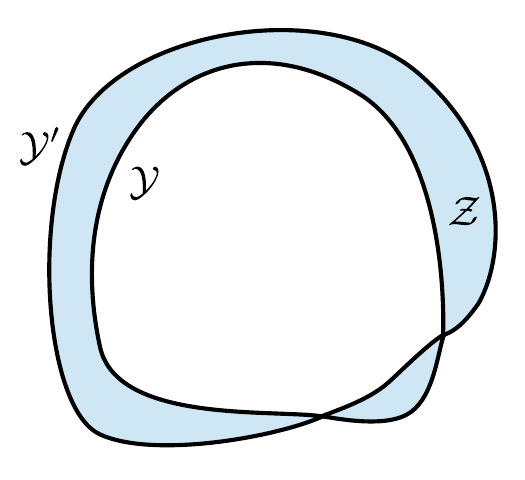}
    \caption{A schematic drawing of a small deformation of a submanifold $\mathcal{Y}\subset M$ to $\mathcal{Y}' \subset M$. 
    The shaded domain depicts $\mathcal{Z}$ such that \cpurple{ $\partial{\mathcal{Z}}=\mathcal{Y}'-\mathcal{Y}$}.}
    \label{fig:defect-deformation}
\end{figure}

The (\ref{j-p1-divergence}) implies that this operator is actually not topological. Namely, consider slightly deformed support $\CY'$. 
By the Stokes theorem, the change of the na\"ive charge operator is the following:
\begin{equation}
U_\alpha(\CY') \,U_\alpha(\CY)^{-1}
=\e^{ \ii \alpha \left(\int_{\CY'} \star j-\int_{\CY} \star j \right)} =
\e^{ \ii \alpha \int_{\CZ}\dd \star j}=\e^{\frac{- \ii k\alpha}{24} \int_{\CZ}p_1}
\label{eq:fig:U'Up1}
\end{equation}
where $\CZ$ is the 4-chain such that \cpurple{$\partial \CZ=\CY'-\CY
$, where, as usual, the minus sign in front of a chain corresponds to its orientation reversal
(see Fig. \ref{fig:defect-deformation}).} 
\cred{This equality \eq{eq:fig:U'Up1} shall be regarded as \cpurple{equality between operators} inserted in the path integral.}
The topological noninvariance then can be fixed using the fact that locally 
\bea
p_1
=-\frac{1}{8\pi^2}\Tr[ R\wedge R]
=-\dd \GCS/(2\pi),
\eea
where $\GCS$ is the gravitational Chern-Simons 3-form
\begin{equation}
    \GCS \coloneqq  \frac{1}{4\pi}\Tr[\omega\wedge \dd\omega+
    \frac{2}{3}\,\omega\wedge \omega\wedge \omega].
\end{equation} 
It is the Chern-Simons 3-form of the Levi-Civita connection 1-form $\omega$ 
\cred{(called the spin connection)}
on the \cpurple{frame bundle of the} spacetime tangent bundle $TM^4$. 

This noninvariance can be fixed by modifying the charge operator as follows\footnote{\cred{Note that for a different choice of spacetime metric, Lorentzian in particular, the Pontryagin density 4-form $p_1$ would be different. Denote by $p_1'$ the Pontryagin 4-form for such a different choice. It is known that the cohomology class of the Pontryagin 4-form is the same for any \textit{complex} non-degenerate metric (cf. \cite{Witten:2021nzp}). So that $p_1'=p_1
+\dd L$ for some globally defined 3-form $L$. The defect then would be simply modified by the extra $\exp \ii\frac{\alpha\,k}{24}\int_{\mathcal{Y}}L$ phase.}}:
\begin{equation}
    \tilde{U}_\alpha(\cred{\CY})=\e^{ \ii\alpha \int_{\cred{\CY}}(\star j-\frac{k\,\GCS}{24\cdot 2\pi})}.
    \label{U-GSC-mod}
\end{equation}
Note that in order to define $\GCS$, 
we have to make a choice of 4d vielbein \cred{(in order to define $\omega$ and $R$)}, 
that vielbein is a particular trivialization of the tangent bundle $TM^4$. Since we only need to integrate $\GCS$ over $\CY$, we need to choose the vielbien over (a small neighborhood of) $\CY$. Such a choice of 4d vielbein can be made by choosing a 3d vielbein on $\CY$ and supplementing it with the normal unit vector.\footnote{More formally this can be described as follows. First note that $N\CY=TM|_{\CY}/T\CY$, where $N\CY$ is the rank one normal bundle over $\CY$, $TM|_{\CY}$ is the restriction of the rank 4 tangent bundle of $M$ to the submanifold, and the quotient if performed fiberwise. It is known that $T\CY$ can be always trivialized globally over the 3-manifold $\CY$. Let us choose a particular trivialization isomorphism $\varphi_T:T\CY\rightarrow \R^3\times \CY$. Moreover, since \cred{both $\CX$ and $\CY$} 
are oriented, $N\CY$ can be also globally trivialized, $\varphi_N:N\CY\rightarrow \R\times \CY$, and there is a canonical choice. Therefore $TM|_\CY$ itself can be also trivialized with trivialization $(\varphi_T,\varphi_N):TM|_{\CY}\rightarrow (\R^3\times \R)\times \CY$.}

The modification of the charge operator (\ref{U-GSC-mod}) is very similar to the modification considered in \cite{Choi2205.05086:2022jqy, Cordova2205.06243:2022ieu} in the case of ABJ type anomaly between 
\cred{a global $\U(1)$ and a gauge $\U(1)'$ symmetries}. There,
the \cred{abelian} Chern-Simons term of the bulk \cred{$\U(1)'$ gauge field} was considered \cite{Choi2205.05086:2022jqy, Cordova2205.06243:2022ieu}; 
here, instead we have a gravitational \cred{nonabelian} Chern-Simons term. 

For a small deformation of $\CY$ to $\CY'$, 
we can consider the extension of the trivialization of the tangent bundle to $\CZ$. 
By Stokes theorem, we then have
\begin{equation}
   \int_{\CY'}\GCS-\int_{\CY}\GCS=
   \int_\CZ \dd\GCS=-2\pi\int_\CZ p_1
   = \frac{1}{4\pi} \int_\CZ \Tr[ R\wedge R]
\end{equation}
and therefore 
\begin{equation}
    \tilde{U}_\alpha(\CY') \cred{=}\tilde{U}_\alpha(\CY).
\end{equation}
{in the correlation functions, as long as there are no insertions of other operators with support inside $\mathcal{Z}$.}

\subsection{Framing vs Atiyah's 2-framing}

However, the definition of the $\tilde{U}(\CY)$ above requires a choice of dreibein (vielbein in 3 dimensions), namely a choice of trivialization of $T\CY$, and there is no canonical choice. 
A change of the trivialization of $T\CY$ corresponds to an 
$\SO(3)$ gauge transformation of the spin-connection 1-form $\omega$ used to define the gravitational Chern-Simons 3-form. Although the integral $\int_\CY \GCS$ is invariant under continuous changes of the trivialization 
(i.e. ``small gauge transformations''), it can change under arbitrary changes of the trivialization (``large gauge transformations''). In other words, the gravitational Chern-Simons action depends on the choice of framing of the tangent bundle $T\CY$ --  the homotopy class of its trivialization \cite{Witten1988hfJonesQFT}. In particular, one can consider a change of framing by a ``twist'' corresponding to a gauge transformation $Y^3\mapsto \SO(3)$ which is different from the identity only inside some ball in $Y^3$. Homotopy class of such a map is classified by $f\in \pi_3(\SO(3))\cong \Z$. The value of the gravitational Chern-Simons action then changes by $2\pi f$ with $f\in \Z$.

\begin{enumerate}
\item {\bf Atiyah's 2-framing structure}:
As was argued by Atiyah in \cite{atiyah1990framings}, $\int_\CY \GCS$ actually depends \emph{only} on the so-called \textit{2-framing}. The 2-framing can be defined as the homotopy class of the trivialization of the spinor bundle of the direct sum of the two copies of the tangent bundle
$2 T\CY = T\CY\oplus T\CY$. Its spin structure is the canonical spin structure induced by the diagonal embedding $\Spin(3)\rightarrow \Spin(3)\times \Spin(3)\subset \Spin(6)$ of the structure group. Namely, under this map, any choice of spin structures on $TY$ defines a spin structure on $2T\CY$, which is independent of that choice. 

In general, for a vector bundle $V$ with a chosen spin structure, there is a well-defined integral characteristic class $\frac{1}{2}p_1(V)$.\footnote{This follows from the fact 
that $p_1(V)=w_2(V)^2\mod 2$, where $w_2$ is the second Stiefel-Whitney class, and that the choice of spin structure on $V$ provides a trivialization of $w_2(V)$. So $\frac{1}{2}p_1(V) \in \Z$ is in integer.}  
Applying this statement to the case when $V$ is the direct sum of two copies of a tangent bundle, one can then define the value of the gravitational Chern-Simons action as follows. Consider a compact oriented Riemannian 4-manifold $\CX$ which has $\CY$ as its boundary (i.e. $\CY=\partial \CX$) and with metric near the boundary being the product metric on $(-\epsilon,0]\times \CY$. Choosing a trivialization of $2T\CY$, that is a 2-framing $\beta$, provides a trivialization of $2T\CX$ near the boundary. The spin structure on $2T\CX$ can be defined in the same way as for $2T\CY$. Therefore one can define a relative characteristic number $\frac{1}{2}p_1(2T\CX,\beta)\in \Z$. Using this number, one can then define the value of gravitational Chern-Simons action (in $\R$ instead of just in $\R/2\pi\Z$), by the following formula:
\begin{equation}
\int_{\CY} \GCS=  2\pi\big(\frac{1}{8\pi^2} \int_\CX \Tr [R\wedge R] - \frac{1}{2} p_1(2T\CX,\beta)\big).
\end{equation}
Changing $\CX$ while keeping both $\CY$ and $\beta$ the same results in a change of the two terms in the right-hand side by the same number (with opposite signs). Therefore the whole expression depends only on $\CY$ and $\beta$.

\item {\bf $p_1$ structure}:
Equivalently, instead of 2-framing, one can use a $p_1$-\textit{structure} $\beta'$ on $\CY$, that is a choice of the trivialization of the first Pontryagin class $p_1(T\CY)=0$ (which vanishes identically by dimensional reasons). Similarly extending $\CY$ to a 4-manifold $\CX$, we have an integral relative characteristic number $p_1(T\CX,\beta')\in \Z$, which can be used to define the value of the gravitational Chern-Simons action in $\R$ as follows:
\begin{equation}
\int_{\CY} \GCS= 2 \pi \big(\frac{1}{8\pi^2} \int_{\CX} \Tr [R\wedge R] -  p_1(T \CX,\beta') \big).
\end{equation}

\end{enumerate}

The \textit{changes} of $p_1$-structures or 2-framings are in one-to-one correspondence with the group\footnote{\cpurple{For a pair $\beta_1',\beta_2'$ of $p_1$-structures (resp. a pair $\beta_1,\beta_2$ of 2-framings),  a \textit{difference} can be defined as the corresponding relative charecteristic class on the cylinder $[0,1]\times \CY$ with the trivializations $\beta_{i}'$ (resp. $\beta_i'$) at the boundaries: $\beta_1'-\beta_2'=p_1(T([0,1]\times \CY),\beta_1',\beta_2')\in \H^4([0,1]\times \CY,\partial([0,1]\times \CY),\Z)\cong \H^3( \CY,\Z)$ (resp. $\beta_1-\beta_2=\frac{1}{2}p_1(2T([0,1]\times \CY),\beta_1',\beta_2')\in \H^4([0,1]\times \CY,\partial([0,1]\times \CY),\Z)\cong \H^3( \CY,\Z)$). This is analogous to the classification of the changes of spin structures, that is a trivialization of $w_2(T\CY)\in \H^2(\CY,\Z_2)$, by $\H^1(\CY,\Z_2)$.}} $\H^3(\CY,\Z)\cong \Z$. This group (up to a sign of the generator) can be identified with the group of integral ``twists'' of framing $\pi_3(SO(3))\cong \Z$ considered above. This, in particular, can be seen from the fact that a change of $p_1$-structure $\beta'$ (respectively 2-framing $\beta$) corresponding to an integer $f\in \Z$ shifts the value $p_1(T\CX,\beta')$ (respectively $\frac{1}{2}p_1(2T\CX,\beta)$) by $f$, which, in turn, results with the change of $\int_\CY \mathrm{GCS}$, defined by the formulas above, by $-2\pi f$.

To summarize, the ordinary framing of $T\CY$ provides the trivialization of all characteristic classes: $w_1(T\CY),\,w_2(T\CY),$ and $p_1(T\CY)$. A trivialization of $w_1(T\CY)$ corresponds to a choice of orientation on $\CY$, which was already assumed. A trivialization of $w_2(T\CY)$ corresponds to choice of spin structure on $T\CY$. This is an extra information (in addition to orientation and $p_1$-structure), which is not actually required to define the value of the gravitational Chern-Simons theory. Using Atiyah's 2-framing or $p_1$-structure instead of the ordinary framing is a way to explicitly forget that additional structure.

Note that although for a \textit{closed} 3-manifold $\CY$,
one can define a certain ``canonical'' 2-framing \cite{atiyah1990framings}, such a choice is nonlocal and cannot be respected by ``cutting and gluing''. In particular, for a given rational CFT, although one can define a WRT invariant of a closed 3-manifold unambiguously (without any additional structure), to define a 3d TQFT, one still needs to consider a framed version of bordisms \cite{Bartlett:2015baa}. 
This would be required in a more general setting when the ambient 4-dimensional spacetime $M\supset \CY$ containing the defect $\bar{U}_\alpha(\CY)$ has a boundary (i.e. a fixed time slice) on which the defect can end.

\subsection{Noninvertible symmetry's topological defect and fusion rule}

%\newpage

Under the change of 2-framing by $f\in \cred{\H^3(\CY,\Z)\cong } \Z$ units, the charge operator changes as follows:
\begin{equation}
   \tilde{U}_\alpha(\CY)\;\rightarrow \;
    \tilde{U}_\alpha(\CY)\, \e^{-\frac{\ii\alpha\,k \,f}{24}}.
\end{equation}

%\newpage
If $\alpha/(2\pi)$ is rational, this however can be compensated by putting a 
Witten-Reshetikhin-Turaev (WRT) 3d TQFT $\bT$ associated with a 2d rational CFT of a certain 
\cred{rational chiral central charge $c_- \equiv c_L - c_R \in \Q$}.  
The partition function of such a TQFT on $\CY$ also changes with the change of the framing of $\CY$ \cite{Witten1988hfJonesQFT}
\begin{equation}
      \bZ_\bT[\CY]\;\rightarrow \;
    \bZ_\bT[\CY]\,\e^{\frac{2\pi \ii\,f\,c_-}{24}}.
\end{equation}
\cpurple{Note that the large gauge transformations corresponding to changes of 2-framing are classified by $f \in \Z$, as explained above,
while the framing anomaly is classified by $c_- \in \Q/24\Z$ which contains a subgroup $\Z/24\Z = \Z_{24}$.}

We then can consider the following family of topological operators:
\begin{equation}
    D_{(c_-,\bT)}(\CY)\coloneqq \e^{\ii\,c_-\int_\CY\left({\frac{2\pi}{k}\,\star j -\frac{1}{24}\,\GCS}\right)}\cdot \bZ_\bT[\CY]
    \label{Dtop-def}
\end{equation}
labeled by pairs $(c_-,\bT)$ where $c_- \in \Q$ (such that $2\pi\,c_-/k=\alpha \mod 2\pi$ for the original $\alpha\in 2\pi\cdot (\Q/\Z) \equiv \Q/(2\pi \Z)$), and $\bT$ is a TQFT associated \cred{with} a rational CFT with central charge $c_-$. Note that such pairs exist for any $\alpha\in 2\pi \cdot\Q/\Z$. Although for each $\alpha\in 2\pi\cdot (\Q/\Z)$, one could choose specific $c_-(\alpha)$ and $\bT(\alpha)$, for example of a Chern-Simons type, with $c_-(\alpha)$ being the Sugawara chiral central charge.\footnote{
For example, if $\alpha=2\pi p/N\mod 1$, with $p<N$ being positive coprime integers, one can take $\bT(\alpha)$ to be a stack of $p\cdot k$ copies of $\SU(2)$ level $-(6N-2)$ Chern-Simons theory. The total central charge is $c_-(\alpha) =-p\,k\,\frac{(6N-2)\cdot 3}{6N-2+2}
=\frac{pk}{N}-3pk$, which satisfies the condition $c_-(\alpha)/k=\alpha\cred{/(2 \pi)} \mod 1$. \cred{We recall that $\SU({\rm m})_k$ Chern-Simons theory has a central charge $c_- = 
k \frac{\dim \mathfrak{g}}{(|k|+h^{\vee})}
=\frac{k({\rm m}^2-1)}{|k| + {\rm m}}$ for a generic integer level k,
the dimension of Lie algebra $\dim \mathfrak{g}$, and the dual Coxeter number $h^{\vee}$
 \cpurple{computable from Lie algebra commutators: $[\rT^\rb, \rT^\rc]= f^{\rb \rc \rd} \rT^\rd$, $\sum_{b,c} f^{\rm abc} f^{\rm dbc} = 2 h^{\vee} \delta^{ad}$.}}}
\cred{However, such an assignment would not be respected by the fusion of the defects. In particular:
\begin{equation}
 \bT(\alpha_1)\otimes \bT(\alpha_2) \neq  \bT(\alpha_1+\alpha_2).
\end{equation}
} 
\cred{Instead, we will consider all possible  pairs, forming a commutative monoid} 
\begin{equation}
\mathfrak{N}=\{(c_-,\bT)\}
\end{equation}
with the binary operation defined by
\cred{\begin{equation}
 (c_{-,1},\bT_1)+(c_{-,2},\bT_2)\coloneqq
    (c_{-,1}+c_{-,2},\bT_1\otimes \bT_2)
\end{equation}
}
so that it corresponds to the fusion of the defects (cf. \cite{Putrov2208.12071:2022pua}):
\begin{equation}
    D_{(c_{-,1},\bT_1)}(\CY)D_{(c_{-,2},\bT_2)}(\CY)
    = D_{(c_{-,1}+c_{-,2},\bT_1\otimes \bT_2)}(\CY)\equiv D_{(c_{-,1},\bT_1)+(c_{-,2},\bT_2)}(\CY).
\end{equation}
This monoid related to the subgroup $\Q/\Z\subset \R/\Z\cong \U(1)$ of the original invertible symmetry by a surjective morphism of monoids:
\begin{equation}
    \begin{array}{rcl}
        \mathfrak{N} & \longrightarrow & \Q/\Z,  \\
         (c_-,\bT) & \longmapsto &\frac{\alpha}{2\pi}=c_- /k\mod 1. 
    \end{array}
\end{equation}

The operators (\ref{Dtop-def}) are noninvertible when  $c_-\notin\frac{1}{2}\Z$ because $\bT$ is necessarily a noninvertible TQFT. When $c_-\in\frac{1}{2}\Z$, one can choose $\bT$ to be invertible, if one considers it as a spin TQFT. For example, one can take $\bT$ to be a stack of Ising spin-TQFTs
\cred{(also known as the fermionic invertible TQFT or iTFT 
for the low energy theory of the chiral $p_x+ \ii p_y$-wave topological superconductor in condensed matter in a 3d spacetime 
\cite{MooreReadNPB1991,ReadGreen9906453-2000, SeibergWitten1602.04251, Putrov2016qdo1612.09298PWY})}. 
Note that there is a canonical spin structure induced on the codimension one submanifold $\CY$ from the spin structure on $M$. This, however, requires a choice of spin-structure on $M$. The invertability for $c_-\in\frac{1}{2}\Z$ corresponds to the fact that  $\Z_{2k}\equiv \Z/(2k\Z)$ subgroup of $\U(1)$ symmetry can be preserved as an invertible symmetry. 

%\footnote{(1) abelian gauge background in 3d  
%$A \dd A$.
%(2) nonabelian gauge background in 3d 
%$A \dd A + 2/3 A^3$
%(3) nonabelian SO4 background in 3d 
%$\omega \dd  \omega + 2/3 \omega^3$.}

%The $c_1^3$ anomaly by itself does not break the global $\U(1)$ symmetry. It is purely 't Hooft anomaly. It only prevents gauging the symmetry. This anomaly does not violate local conservation of the current in the trivial background $\U(1)$ gauge field. 

Now let us consider the additional effect of the pure $\U(1)$ anomaly. 
To do so, let us turn \cred{on} the nontrivial background $\U(1)$ gauge field. Moreover, if $\Z_2^\rF\subset \U(1)$, \cred{we can now drop the assumption of the spacetime manifold being a spin manifold} (note that although not every 4-manifold admits a spin structure, every 4-manifold does admit a Spin$^c$ structure). The anomaly polynomial of the general form (\ref{I6-lk}) implies that 
\begin{equation}
\label{eq:dj-kandell}
    \dd \star j=\frac{24\ell+k}{6}\,c_1^2\, {-}\, \frac{k}{24}p_1.
\end{equation}
 Because of the first term, the defect defined in (\ref{Dtop-def}) will no longer be topological. Namely, a deformation of $\CY$ into $\CY'$ changes the operator as follows:
\begin{equation}
    D_{(c_-,\bT)}(\CY')D_{(c_-,\bT)}(\CY)^{-1}=
    \e^{2\pi\ii\,c_- \,\frac{24\ell+k}{k}\int_{\CZ}\frac{F^2}{6\cdot4\pi^2}}.
    \label{DcT-defect-change}
\end{equation}

However, this can be fixed as in \cite{Choi2205.05086:2022jqy, Cordova2205.06243:2022ieu}, by putting on top of it an additional abelian TQFT that couples to the bulk $\U(1)$ gauge field. The main difference is that in our setup, this $\U(1)$ gauge field is not dynamical in the bulk. 

Such an abelian TQFT can be always realized by a $\U(1)^L$ Chern-Simons theory with a certain $L\times L$ symmetric level matrix $K$ with integral elements $K_{ij}\in\Z$. The coupling to the external 4d $\U(1)$ gauge field $A$ then can be described by a choice of an integral vector $n\in\Z^L$. 
The matrix $K$ can be thought of as defining an integral rank $L$ lattice $\Lambda\cong \Z^L$ equipped with the quadratic form given by $K$: 
$({\rm a,b})_\Lambda={\rm a}^TK {\rm b},\;
{\rm a,b}\in\Lambda$. Then one can consider $n\in \Lambda^*$ as an element of the dual lattice in a basis-independent way.  \cpurple{The path integral for the partition function of such abelian (ab) Chern-Simons theory with level matrix $K$} defined on a 3-manifold $\CY$ reads:
\begin{equation}
    \bZ^\text{ab}_{(\Lambda,n\in \Lambda^*)}[\CY;A]=
    \int \prod_{i=1}^L [\mathcal{D}a_i]\,
    \e^{\frac{\ii}{2\pi}\int_\CY\left(\frac{1}{2}\sum_{i,j=1}^L K_{ij}a_i\wedge \dd a_j+\sum_{i=1}^L n_ia_i\wedge \dd A\right)}
    \label{Z-abelian-TQFT}
\end{equation}
where $a_i$ are internal 3d dynamical $\U(1)$ gauge field. The theory depends only on the isomorphism class of the lattice, since the theories with equivalent matrices are related by field redefinition. The classification of the theories  inequivalent on the quantum level, without coupling to the external field, is given in 
\cite{Belov:2005ze,Kapustin:2010hk}.\footnote{{Namely, the invariant data of the theory on the quantum level is $\sigma=\mathrm{sign}\,\Lambda$, the signature of the Lattice, and the discriminator group $\mathsf{D}=\Lambda^*/\Lambda$ together with a quadratic refinement $\mathsf{q}:\mathsf{D}\rightarrow \Q/\Z$ of the bilinear form $\mathsf{D}\times \mathsf{D}\rightarrow \Q/\Z,\; (a,b)\mapsto (a,b)_\Lambda\mod 1$. The element $n\in\Lambda^*$ defining the coupling to the external field $A$ then descends to an element $[n]\in \mathrm{D}$. For a given $[n]$, however, the partition function is independent of the choice of representative $n$ only up to gauge invariant counterterm, namely a factor of the form $\exp(\frac{\ii\,r}{4\pi}\int_\CY A\wedge dA)$ for an \textit{integer} $r\in \Z$}.}

By performing the Gaussian integration over $a_i$ in (\ref{Z-abelian-TQFT}) one can see that the change in (\ref{DcT-defect-change}) will be canceled if
\begin{equation}
    (n,n)_{\Lambda}\equiv \,n^TK^{-1}n
    =c_-\,\frac{24\ell+k}{3k}
    \label{Kn-c-condition}
\end{equation}
which can be always satisfied by an appropriate choice\footnote{The right-hand side of 
(\ref{Kn-c-condition}) is a certain fraction that can be always represented by a pair of integers $p'\geq 0$ and \cred{$N'\neq 0$}:
\begin{equation}
   c_- \,\frac{24\ell+k}{3k} = 
   \frac{p'}{N'}.
\end{equation}
To satisfy (\ref{Kn-c-condition}) one can take for example $L=2p'$, $K=\mathrm{diag}(2N',2N',\dots,2N')$, and $n=(1,1,\ldots,1)$.\label{foot:lattice-example}} of $K$ and $n$. 

This extra abelian TQFT, however, also has a framing anomaly corresponding to the chiral central charge equal to the signature of the lattice, $\mathrm{sign}\,\Lambda$. Therefore one should adjust the TQFT $\bT$ that appeared above to have \cred{central charge} $(c_- -\mathrm{sign}\,\Lambda)$ instead. Note that if the ambient spacetime $M^4$ is not considered to be spin (with a chosen spin structure), there is no canonically induced spin structure on $\CY \subset M^4$ and one has to stay in the realm of bosonic TQFTs, meaning that 
\cred{the lattice $\Lambda$ must be even}. The condition (\ref{Kn-c-condition}) still can always be satisfied, for example by the choice as in Footnote \ref{foot:lattice-example}. Unlike in the case of fermionic theories, however, it is  not possible to adjust the framing anomaly of the abelian TQFT to be zero by coupling it with invertible TQFTs. In the bosonic case, the central charge can be only shifted by invertible TQFT by multiples of 8 (which corresponds to adding to $\Lambda$ the even self-dual E$_8$ lattice, whose invertible TQFT
corresponds to the E$_8$ quantum Hall state written as an abelian Chern-Simons theory with symmetric bilinear form $K$ matrix given by the rank-8 Cartan matrix of E$_8$). 
Therefore we will keep $\mathrm{sign}\,\Lambda$ to be a generic integer. 

%\cred{JW: Shall we comment on the noninvertible subgroup for bosonic case? So $\Z_{\frac{k}{8}}\equiv \Z/(\frac{k}{8}\Z)$ subgroup of $\U(1)$  symmetry can be preserved as an invertible symmetry?}

Finally, to take into account the pure $\U(1)$ anomaly we redefine the topological defects as 
follows:\footnote{\cred{Here and earlier, n\"aively we have included
the improperly quantized gravitational Chern-Simons (GCS) with $\frac{c_-}{24} \GCS$ 
without introducing additional dynamical fields to be integrated out to produce $\frac{c_-}{24} \GCS$.
Our approach seems to contrast with 
\cite{Choi2205.05086:2022jqy}, 
where the improperly quantized Chern-Simons (CS) $\frac{1}{4 \pi N} A \dd A$ is obtained from
integrating out the dynamical field $\int [\CD a]$ 
with a fractional quantum Hall term $\frac{N}{4 \pi } a \dd a + \frac{1}{2 \pi } a \dd A$.
However, we can redefine
\bea
\bZ_{\bT'}[\CY] \coloneqq \e^{-\ii \frac{c_-}{24} \int_\CY \GCS} \cdot \bZ_\bT[\CY],
\eea
where the new Witten-Reshetikhin-Turaev (WRT) type 3d TQFT 
$\bT'$ differed by the old $\bT$ by a local counter term.
The $\bT'$ can be obtained from Witten's original work \cite{Witten1988hfJonesQFT},
such that \eq{Dtop-def-full} becomes
\begin{equation}
    D_{(c_-,\bT',\Lambda,n)}(\CY)\coloneqq\e^{\ii\,c_- \int_\CY\left({\frac{2\pi}{k}\,\star j 
   }\right)}
    \cdot \bZ_{\bT'}\cred{[\CY]}
    \cdot \bZ^\text{ab}_{(\Lambda,n)}\cred{[\CY;A]}.
    \label{Dtop-def-full-proper}
\end{equation}
The ${\bT'}$ depends on the metric, while the ${\bT}$ depends on the framing,
which is ``more topological.'' 
Thus, an improperly quantized GCS attached to ${\bT}$ appears as the metric-dependent phase as we
perform the path integral over internal fields in ${\bT'}$. This is 
analogous to the appearance of $\frac{1}{4 \pi N} A \dd A$ in \cite{Choi2205.05086:2022jqy}
after integration over their $a$.\\
In summary, one can also first
add an improperly quantized CS term $\frac{1}{4 \pi N} A \dd A$ or 
an improperly quantized GCS term $\frac{c_-}{24} \int_\CY \GCS$
by hand to make the defect topological, 
but then it will be non-invariant under large gauge-diffeomorphism transformations
(which is the framing dependence of GCS in our story).
For the former CS $\frac{1}{4 \pi N} A \dd A$, 
it can be fixed by considering a 3d TQFT with anomalous 
discrete magnetic 1-form $\Z_N$ symmetry \cite{Cordova2205.06243:2022ieu}.
For the later $\frac{c_-}{24} \int_\CY \GCS$,
it can be fixed by a 3d WRT TQFT with an opposite framing anomaly.
}}
\begin{equation}
    D_{(c_-,\bT,\Lambda,n)}(\CY)\coloneqq\e^{\ii\,c_- \int_\CY\left({\frac{2\pi}{k}\,\star j -\frac{1}{24}\,\GCS}\right)}
    \cdot \bZ_\bT\cred{[\CY]}
    \cdot \bZ^\text{ab}_{(\Lambda,n)}\cred{[\CY;A]}.
    \label{Dtop-def-full}
\end{equation}
The defects are now labeled by quadruples $(c_-,\bT,\Lambda,n)$ where $c_- \in \Q$, 
the $\Lambda$ is an integral lattice with $n\in \Lambda^*$ such that\footnote{Note that unless $24\ell+k=0$,  
\cred{the first entry of the quadruple $c_-$} is completely determined by the pair $(\Lambda,n)$.} $(n,n)_\Lambda=-c_-\,(24\ell+k)/(3k)$, and $\bT$ is a Witten-Reshetikhin-Turaev-type 3d TQFT corresponding to a 
\cred{2d rational CFT} with central charge $(c-\mathrm{sign}\,\Lambda)$. The relation to the original $\alpha$ in the na\"ive undressed defect is as before: $\alpha =c_-/k\mod 1$. 

%\newpage

{Moreover, two quadruples define the same defect if they satisfy the equivalence relation
\begin{equation}
    (c_-,\bT,\Lambda\oplus \Lambda',n\oplus 0)
    \; \sim \; 
    (c_-,\bT\otimes \bT^\text{\cred{ab}}_{\Lambda'},\Lambda,n)
\end{equation}
where $\bT^\text{\cred{ab}}_{\Lambda'}$ is the abelian TQFT associated with the lattice $\Lambda'$. The equivalence relation corresponds to absorbing $\bT^\text{\cred{ab}}_{\Lambda'}$, a part of the abelian TQFT which is not coupled to the bulk field $A$, into the TQFT $\bT$.}

%\newpage

As before, the quadruples form a commutative monoid 
$$
\mathfrak{N}'=\{(c_-,\bT,\Lambda,n)\}
$$ 
with the binary operation:
\begin{equation}
    (c_{-,1},\bT_1,\Lambda_1,n_1)+(c_{-,2},\bT_2,\Lambda_2,n_2)\coloneqq
    (c_{-,1}+c_{-,2}, \bT_1\otimes \bT_2,\Lambda_1\oplus \Lambda_2,n_1\oplus n_2)
\end{equation}
corresponding to the fusion of the operators (cf. \cite{Putrov2208.12071:2022pua}):
\begin{equation}
    D_{(c_{-,1}, \bT_1,\Lambda_1,n_1)}(\CY)D_{(c_{-,2},\bT_2,\Lambda_2,n_2)}(\CY)
    =  D_{(c_{-,1},\bT_1,\Lambda_1,n_1)+(c_{-,2},\bT_2,\Lambda_2,n_2)}(\CY).
\end{equation}
The relations between the elements of  quadruples are respected by the binary operation because $\mathrm{sign}\,(\Lambda_1\oplus \Lambda_2)=\mathrm{sign}\,\Lambda_1+ \mathrm{sign}\,\Lambda_2$, $(n_1\oplus n_2,n_1\oplus n_2)_{\Lambda_1\oplus\Lambda_2}=(n_1,n_1)_{\Lambda_1}+(n_2,n_2)_{\Lambda_2}$.

This monoid is related to the subgroup $\Q/\Z\subset \R/\Z\cong \U(1)$ of the original invertible symmetry by a surjective morphism of monoids:
\begin{equation}
    \begin{array}{rcl}
        \mathfrak{N}' & \longrightarrow & \Q/\Z,  \\
         (c_-,\bT,\Lambda,n) & \longmapsto &\frac{\alpha}{2\pi}=c_-/k\mod 1. 
    \end{array}
\end{equation}
Note that for a given $c_-$, the defect can be made invertible if $c_-(24\ell+k)/(6k)\in\Z$ and also $c_-\in 8\Z$.

%with global $\Z_N$ 1-form symmetry that has a given 't Hooft anomaly. Namely, the one corresponding to the invertible 4d TQFT  with the effective action $\frac{\pi \ii p(24\ell+k)}{3N}\int \mathcal{P}_2(B)$, where $B$ is the background gauge field for $\Z_N$ 1-form symmetry (an element of degree 2 cohomology with $\Z_N$ coefficients), $\mathcal{P}_2$ is the Pontryagin square operation, and $p,N$ are integers such that $p/N=-c/k\mod 1$.

\section{Categorical Symmetry from Mixed $\Z_{4}$-Gravitational Anomaly}
\label{sec:Z4-grav}

Let $\upnu \in \Z_{16}$ be the anomaly index of the $\Z_4\supset \Z_2^\rF$ symmetry. In the Standard Model setup considered in Section \ref{sec:SM-Polynomial}, this symmetry is $\Z_{4,X}$ and $\upnu=-N_f+n_{\nu_R}$. Let us start with the na\"ive (i.e. classically defined) network\footnote{The reason we consider a network of charge operators rather than a single charge operator supported will become apparent later.} $U(\tilde{\CY})$ of charge operators supported on a 3-cycle $\tilde{\CY}$ with $\Z_4$ coefficients (corresponding to the charges assigned to the individual operators in the networks). Note that it is not always possible to resolve $\tilde{\CY}$ into a submanifold. Because the symmetry group involves fermion parity the definition of the charge operator is rather subtle already on the classical level. What we mean by it is the following. A choice of the background $\Z_4$ field corresponds to choosing $\Spin\times_{\Z_2^\rF}\Z_4$ structure on the spacetime 4-manifold $M^4$. Let us fix one such structure. The \textit{changes} of structures are in one-to-one correspondence with the elements of $\H^1(M^4,\Z_4)$ (meaning that the space of structures is a torsor over this group). The insertion of $U(\tilde{\CY})$ then implements the \textit{change} of the structure corresponding to the Poincar\'e dual of $[\tilde{\CY}]\in \H_3(M^4,\Z_4)$. 

On the quantum level, the theory has an anomaly corresponding to the 5d iTFT with the following effective action $S_{5}$ on a 5d spacetime manifold $M^5$:
\begin{equation}
    S_{5}=\upnu \frac{2\pi \,\eta(\PD(A))}{16}\big\vert_{M^5}
    \label{S5d-eta-A}
\end{equation}
where $A$ is the $\Z_2$ background gauge field (i.e. the element of $\H^1(M,\Z_2)$) defined from the 
\begin{equation}
    \Spin \times_{\Z_2^\rF} \Z_4  
    \longrightarrow \Z_4/\Z_2^\rF\equiv \Z_2
\end{equation}
and $\eta$ is the eta-invaraint normalized such that $\eta \in \Z_{16}$ is an integer well-defined modulo 16 on a closed $\Pin^+$ 4-manifold. The expression (\ref{S5d-eta-A}) is not quite mathematically precise. What it actually means is the following. The Poincar\'e dual of $A\in \H^1(M^5,\Z_2)$ can be represented by an \emph{unoriented} closed codimension-1 submanifold in $M^5$. The $\Spin \times_{\Z_2^\rF} \Z_4$ structure on $M$ induces $\Pin^+$ structure on it \cite{Hason1910.14039}. The $\eta(\PD(A))$ in (\ref{S5d-eta-A}) is defined as the eta-invariant $\eta$ of this $\Pin^+$ 4-manifold, which we denote by $\PD(A)$. 

The anomalous 4d theory is unambiguously defined in a general background only if considered as the theory on the boundary of the 5-dimensional spacetime $M^5$. That is when $M^4$ is considered to be one of the boundary components of $M^5$ with $\Spin \times_{\Z_2^\rF}\Z_4$ structure on $M^4$ induced from $M^5$. In particular, the boundary components of the submanifold $\PD(A)$ that lie in $M^4$, that is $\partial\PD(A)\cap M^4 = \PD(A|_{M^4})\subset M^4$, represent the Poincar\'e dual of $A|_{X}\in \H^1(M^4,\Z_2)$ which is defined by the  $\Spin \times_{\Z_2^\rF} \Z_4$ structure on $M^4$.
\begin{figure}[h]
    \centering
\includegraphics[scale=0.5]{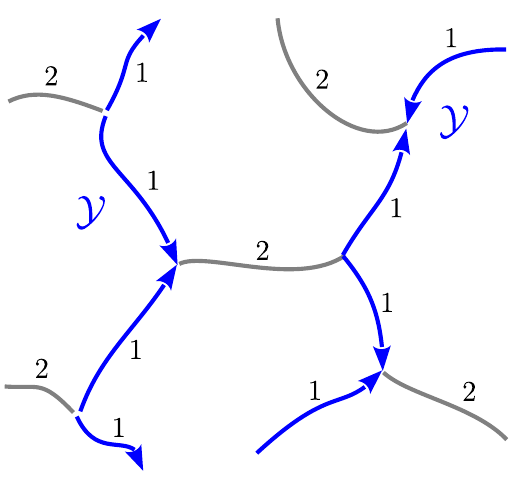}
    \caption{A schematic drawing of how locally a $\Z_4$-cycle $\tilde{\CY}$ inside $M^4$ looks like. The cycle  corresponds to a network of charge operators on the classical level. The numbers denote the $\Z_4$ charges of the operators in the network. Note that  operators of charge $3=-1\mod 4$ is equivalent to operators of charge $+1$ but with reversed orientation. Therefore we can assume that there are only two types of nontrivail operators in the network: of charge $1$ and $2$. Since $2=-2\mod 4$, the operators of charge $2$ do not require a choice of orientation. In blue we depict $\CY$, the mod 2 reduction of $\tilde{\CY}$. It is realized by forgetting in the network all operators of charge $2$ and also forgetting the orientation of operators of charge $1$. It can always be deformed into a smooth \emph{unoriented} 3-submanifold inside $M^4$, which we will denote by the same symbol, $\CY$.}
    \label{fig:defect-Z4-to-Z2}
\end{figure}
The insertion of the operator network $U(\tilde{\CY})$ has an effect of changing $\PD(A|_{M^4})$ by the union with $\CY=\tilde{\CY}\mod 2$, the  cycle in $M^4$ with $\Z_2$ coefficients obtained by mod 2 reduction of 
$\tilde{\CY}$ (see Fig. \ref{fig:defect-Z4-to-Z2}). It can always be resolved into a smooth \emph{unoriented} 
3-manifold by a small deformation.\footnote{Note that if we started with  $\tilde{\CY}$ being a manifold, the resulting $\CY$ would be either empty or an orientable manifold. This would be quite restrictive for the analysis below. This is the reason why we have started with a nontrivial network of charge operators.} This means that, on the quantum level, the operator network supported on $\tilde{\CY}$ by itself is not well defined, but becomes so if we extend $\CY=\tilde{\CY}\mod 2$ to a 4-dimensional hypersurface in the 5d bulk $M^5$. The effective action, $\pi\upnu\eta/8$, supported on the hypersuface is only topological inside the bulk, meaning it is invariant under deformations that preserve the boundary. 

\begin{figure}[h]
    \centering
\includegraphics[scale=0.5]{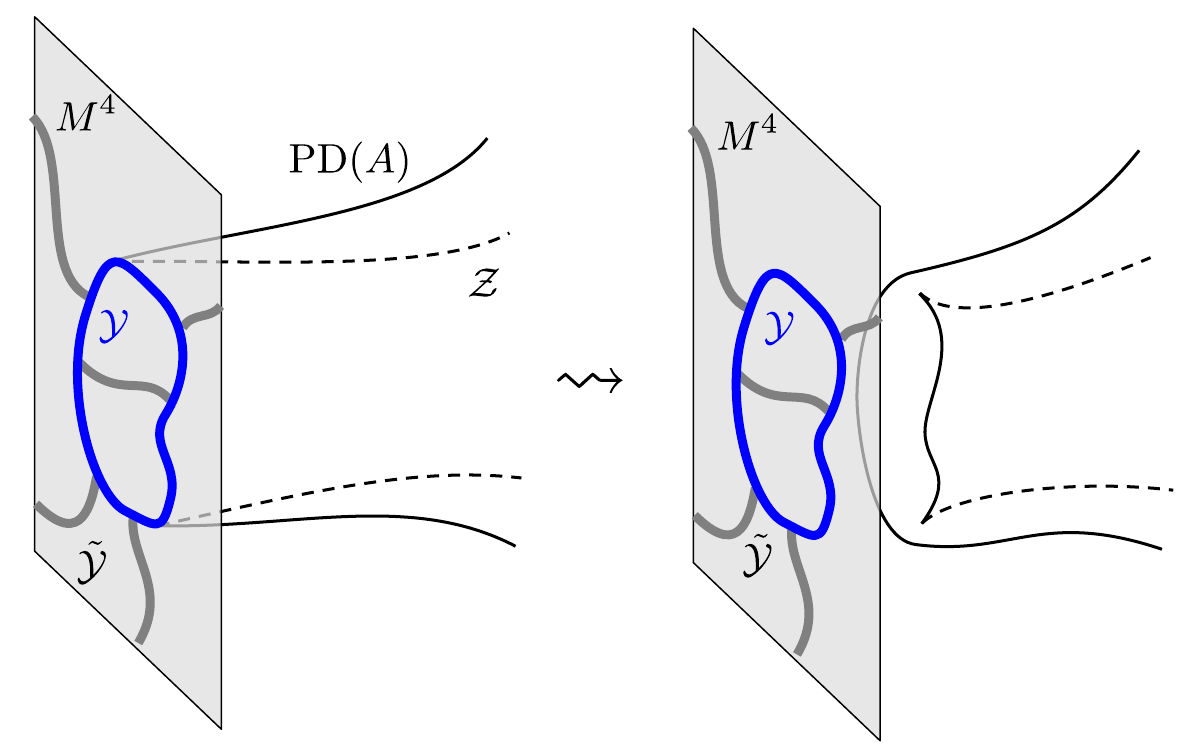}
    \caption{The 4d hypersurface $\PD(A)$ inside the 5d bulk, with the action $\pi \ii \upnu\eta/8$ supported on it and ending on $\CY\subset M^4$, is needed to unambiguosly define the classical charge operator network $U(\tilde{\CY})$, with $\CY=\tilde{\CY}\mod 2$. The 4d hypersurface $\CZ$, with the action $-\pi \ii \eta/8$, and also ending on $\CY$, is needed to unambiguosly define the $\bZ_\bT[\CY]$, the partition function of an anomalous $\Pin^+$ TQFT $\bT$. The union $\PD(A)\cup (-\CZ)$ can be deformed into a smooth hypersurface and pushed inside the 5d bulk, with the total action unchanged. This means that the product (\ref{DZ4-def}) is a well-defined topological defect in 4d, not requiring a choice of extension of $\CY$ into the bulk. }
    \label{fig:defect-bulk-push}
\end{figure}

Assume there is a 3d \cred{$\Pin^+$} TQFT $\bT$ with anomaly described by the 4d effective action 
\cred{$S_{4d}=-\upnu\pi \eta/8$}, 
that is, it has anomaly $-\upnu\in \Z_{16}=\Hom(\Omega_{\Pin^+},\U(1))$. Such TQFTs were considered in 
\cite{Fidkowski1305.5851:2013jua, 
Metlitski1406.3032:2014xqa,
Wang1610.04624:2016qkb,
Tachikawa:2016cha,
Tachikawa:2016nmo,cheng2018microscopic1707.02079, Tata2104.14567:2021jwp}. We can then get rid of the dependence of the extension of $\CY$ into the 5d bulk by supplementing the defect network with TQFT $\bT$ supported on $\CY$:
\begin{equation}
    D_\bT(\tilde{\CY})\coloneqq U(\tilde{\CY})\, \bZ_\bT[\CY].
    \label{DZ4-def}
\end{equation}
Due to the anomaly of $\bT$, the $\bZ_\bT[\CY]$ itself is only unambiguously defined if considered on the boundary of the 4d TQFT describing the anomaly. We can choose to put this TQFT on another 4-dimensional $\Pin^+$ submanifold $\CZ\subset M^5$, such that its ends on $M^4$ along $\CY$, that is 
$\partial \CZ \cap {M^4}=\CY$.  
The total effective action of the bulk 5d TQFT on $M^5$ and the bulk 4d TQFT on $\CZ\subset M^5$ is then
\begin{equation}
    S_{5}+S_{4}=\pi \upnu\,\eta(\PD(A))-
    \pi \upnu\,\eta(\CZ)=\pi \upnu\eta(\PD(A)\cup (-\CZ)).
    \label{S-eta-5d-4d}
\end{equation}
\cred{The $(-\CZ)$ means the orientation reversal of $\CZ$.}
Since $\partial \PD(A)=\CY\sqcup \ldots$ and $\partial (-\CZ)=-\CY\sqcup \ldots$, we can deform $\PD(A)\cup (-\CZ)$ into a smooth hypersurface and push it inside the bulk so that it does not intersect with $\CY$ anymore (see Fig. \ref{fig:defect-bulk-push}).

This will not change the total action (\ref{S-eta-5d-4d}), since $\eta$-invariant is topological in the bulk. But this implies that the operator (\ref{DZ4-def}) is well defined by itself as a topological operator in $M^4$, without the need of specifying  the 4d extension of $\CY$ into 5d bulk. As before, the defects become noninvertible because $\bT$ is a noninvertible TQFT.

Note that the embedding $\Z_4\hookrightarrow \U(1)$ induces the map between the corresponding anomalies in the opposite direction (see App. \ref{app:anomaly-polynomial-quantization}):
\begin{equation}
    \begin{array}{rcl}
        \Z^2 & \longrightarrow & \Z_{16},  \\
         (k,\ell) & \longmapsto & \upnu=k-4\ell.
    \end{array}
    \label{U1-Z4-anomaly-map}
\end{equation}
Under such embedding the na\"ive charge operator corresponding to the generator of $\Z_4$ can be realized as the charge operator $U_\alpha$ of $\U(1)$ symmetry with $\alpha=\frac{1}{4}$, considered in Section \ref{sec:U1-grav}. There it was shown that on the quantum level the operator can be made topological by supplementing it with the gravitational Chern-Simons term and a TQFT with the corresponding central charge satisfying $c_-=k\alpha=k/4\mod k$. Using the anomaly map (\ref{U1-Z4-anomaly-map}) and taking into account that $\ell\in \Z$, we then get relation $c_-=\upnu/4\mod 1$. This is consistent with the fact that, the 3d TQFT realizing the $\Pin^+$ anomaly must be supplemented with the 4d bulk action term $2\pi \,c_-\, \frac{p_1}{24}$ with $c_-$ satisfying the condition above \cite[Footnote 3 in particular]{Tachikawa:2016cha}. The half-integer shifts of the central charge $c_-$ can be implemented by stacking with invertible spin-TQFTs.

%See footnote 3 of 1610.07010 for the gravitational Chern-Simons term in addition to the 3d noninvertible TQFT.

\section{Conclusion}
\label{sec:Conclusion}

In this work. we have shown that although 
an \emph{invertible} symmetry can suffer from mixed gravitational anomalies
under gravitational backgrounds (such as gravitational instantons),
still a certain \emph{noninvertible} counterpart of an infinite discrete subgroup of 
this original broken symmetry can be revived as a noninvertible categorical symmetry. %
We have constructed the noninvertible symmetry charge operators as topological defects,
specifically for the case of a mixed $\U(1)$-gravitational anomaly \cite{Eguchi:1976db, AlvarezGaume1983igWitten1984}
and a mixed $\Z_{4}$-gravitational anomaly  \cite{2018arXiv180502772T, GarciaEtxebarriaMontero2018ajm1808.00009, Hsieh2018ifc1808.02881, GuoJW1812.11959, Hason1910.14039, WW2019fxh1910.14668}. 
Built upon the previous construction based on the mixed gauge anomaly pioneered in \cite{Choi2205.05086:2022jqy, Cordova2205.06243:2022ieu},
our construction can be regarded as a natural extension to the mixed gravitational anomaly counterpart.
Meanwhile, thanks to the previous systematic classification of the anomalies and the corresponding cobordism class of the Standard Model (SM) \cite{GarciaEtxebarriaMontero2018ajm1808.00009, WangWen2018cai1809.11171, 
DavighiGripaiosLohitsiri2019rcd1910.11277, WW2019fxh1910.14668, 
JW2006.16996, JW2012.15860, WangWanYou2112.14765, WangWanYou2204.08393},
we implement the aforementioned mixed anomalies in the SM naturally with 
the baryon {\bf B} minus lepton {\bf L} number symmetries, such as 
$\U(1)_{}=\U(1)_{{\bf Q} - N_c {\bf L}}$
and $\Z_{4}=\Z_{4,X\equiv 5({ \mathbf{B}-  \mathbf{L}})-\frac{2}{3} {\tilde Y}}$.
The anomaly coefficients crucially depend on the difference between the family number
and the total ``sterile right-handed'' neutrino number: $(-N_f+n_{\nu_R})$.

 We conclude with some final remarks, connections to other works, and open puzzles:
 
\begin{enumerate}

\item {\bf Interpretation}: 
In the main text, {\cpurple{we} have shown that the noninvertible symmetry defect 
can be constructed out of 
an invertible symmetry suffering from
the pure $\U(1)$-anomaly (namely $\U(1)^3$), 
a mixed $\U(1)$-gravitational anomaly (namely $\U(1)$-{gravity$^2$}), 
and a mixed $\Z_{4}$-gravitational anomaly.}
See \Table{table:sym-breaking} for a summary on whether  
invertible vs noninvertible symmetries are broken, preserved, or dynamically gauged,
under background gravity, 
semiclassical dynamical gravity, 
or UV-complete full quantum gravity:

\begin{enumerate}

\item  If an invertible $\U(1)$ symmetry 
only suffers from a pure $\U(1)$, 
then this $\U(1)$ symmetry is \emph{not} broken but only has a 't Hooft anomaly that prevents it to be consistently dynamically gauged.

\item  If an invertible $\U(1)$ symmetry 
only suffers from a $\U(1)$-gravitational anomaly, 
then this $\U(1)$ symmetry is \emph{not} broken when no gravitational background is turned on
 -- the typical viewpoint that this $\U(1)$ symmetry is anomalous but not broken in the flat spacetime.
However,

$\bullet$ When the background gravitational field is turned on, 
under curved spacetime or gravitational instanton effects, 
the $\U(1)$ symmetry current conservation is violated thus nonconserved.
Arguably, one may regard this $\U(1)$ symmetry is broken by merely \emph{background} gravity 
without even including \emph{dynamical} gravity. 

$\bullet$  When gravity becomes \emph{dynamical} but only fluctuates the spacetime \emph{semiclassically},
we have a quantum effective field theory coupled to semiclassical dynamical gravity
valid below a cutoff energy scale $\Lambda_{{\rm EFT-cutoff}}$ 
(much below the full UV-complete quantum gravity).
Unambiguously, this $\U(1)$ symmetry is broken by semiclassical \emph{dynamical} gravity.

$\bullet$  Definitely, for both cases (background gravity and semiclassical dynamical gravity), 
we can say the original $\U(1)$ symmetry charge operator becomes nontopological. 
But the noninvertible counterpart symmetry charge operator can be topological,
while its noninvertible global symmetry still survives and 
makes sense in this effective field theory coupling to gravity.

\begin{table}[!h]
\hspace{0.mm}
    \setlength{\extrarowheight}{2pt}
    \begin{tabular}{|c|c|c|}
    \hline
    & U(1) invertible symmetry & $2\pi\cdot\mathbb{Q}/\mathbb{Z}$ noninvertible symmetry \\
     \hline
  Background Grav  & Ambiguous & Preserved \\
     \hline
  Semiclassical Dynamical Grav  & Broken$_{\text{(by Grav Anom)}}$ & Preserved \\
     \hline
  UV-Complete Full Quantum Grav  & Broken$_{\text{(by Grav Anom)}}$ & Broken$_{\text{(e.g. by wormhole)}}$ or Dynamically Gauged \\
    \hline
     \end{tabular}
\caption{The fate of 
U(1) invertible symmetry and $2\pi\cdot\mathbb{Q}/\mathbb{Z}$ noninvertible symmetry (where $2\pi\cdot\mathbb{Q}/\mathbb{Z} \subset \U(1)$)
under different gravitational (Grav) effects: background, semiclassical dynamical, 
or UV-complete full quantum gravity. 
Here ``Ambiguous'' means that one can either regard 
U(1) invertible symmetry is preserved but with 't Hooft anomaly,
or it is broken by nontrivial background gravity. For all those broken cases,
the original U(1) invertible symmetry charge operator becomes nontopological.
For noninvertible symmetry being ``Preserved,''
their charge operators are topological.
``Broken$_{\text{(e.g. by wormhole)}}$'' means the quantum gravity effect to break noninvertible symmetry,
making the charge operator nontopological again.
``Dynamically Gauged'' means to condense the charge operators, which gives rise to 
a new phase of ground state vacuum.
}
\label{table:sym-breaking}
\end{table}

\item  If an invertible $\U(1)$ symmetry 
 suffers from both mixed $\U(1)$-gravitational and pure $\U(1)$ anomalies,
 such as the SM's ${\bf B - L}$  
 $\U(1)_{{\bf Q} - N_c {\bf L}}$ symmetry, 
 in \Sec{sec:U1-grav},
we have constructed the counterpart noninvertible symmetry that survives from
these anomalies ---
The subgroup of rotations by angles of the form $\alpha = 2\pi p/N$, that is $2\pi\cdot\mathbb{Q}/\mathbb{Z}$  subgroup
of the original 
invertible symmetry $2\pi\cdot\mathbb{R}/\mathbb{Z} \cong \U(1)_{{\bf Q} - N_c {\bf L}} $, can be revived as a noninvertible symmetry.\footnote{{The maximal \textit{invertible} symmetry $\Z_{2m}\subset \U(1)_{\mathbf{B-L}}$ which is free of any self- and gravitational anomalies can be determined as follows for a given $-N_f+n_{n_{\nu_R}}$, or, more generally given integer anomaly coefficients $k$ and $\ell$ (as in Section \ref{sec:U1-grav}). The $m$ is the maximal number such that the image of the anomaly $(k,\ell)\in \Z^2 \cong \Hom(\Omega_6^{\Spin^c},\Z)$ under the pullback map to $\Hom(\Omega_5^{\Spin\times_{\Z_2^\rF}\Z_{2m}},\U(1))$ is zero. Using the results of \cite{Hsieh2018ifc1808.02881}, this condition explicitly reads as the following system of equations:
\begin{equation}
    \left\{
    \begin{array}{rl}
        (2m^2+m+1)(24\ell+k)-(m+3)k & =0\mod 48m,  \\
         m(24\ell+k)+k&=0\mod 2m. 
    \end{array}
    \right.
\end{equation}
For the standard model setup, with $N_c=3$, let $|-N_f+n_{\nu_R}|=2^p\cdot r$ for some odd $r$. Then $m=2^{\max\{p-3,0\}}\cdot 3r$.}}
This procedure requires several steps, enlisted below and shown also in \Table{table:noninvertible-sym}: 
\begin{table}[!h]
\hspace{-10.mm}
    \setlength{\extrarowheight}{2pt}
    \begin{tabular}{|c|c|c|c|c|}
    \hline
        & $U_\alpha(M)$ & $\tilde{U}_\alpha(M)$ & $D_{(c_-,{\bf T})}(M)$ & $ D_{(c_-,{\bf T},\Lambda,n)}(M)$  \\
        \hline
         Topological (w/ Grav) & \xmark & \cmark & \cmark & \cmark \\
        \hline
          Topological (w/ Grav $+$ U(1)) & \xmark & \xmark & \xmark & \cmark \\
        \hline
        Grav general-covariant & \cmark & \xmark & \cmark & \cmark \\
        \hline
        U(1) gauge-invariant & \cmark & \xmark & \xmark & \cmark  \\
         \hline
        Unitary & N/A  & \cmark & \xmark & \xmark  \\
        \hline
        Invertible & N/A & \cmark & \xmark & \xmark  \\
        \hline
    \end{tabular}
\caption{Here ``Topological'' means the operator is of the same form
under arbitrary deformation depending only on a general closed 3-manifold topologically, 
``w/ Grav'' means deformation \cpurple{in the presence of non-trivial background} gravity, and
``w/ Grav $+$ U(1)'' means deformation \cpurple{in the presence of both background gravity and background global U(1)} in Spin$^c$. 
Because the $\U(1)^3$ and $\U(1)$-gravitational
anomalies of SM are more precisely \cpurple{the anomalies of QFTs defined on manifolds with Spin$^c \equiv \frac{\Spin\times \U(1)}{{\Z_2^\rF}}$ structure, 
the separation of gravitational general covariance and U(1) gauge invariance into two parts is schematic}.
The ``\cmark'' means yes, ``\xmark'' means no, ``N/A'' means not available. 
\cred{The ``N/A'' for unitary and invertible non-availability 
is due to \cpurple{the fact that} the operator product expansion (OPE)
has singularities for those charge operators,
thus those charge operators' fusion has a position dependence.}  
}
\label{table:noninvertible-sym}
\end{table}

\item The full $\Z_{4,X}$ symmetry suffers from the mixed $\Z_4$-gravitational anomaly
can also be revived, such that the modified $\Z_{4,X}$ charge operator generates a noninvertible symmetry,
while the original normal subgroup $\Z_2^\rF$ fermion parity charge operator 
still generates an invertible symmetry.

\end{enumerate}

\cred{Quantum theory coupling to dynamical gravity semiclassically 
can still be regarded as an effective field theory valid below a cutoff 
energy scale $\Lambda_{{\rm EFT-cutoff}}$ below the full UV-complete quantum gravity.}

\item {\bf Comparison with other noninvertible categorical symmetries in the SM-related models}:

A few prior works had studied other types of noninvertible categorical symmetries closely related to the
SM or Grand Unified Theories (GUTs) \cite{WangYou2111.10369GEQC, Choi2205.05086:2022jqy, Cordova2205.06243:2022ieu, Cordova2211.07639:2022fhg, CordovaKoren2212.13193:2022qtz}. 

$\bullet$ 
\Refe{WangYou2111.10369GEQC} studies 
the compatible higher-form electric and magnetic symmetries of the SM and GUTs,
and then finds that there are two 1-form magnetic symmetries, $\U(1)_{[1]}^{m_{X_1}}$ and $\U(1)_{[1]}^{m_{X_2}}$,
within the $\U(1)_{X_1} \times \U(1)_{X_2}$ gauge subgroups of
the Georgi-Glashow (GG) U(5) and flipped U(5) GUT models respectively.
Moreover, there is a $\Z_2^{\rm flip}$ flipping symmetry that exchanges the GG U(5) and the flipped U(5). So upon dynamically gauging $\Z_2^{\rm flip}$,
within the $\big[ (\U(1)_{X_1} \times_{\Z_{4,X}} \U(1)_{X_2}) \rtimes \Z_2^{\rm flip} \big]$
gauge sector, indeed, 
the 2d charge operators of the 1-form magnetic symmetries $(\U(1)_{[1]}^{m_{X_1}} \times \U(1)_{[1]}^{m_{X_2}})$ have noninvertible fusion rules and thus become 
the 2d charge operators of noninvertible
1-form magnetic symmetries. (In contrast, many other closely related to the SM examples 
in Ref.~\cite{Choi2205.05086:2022jqy, Cordova2205.06243:2022ieu, 
Cordova2211.07639:2022fhg, CordovaKoren2212.13193:2022qtz}
have 3d charge operators of the noninvertible 0-form symmetries.)

$\bullet$ Ref.~\cite{Choi2205.05086:2022jqy, Cordova2205.06243:2022ieu}
finds the noninvertible symmetry interpretation of the $\U(1)$-$G^2$ ABJ anomaly
when $G$ is also an abelian group $G=\U(1)_V$, so that 
$G$ is meant to be a dynamically gauged group in
an abelian gauge theory with matter, such as QED. Thus, this has applications to
noninvertible symmetries in QED, and the re-derivation of a pion decay into two photos $\pi^0 \to \gamma\gamma$ \cite{Choi2205.05086:2022jqy}. Ref.~\cite{Cordova2205.06243:2022ieu}
also studies the case for the $\U(1)$-$G^2$ ABJ anomaly
with a nonabelian $G=\PSU(N)=\frac{\SU(N)}{\Z_N}$ with the matters in the adjoint representation of $G$.
Their constructions focus on the 3d charge operators of the noninvertible 0-form chiral symmetries.

$\bullet$ Ref.~\cite{Cordova2211.07639:2022fhg} applies the $\U(1)$-$\U(1)'^2$ ABJ anomaly
to the case of lepton number difference symmetries crossing different families:
the electron minus muon number
$\U(1)=\U(1)_{{\bf L}_e - {\bf L}_\mu}$
and the muon minus tau number $\U(1)'=\U(1)_{{\bf L}_\mu - {\bf L}_\tau}$.
It is easy to read from the formula $\dd \star j_{\bf L}$
in \eq{eq:dJQL} that as long as there are the same number of leptonic Weyl fermions in each family
(either fixing 3 or 4 leptons for all families, so fixing 15 or 16 Weyl fermions for all families),
the anomaly-free cancellation holds among
$\U(1)_{{\bf L}_e - {\bf L}_\mu}^3$, 
$\U(1)_{{\bf L}_\mu - {\bf L}_\tau}^3$,
$\U(1)_{{\bf L}_e - {\bf L}_\mu}$-gravity$^2$,
and
$\U(1)_{{\bf L}_\mu - {\bf L}_\tau}$-gravity$^2$ local anomalies.
However, the nonvanishing mixed anomaly 
$\U(1)_{{\bf L}_e - {\bf L}_\mu}$-$\U(1)_{{\bf L}_\mu - {\bf L}_\tau}^2$ implies that
in the case of dynamically gauged $\U(1)_{{\bf L}_\mu - {\bf L}_\tau}$ with the so-called BSM $Z'$ gauge boson,
the invertible $\U(1)_{{\bf L}_e - {\bf L}_\mu}$ is broken. 
But Ref.~\cite{Cordova2211.07639:2022fhg} revives the noninvertible counterpart 
of $\U(1)_{{\bf L}_e - {\bf L}_\mu}$ symmetry, and uses this noninvertible symmetry
to protect neutrino masses as well as to generate small neutrino masses through the quantum effect
of the instantons of the non-abelian `horizontal' lepton-symmetry gauged group at UV.

$\bullet$ Ref.~\cite{CordovaKoren2212.13193:2022qtz} studies the flavor symmetries between different families and the aspects of 
their higher-group symmetries or noninvertible symmetries.

$\bullet$ In contrast, in our work, we do not look at the low energy QED or QCD below the electroweak scale \cite{Choi2205.05086:2022jqy, Cordova2205.06243:2022ieu}. 
We also do not assume any additional BSM $Z'$ gauge boson \cite{Cordova2211.07639:2022fhg}, nor do we require any hypothetical GUT structure \cite{WangYou2111.10369GEQC} or any approximate flavor symmetry at UV \cite{CordovaKoren2212.13193:2022qtz}.
Instead, we only implement the full honest SM gauge structure and matter content,
\eq{eq:SMLieAlgebra} and \eq{eq:SMrep},
given by the confirmed experiments. 
Therefore, by the completeness of anomalies examined in 
\cite{WW2019fxh1910.14668, WangWanYou2112.14765, WangWanYou2204.08393},
what we obtain is really 
a \emph{noninvertible categorical symmetry} of the \emph{minimal} SM from the mixed gravitational anomaly. 
It is possible that there are other new types of constructions of {noninvertible symmetries}
beyond what we know of at this moment. But as long as the index $(-N_f +n_{{\nu}_R}) \neq 0$
in our SM vacuum, then our noninvertible symmetry charge operators are valid topological 
defects in the SM.

\item {{\bf Global symmetry in dynamical gravity vs No global symmetry in quantum gravity}}:

\cred{When the gravity becomes \emph{dynamical} but only fluctuates the spacetime \emph{semiclassically},
it still makes sense to discuss the global symmetry in the quantum theory coupling to 
semiclassical dynamical gravity. Thus, the invertible $\U(1)$ symmetry is broken by gravitational anomaly, 
but the noninvertible counterpart still survives,
at least within an effective field theory below a cutoff energy scale $\Lambda_{{\rm EFT-cutoff}}$, 
further below the full UV-complete quantum gravity.} 

In contrast, due to the absence of global symmetries in \cred{the full}
quantum gravity \cite{KalloshKLLS9502069,Banks2010zn1011.5120,HarlowOoguri1810.05338, MonteroRudelius2104.07036,McNamara1909.10355:2019rup}, the revived noninvertible symmetry
must be either (1) completely broken (so the conservation law
of revived noninvertible symmetry's charge operators must be violated again by another mechanism 
beyond the original mixed gravitational anomaly, such as through wormholes [e.g.,\cite{AbbottWiseWormholes1989}
and more recent references in \cite{BahChenMaldacena2212.08668}])
or (2) dynamically gauged in the UV-complete theory at 
\cred{$\Lambda_{{\rm EFT-cutoff}}$ or higher energy} 
(by gauging, we require to condense the topological defects ---
namely the noninvertible symmetry's charge operators 
must be absorbed and become part of the vacuum). 
In either case, it will be interesting to work out the details in the future.

\item {\bf Leptogenesis and Baryogenesis, Dirac vs Majorana masses 
vs exotic-BSM TQFT/CFT sectors}:

$\bullet$ \emph{Leptogenesis} \cite{Davidson0802.2962:2008bu}
concerns generic hypothetical physical processes that produced the lepton asymmetry (an asymmetry between the numbers of leptons and antileptons) in the very early universe, resulting in the present-day dominance of leptons over antileptons. 
In particular, 
a class of scenarios proposes 
that the baryon asymmetry of the universe is produced from 
the lepton asymmetry, e.g. generated in the decays of heavy sterile neutrinos.

$\bullet$ \emph{Gravitational leptogenesis} \cite{PhysRevLett.96.081301}
provides the lepton asymmetry based on the gravitational anomaly
\eq{eq:dJQL} so that lepton number violation
$\dd \star j_{\bf L} =     
(-N_f+n_{\nu_R})\,\frac{p_1}{24}=
(-N_f+n_{\nu_R})\,\frac{1}{24}\frac{\Tr[R \wedge R]}{8 \pi^2}$
comes from the gravitational background and curved spacetime.

$\bullet$ \emph{Baryogenesis} and the baryon asymmetry 
can follow from the sphaleron process,
once the lepton asymmetry is produced. 
The sphaleron converts between
$\dd \star j_{\bf L}$ and $\dd \star j_{\bf Q}$
via the SU(2) instanton or even U(1) instanton in \eq{eq:dJQL}. 
The lepton and baryon asymmetries also affect the Big Bang \emph{nucleosynthesis} at later times.

Previously \Refe{Adshead1711.04800:2017znw} studied the gravitational leptogenesis 
based on Dirac or Majorana neutrino mass scenarios. However, 
experiments have not yet confirmed (1) whether heavy sterile neutrinos do exist, 
(2) what is the index $(-N_f +n_{{\nu}_R})$, and (3) what is the mass generating mechanisms
for left-handed neutrinos as well as (if any) sterile right-handed neutrinos.

One interesting future direction is whether the noninvertible symmetry 
topological defects provide any new perspectives on the gravitational leptogenesis.
To recall, when we focus on the mixed $\Z_{4,X}$-gravitational anomaly 
classified by $(-N_f +n_{{\nu}_R}) \in \Z_{16}$, we decorate the generating  
topological defect with the 3d symmetric anomalous boundary topological order of 
the 4d $\Z_4^{\rm TF}$-time-reversal symmetric topological superconductor 
(4d $\Pin^+$ iTFT with $\rT^2=(-1)^\rF$) classified by $\Z_{16}$.

In fact, there has been a proposal to replace the hypothetical heavy sterile neutrino 
with the exotic 4d TQFT or CFT, called Ultra Unification \cite{JW2006.16996, JW2008.06499, JW2012.15860}.
In that case, the SM lives with 
the 4d symmetric anomalous TQFT or CFT on the boundary 
of the 5d bulk $\Z_{4,X}$-symmetric iTFT
(5d ${\Spin \times_{\Z_2^\rF} {\Z_{4,X}}}$ iTFT with $X^2=(-1)^\rF$) classified also by $\Z_{16}$.
It will be illuminating to explore the relations between all these physics better altogether.

\end{enumerate}

%\newpage

\section*{Acknowledgments}

The authors are listed in alphabetical order as a standard convention.
PP would like to thank Po-Shen Hsin, Mehrdad Mirbabayi and Cumrun Vafa for the relevant discussions.
JW appreciates 
Eduardo Garcia-Valdecasas, Justin Kaidi, 
Hotat Lam, Gabi Zafrir, and Yunqin Zheng
for helpful conversations
on the related issues in the past. 
JW thanks Meng Cheng, Jun Hou Fung, Clifford Taubes, Kai Xu, and Zheyan Wan for helpful comments on several clarifying statements in the manuscript.
%for correspondence on the index theorem and anomaly polynomials
We are grateful to the hospitality and inspiring venues provided by the conferences of
Simons Collaboration on Global Categorical Symmetries and Simons Center for Geometry and Physics in 2022. JW is supported by Harvard University CMSA.
%

%\newpage
\appendix

\section{Table of Representations of Quarks and Leptons}
\label{app:rep}

\begin{table}[!h]
$\hspace{-10.mm}
\setlength{\extrarowheight}{1pt}
  \begin{tabular}{ccccccc c c c c c c c c}
    \hline
    $\begin{array}{c}
    \textbf{SM}\\ 
   \textbf{fermion}\\
   \textbf{spinor}\\ 
   \textbf{field}
       \end{array}$
   & ${\SU(3)}$& ${\SU(2)}$ & $\U(1)_{Y}$  %{Y = \frac{2 \sqrt{15} T_{12}}{6}}
   & $\U(1)_{\tilde Y }$
   & $\U(1)_{\rm{EM}}$ 
    & $\U(1)_{{ \mathbf{B}-  \mathbf{L}}}$  & 
    $\U(1)_{{{\bf Q}} - {N_c}{\bf L}}$ 
    & 
    $\begin{array}{c}
     \Z_{ 2N_cN_f, {{\bf Q}} + {N_c}{\bf L}}\\
    \text{ as }\U(1)_{{{\bf Q}} + {N_c}{\bf L}}\\
    \mod 2N_cN_f
       \end{array}$
    & $\U(1)_{X}$
     & 
    $\Z_{5,X}$
 & 
    $\Z_{4,X}$
    & $\Z_{2}^\rF$  & SU(5) & Spin(10) \\
        \hline\\[-4mm]
    $\bar{d}_R$& $\overline{\mathbf{3}}$& $\mathbf{1}$ & 1/3  & 2 & 1/3 & $-1/3$ & $-1$ & $-1$ & $-3$ & $-3$ &1   &   1 &  \multirow{2}{*}{
     $\overline{\bf{5}}$}  &  \multirow{6}{*}{
     ${\bf{16}}$} \\
\cline{1-11} $l_L$& $\mathbf{1}$& $\mathbf{2}$& $-1/2$  & $-3$ & 0 or $-1$ & $-1$ & $-3$ & $+3$ & $-3$& $-3$  &1  &   1 \\
\cline{1-14}  $q_L$& ${{\mathbf{3}}}$& $\mathbf{2}$& 1/6  &1 & 2/3 or $-1/3$ &  1/3 &  $1$ &  $1$ &1&1 & 1 &    1 & \multirow{3}{*}{
     ${\bf{10}}$} \\
\cline{1-11} $\bar{u}_R$& $\overline{\mathbf{3}}$& $\mathbf{1}$& $-2/3$  & $-4$ & $-2/3$ & $-1/3$ & $-1$ & $-1$ &1 & $1$  &1 &   1 \\
\cline{1-11} $\bar{e}_R= e_L^+$& $\mathbf{1}$& $\mathbf{1}$& 1 & 6 & 1 & 1  & $3$ & $-3$  &1 &1 &   1 &    1\\
\cline{1-14} $\bar{\nu}_R= {\nu}_L $& $\mathbf{1}$& $\mathbf{1}$& 0  & 0 & 0 & 1 & $3$ & $-3$  & 5 &0 &   1 &   1 & \;\;${\bf{1}}$ &\\
%\hline
%    \hline $\phi_H$ & $\mathbf{1}$ & $\mathbf{2}$& 1/2 & 1 & 0 & -2 &   2 & 3 & 0 & -6  &   -2\\
    \hline\end{tabular}
$
\caption{Representations of quarks and leptons in terms of  
Weyl fermions 
in various internal symmetry groups.
Each fermion is a spin-$\frac{1}{2}$ Weyl spinor 
${\bf 2}_L$ representation \cpurple{of} the spacetime symmetry group Spin(1,3).
Each fermion is written as a left-handed particle $\psi_L$ or a right-handed anti-particle $\ii \sigma_2 \psi_R^*$.
}
\label{table:SMfermion}
\end{table}

For the reader's convenience, 
we organize the representations of Weyl fermions
with respect to various internal symmetry groups in \Table{table:SMfermion},
including:\footnote{For terminology, 
a \emph{vector} symmetry means that it transforms left-handed particles and right-handed particles equally,\\
a \emph{chiral} symmetry means that it transforms left-handed particles and right-handed particles differently,\\
and an \emph{axial} symmetry means that it transforms left-handed particles and right-handed particles oppositely.} 

\noindent
$\bullet$ SM Lie algebra
$\cG_{\rm SM} \equiv su(3) \times  su(2) \times u(1)_{\tilde Y}$ 
is compatible with four versions of Lie group $G_{\SM_\q} \equiv \frac{\SU(3) \times   \SU(2) \times \U(1)_{\tilde Y}}{\Z_\q}$ with $\q=1,2,3,6$.
In order to have a proper quantization,
we choose the charge of $\U(1)_{\tilde Y}$ as 6 times the charge of the particle physics convention $\U(1)_{Y}$. 
The $\U(1)_{\rm{EM}}$ is a linear combination of the $\U(1)_{T_3} \subset \SU(2)$ weak force gauge 
subgroup and $\U(1)_{\tilde Y}$.
\cred{In the SM, the $su(3)$ and $\U(1)_{\rm{EM}}$ are vector-like gauge symmetries,
the $su(2)$ and $u(1)_{\tilde Y}$ are chiral-like gauge symmetries.}

\noindent
$\bullet$ \cred{The vector} 
$\U(1)_{{\bf Q}-N_c {\bf L}}$ symmetry
(the precise form of $\U(1)_{\bf B - L}$ with properly quantized charges, with the color number 
$N_c=3$). The $\U(1)_{{\bf Q}}$ and $\U(1)_{{\bf L}}$ are also vector symmetries. 

\noindent
$\bullet$ \cred{The vector} $\Z_{2N_cN_f, {{\bf Q} + {N_c} {\bf L}} }$ symmetry
(the precise form of $\Z_{2 N_f,\bf B + L}$ with properly quantized charges).

\noindent
$\bullet$ \cred{The chiral} $X$ symmetry, with $X \equiv 5({ \mathbf{B}-  \mathbf{L}})-4Y \equiv 
5({ \mathbf{B}-  \mathbf{L}})-\frac{2}{3} {\tilde Y}
=\frac{5}{N_c}({ \mathbf{Q}-  N_c \mathbf{L}})-\frac{2}{3} {\tilde Y}$,
including $\U(1)_{X}$, $\Z_{5,X}$, and $\Z_{4,X}$.

\noindent
$\bullet$ Fermion parity $\Z_2^\rF$ symmetry. Note that $G_{\SM_\q}$ does not contain $\Z_2^\rF$.
So the fermion parity is not dynamically gauged within the $G_{\SM_\q}$. 
The SM requires a spin structure to have fermions. The quotient group 
$\frac{\Spin}{\Z_2^\rF}=\SO$ gives rise to the bosonic $\SO$ special orthogonal group of (local) spacetime rotations.

\noindent 
$\bullet$ SU(5): The multiplet
$\overline{\bf{5}}$, ${\bf{10}}$, and ${\bf{1}}$ structure
of the Georgi-Glashow SU(5) grand unified theory.
Note that the $\U(1)_{X}$ is compatible with the SU(5) multiplet
structure, so together they combine to form a $u(5)$ or $su(5)\times u(1)$ structure.
More precisely, it is compatible with the refined Lie group 
$\U(5)_{\hat \q}\equiv \frac{\SU(5) \times_{\hat \q} \U(1)_X}{\Z_{5,X}}$ defined in \cite{WangYou2111.10369GEQC},  with $\hat \q =2$ or $3$.
Both $\SU(5)$ and $\U(1)_X$ share the $\Z_{5,X}$ center normal subgroup, 
so that it is quotient over to define $\U(5)_{\hat \q}$.

\noindent 
$\bullet$ Spin(10): The multiplet {\bf 16} of Spin(10).
Note that $\Spin(10) \supset Z(\Spin(10)) =\Z_{4,X} \supset \Z_2^\rF$,
namely the Spin(10)  center $Z(\Spin(10))=\Z_4$ can be identified with $\Z_{4,X}$
which also contains a $\Z_2^\rF$ normal subgroup.

Note that a ``sterile right-handed'' neutrino (written as a right-handed anti-neutrino 
$\bar{\nu}_R$ and regarded a left-handed Weyl spinor here in \Table{table:SMfermion}) 
is \emph{only sterile} to the SM's 
strong and electroweak forces in $\cG_{\rm SM}$, and \emph{sterile} to the Georgi-Glashow SU(5) gauge force. However, the ``sterile right-handed'' neutrino is \emph{not sterile} to but 
charged under the ${\bf B \pm L}$ (more precisely $\U(1)_{{\bf Q} \pm N_c {\bf L}}$), 
$\U(1)_{X}$, $\Z_{5,X}$, and $\Z_{4,X}$, and $\Z_2^\rF$.

\section{Notations and Conventions}
\label{app:Notations-and-Conventions}

In this appendix, we explain the notations of the SM action \eq{eq:SM-action} in a curved spacetime,
and we convert some of the formal characteristic class or cohomology class expression\cpurple{s}
to the more friendly differential form or differential calculus expression\cpurple{s}. 
Note however the topological\cpurple{y} invariant data from \cpurple{a} characteristic class is \emph{not} captured by \cpurple{its local expression in} a single patch,
but instead is typically captured by the transition functions between different overlapping patches. 
So in order to define \cpurple{characteristic classes differential-geometrically}, 
we \emph{cannot} just use differential forms \emph{locally}, but we need to define \cpurple{them} \emph{globally}.
In any case, to be explicit, we will still write down the local data on a single patch of the manifold.

\begin{enumerate}

\item The completely antisymmetric Levi-Civita symbol $\tilde{\epsilon}_{\mu_1 \mu_2 \dots \mu_d}= \pm 1$
where $+1$ or $-1$ corresponds to even or odd permutation of the standard ordering $12\dots d$.
The Levi-Civita tensor is
${\epsilon}_{\mu_1 \mu_2 \dots \mu_d} \equiv \sqrt{|{\rm g}|}\tilde{\epsilon}_{\mu_1 \mu_2 \dots \mu_d}$.
The $\tilde{\epsilon}^{\mu_1 \mu_2 \dots \mu_d}\equiv \sgn({\rm g})\tilde{\epsilon}_{\mu_1 \mu_2 \dots \mu_d}$,
and 
${\epsilon}^{\mu_1 \mu_2 \dots \mu_d} \equiv \frac{1}{\sqrt{|{\rm g}|}}\tilde{\epsilon}^{\mu_1 \mu_2 \dots \mu_d}$.

\item
The volume $d$-form element $\dd^d x \equiv 
\dd t \wedge \dd x_1 \wedge \dd x_2 \wedge \dots \wedge \dd x_{d-1}
=\frac{1}{d!} \tilde{\epsilon}_{\mu_1 \mu_2 \dots \mu_d} {\dd x^{\mu_1} \dd x^{\mu_2} \dots \dd x^{\mu_d}}$ 
transforms as a density not as a tensor, 
but $\sqrt{|{\rm g}|}\dd^d x$ is an invariant volume element.

\item Given a metric of curved (pseudo-)Riemannian manifold
with a metric ${\rm g}_{\mu \nu}$ and its determinant ${\rm g}$,
we have the inverse ${\rm g}^{\nu \rho}$ so 
${\rm g}_{\mu \nu} {\rm g}^{\nu \rho}=\delta_\mu^\rho$.
\cpurple{A} vielbein diagonalizes the metric ${\rm g}_{\mu \nu}$ to the flat metric $\eta_{ab}$,
so ${\rm g}_{\mu \nu}(x) e^\mu{}_a e^\nu{}_b=\eta_{ab}(x)$
and ${\rm g}_{\mu \nu}(x) =\eta_{ab}(x)  e_\mu{}^a e_\nu{}^b$,
so the vielbeins $e_\mu{}^a$ are the ``square root'' of the metric ${\rm g}_{\mu \nu}$.
We use Greek indices $\mu,\nu,\dots$ for coordinates \cpurple{on} a curved manifold,
we use Roman Latin indices $a,b,\dots$  \cpurple{for coordinates on a flat space with the standard metric associated to a given point $x\in M$}. Note that $e^\mu{} _a e_\nu {}^a =\delta^\mu_\nu$ and $e^\mu {}_a e_\mu {}^b=\delta^b_a$
give the Kronecker delta.

The $\{\dd x^\mu\}$ is a set of \cpurple{basis 1-forms on} the \emph{cotangent space} $T^*M$, known as \cpurple{\emph{covariant vector space
} which is \emph{dual vector} space
 to the \emph{tangent space} $TM$ known as \emph{contravariant vector space}}.\footnote{Recall that the vector $V^\mu \prt_\mu$ in the vector space $TM$ has
the vector component $V^\mu$ and the basis $\prt_\mu$. The covariant vector $V_\mu \dd x^\mu$ in the dual vector space $T^*M$
has the component $V_\mu$ and the basis $\dd x^\mu$.} 
\cpurple{Let $\{e^a\}$ be the basis of 1-forms froming an orthonormal frame in $T^*M$}.
Both $\{\dd x^\mu\}$ and $\{e^a\}$ span the \emph{cotangent space} $T^*M$, they are related by \cpurple{a vielbein $e_\mu{}^a$
via}
$$
\text{$e^a(x)=e_\mu{}^a (x)\dd x^\mu$,
typically equivalently written as $\hat{\theta}^{(a)}= e_\mu{}^a \hat{\theta}^{(\mu)}$}.
$$
\cpurple{in terms of basis 1-forms $\hat{\theta}$}. The vielbein \cpurple{also relates two bases in $TM$}:
$$
\text{$\prt_\mu=e_\mu{}^a (x) \hat{e}_{(a)}$,
typically equivalently written as $\hat{e}_{(\mu)} = e_\mu{}^a \hat{e}_{(a)}$.}
$$
Thus, the vielbein $e_\mu{}^a$ plays double duties,
as the components of the orthonormal basis 1-form\cpurple{s} ($\hat{\theta}^{(a)}=e^a$)
in terms of the coordinate basis 1-form\cpurple{s} ($\hat{\theta}^{(\mu)}=\dd x^\mu$),
also 
as the components of the coordinate basis vector\cpurple{s}  ($\hat{e}_{(\mu)}=\prt_\mu$)
in terms of the orthonormal basis vector\cpurple{s}  ($ \hat{e}_{(a)}$).

In contrast, the inverse vielbein $e^\mu{}_a$ relates the basis 1-form\cpurple{s} $\hat{\theta}$
via
$$
\text{$\dd x^\mu = e^\mu{}_a(x)  e^a(x)$,
typically equivalently written as $\hat{\theta}^{(\mu)}= e^\mu{}_a \hat{\theta}^{(a)}$}.
$$
The inverse vielbein relates the basis vector\cpurple{s} $\hat{e}$ via
$$
\text{$\hat{e}_{(a)} =e^\mu{}_a (x) \prt_\mu$,
typically equivalently written as $\hat{e}_{(a)} =e^\mu{}_a \hat{e}_{(\mu)}$.}
$$
Thus, the inverse vielbein $e^\mu{}_a$ also plays double duties,
as the components of the coordinate basis 1-form\cpurple{s} ($\hat{\theta}^{(\mu)}=\dd x^\mu$) 
in terms of the orthonormal basis 1-form\cpurple{s} ($\hat{\theta}^{(a)}=e^a$),
also 
as the components of the orthonormal basis vector\cpurple{s} ($ \hat{e}_{(a)}$) 
in terms of the coordinate basis vector\cpurple{s} ($\hat{e}_{(\mu)}=\prt_\mu$).

The \cpurple{square of the} line element is given by $\dd s^2={\rm g}_{\mu \nu}(x) \dd x^\mu \dd x^\nu
=\eta_{ab} e^a(x) e^b(x)$.
We also have 
$\hat{\theta}^{(\nu)}(\hat{e}_{(\mu)})
=\frac{\prt x^{\nu}}{\prt x^{\mu} } =\delta^\nu_\mu$
and 
$\hat{\theta}^{(b)}(\hat{e}_{(a)})
=\delta^b_a$.

\item \cred{To formulate the Standard Model on a curved spacetime manifold mathematically,
we require the \cpurple{mathematical notions of fiber bundles and connections on them} (see
an introduction in \cite{nakahara2018geometry}).} 

\begin{enumerate}

\item  {\bf Fiber bundle}
consists of the data $(E,\pi,M,F,G)$, 
where the total space $E$, 
the base space $M$, and fiber $F$ all are differentiable manifolds;
the projection $\pi: E \to M$ is a surjection, and the structure group $G$ is a Lie group 
acting on the fiber $F$ from the left. 

\item For {\bf SM's gauge bundle},
we require \cpurple{a} {\bf principal $G$-bundle} such that 
$G=G_{\SM_\q} \equiv \frac{\SU(3) \times   \SU(2) \times \U(1)_{\tilde Y}}{\Z_\q}$ 
and the fiber is diffeomorphic to the structure group $G$.
\cpurple{A} principal $G$-bundle, denoted $P(M,G)$,
is a fiber bundle obeying an extra condition $(E,\pi,M,F,G)=(P(M,G),\pi, M,G,G).$

\item For the {\bf spacetime tangent bundle} $TM$ of the base manifold $M$,
the $TM$ is a special case of the {\bf vector bundle}.
The vector bundle is a fiber bundle obeying an extra condition
\cpurple{$(E,\pi,M,F,G)=(E,\pi, M, \rF^n ,\GL(n,\rF))$
such that the fiber $F=\rF^n$} is 
an $n$-dimensional vector space \cpurple{over} a field $\rF$ 
(namely, a commutative ring with multiplicative inverse),
and the structure group is a rank-$n$ general linear group $\GL(n,\rF)$ with 
a field coefficient $\rF$.
The tangent bundle is a vector bundle on a $d$-dim manifold $M=M^d$:
such that 
\cpurple{$(E,\pi,M,F,G)=(TM,\pi, M, \R^d ,\GL(d,\R))$ and $\pi^{-1}(x)=T_xM$}.

\item {\bf Spinor bundle} is required to describe the spinor field 
(in particular Weyl spinor in SM) as the section $s: M \to E$ 
of the spinor bundle.
A spinor bundle is a fibre bundle obeying 
$(E,\pi,M,F,G)=(E,\pi, M,V_{\rm S},\Spin\cpurple{(d)}).$
The spinor bundle is also a vector bundle with
a spinor representation $V_{\rm S}$ as a vector space:
A representation of the group $G$ group 
is a vector space $V$ together 
with a group homomorphism $G \to \GL(V)$ (here as $\Spin\cpurple{(d)} \to \GL(V_{\rm S})$),
where $\GL(V)$ is the general linear group of the vector space $V$.

\end{enumerate}

%When do we have a connection? or local 1-form gauge field? a curvature? or extension to a bulk?

\item 
In general, there is an infinite number of metric connections written \cpurple{as} Christoffel symbol 
$\Gamma^{\al}{}_{\bt \ga}$ 
(which is not a tensor by itself, but the difference of two connections 
$\Gamma^{\al}{}_{\bt \ga} - \Gamma'^{\al}{}_{\bt \ga}$ is a tensor) 
for a given metric tensor ${\rm g}_{\mu \nu}$; 
however, there is a unique connection that is torsion free 
$\Gamma^{\al}{}_{\bt \ga}=\Gamma^{\al}{}_{\ga \bt}$
and metric compatible $\nabla_\al {\rm g}_{\bt \ga}=0$, namely the Levi-Civita connection:
$$
 \Gamma^{\al}{}_{\bt \ga}=
 {\tfrac {1}{2}}{\rm g}^{\al \mu}
 \left(\partial_{\bt}{\rm g}_{\mu \ga}
 +\partial_{\ga}{\rm g}_{\mu \bt}
 -\partial_{\mu}{\rm g}_{\bt \ga}\right)
= \Gamma^{\al}{}_{\ga \bt}.
$$
The covariant derivative $\nabla_\mu$ on a tensor $V^{\nu \dots}{}_{\lambda}$ is given by
$\nabla_\mu V^{\nu \dots}{}_{\lambda \dots}=
\prt_\mu V^{\nu \dots}{}_{\lambda \dots}
+\Gamma^{\nu}{}_{\mu \nu'} V^{\nu' \dots}{}_{\lambda \dots} +\dots
-\Gamma^{\lambda'}{}_{\mu \lambda} V^{\nu \dots}{}_{\lambda' \dots} - \dots$.
Note that
$\nabla_\mu V^{\mu}=
\prt_\mu V^{\mu}
+\Gamma^{\mu}{}_{\mu \al} V^{\al} 
= \prt_\mu V^{\mu} + (\frac{1}{\sqrt{|g|}}\prt_\al \sqrt{|g|}) V^{\al} 
= \frac{1}{\sqrt{|g|}}\prt_\mu (\sqrt{|g|} V^{\mu})$.

The vielbein $e_\mu{}^a$ transforms as a covariant vector under the general coordinate  on $\mu$,
so combining with the basis $\dd x^\mu$,
there is a 1-form $e^a \coloneqq e_\mu{}^a  \dd x^\mu$.
The torsion-free spin connection is obtained from the covariant derivative under Levi-Civita connetion
on the vielbein only on ``the general coordinate curved spacetime index (here $\al$)'' 
$$
\omega_{\mu}{}^a{}_b  \coloneqq 
e_\al {}^a   (\prt_{\mu}  e^\al {}_b  +\Gamma^{\al}{}_{\mu \bt}  e^\bt {}_b) \equiv 
e_\al {}^a  \prt_{;\mu}  e^\al {}_b
= - e^\bt {}_b  \prt_{;\mu}  e_\bt {}^a
\equiv   e^\bt {}_b (-\prt_{\mu}  e_\bt {}^a   +\Gamma^{\al}{}_{\mu \bt}  e_\al {}^a ).
$$
Define
${e^{\mu a}=g^{\mu \nu }e_{\nu}{}^{a}}$ and $e_{\nu a}=\eta _{ab}e_{\nu}{}^{b}$,
so the Greek index can be raised or lower by $g^{\mu \nu }$ or $g_{\mu \nu }$,
the Latin / Lorentz index can be raised or lower by $\eta^{ab}$ or $\eta _{ab}$. 
Then we have ${\omega _{\mu }{}^{ ab}=-\omega _{\mu }{}^{ba}}$,
\bea
&&\omega _{\mu }{}^{ ab}
\coloneqq  e_{\al } {}^{a}(\partial_{\mu }e^{\al b}+\Gamma^{\al }{}_{\mu  \bt }e^{\bt  b })
\equiv e_{\al }{}^{a} \partial _{;\mu }e^{ \al b
}
={e^{\nu [a}({{e_{\nu }}^{b]}}_{;\mu }-{{e_{\mu }}^{b]}}_{;\nu }+e^{\sigma |b]}{e_{\mu }}^{c}e_{\nu c;\sigma })}
\cr
&&\equiv {{\frac {1}{2}}e^{\nu a}(\partial _{\mu }e_{\nu }^{\ b}-\partial _{\nu }e_{\mu }^{\ b})-{\frac {1}{2}}e^{\nu b}(\partial _{\mu }e_{\nu }^{\ a}-\partial _{\nu }e_{\mu }^{\ a})
+{\frac {1}{2}}e^{\nu a}e^{\sigma b}e_{\mu }^{\ c}(\partial _{\sigma }e_{\nu c}- \partial _{\nu }e_{\sigma c})}
=-\omega _{\mu }{}^{ba}.
\eea
The covariant derivative  $\nabla_{\mu}$ on the vielbein $e_{\nu}^a$, 
taking care of both the general coordinate transformation on the curved basis $\nu$ 
and the local Lorentz transformation on the flat basis $a$, shows that
\bea \label{eq:nabla-e}
\nabla_{\mu} e_{\nu}{}^a=\prt_{\mu} e_{\nu}{}^a - \Gamma^\al{}_{\mu \nu} e_{\al}{}^a + \omega_{\mu}{}^a{}_b  e_{\nu}{}^b=0.
\eea
The spin-connection 1-form is
\bea
\omega^a{}_b \coloneqq \omega_{\mu}{}^a{}_b  \dd x^{\mu}.
\eea
The torsion-free connection ($\Gamma^{\al}{}_{\bt \ga}=\Gamma^{\al}{}_{\ga \bt}$) condition is:
$$
\dd e^a + \omega^a{}_b e^b =0.
$$

The spin connection is used for taking the covariant derivative on the spinor.
In \eq{eq:SM-action},  the Weyl spinor lagrangian 
${\psi}^\dagger_L  (\ii \bar  \sigma^\mu {D}_{\mu, A} ) \psi_L$ in the curved spacetime
is the projection of the Dirac spinor lagrangian 
$\bar{\Psi}  (\ii   \gamma^\mu {D}_{\mu, A} ) {\Psi}$ in the curved spacetime.
\ccred{\cpurple{They contain} the 
generalized gamma matrix $\gamma^\mu= \hat \gamma^a e^\mu{}_a$
or
generalized sigma matrix $\sigma^\mu= \hat \sigma^a e^\mu{}_a$ in the curved spacetime
with the vielbein $e^\mu{}_a$;
however $\bar{\Psi}\equiv {\Psi}^\dagger \hat \gamma^0$.}
The flat tangent spacetime gamma matrices obey the anti-commutator relation:
$\{\hat \gamma^a, \hat \gamma^b\}= 2 \eta^{ab}$,
and the generalized gamma matrices obey: $\{ \gamma^\mu,  \gamma^\nu\}= 2 {\rm g}^{\mu \nu}$.

The ${D}_{\mu, A}$ contains
\bea \label{eq:derivative}
{D}_{\mu, A } \equiv \nabla_{\mu} + \,  q_{{\bf R}} \,  A_\mu + q_{{X}} \cA_{\mu}.
\eea
Here we 
choose to write the gauge field $A$ as a 
\emph{Lie-algebra valued differential 1-form}
\footnote{In \eq{eq:derivative}, we have the 
\emph{Lie-algebra valued differential 1-form} gauge field 
$A \equiv A_\mu \dd x^\mu = A_\mu^{\ra} \rT^{\ra} \dd x^\mu$. 
Lie algebra is a vector space ${\mathfrak {g}}$ over some field 
together with a binary operation 
$[\,\cdot \,,\cdot \,]:{\mathfrak {g}}\times {\mathfrak {g}}\to {\mathfrak {g}}$
that satisfies four axioms 
bilinearity, alternativity, Jacobi identity, and anticommutativity.
Lie-algebra 
commutator \cpurple{of generators} is
$[\rT^\rb, \rT^\rc]= f^{\rb \rc \rd} \rT^\rd$;
for \cpurple{a} real Lie algebra such as $u(1), su(n), so(n)$,
it has a {real-valued} structure constant $f^{\rb \rc \rd} \in \R$.
For $su(n)$ \cred{fundamental representation}, we have 
{anti-hermitian} (skew-self-adjoint)
Lie algebra generator\cpurple{s} {$\rT^a$} labeled by the gauge index ``$\ra$,'' such that {anti-hermitian} \cpurple{condition} requires
{$\rT^{\ra}= - {\rT}^{a \dagger}$}.
For $so(n)$ \cred{vector representation}, we have 
{anti-symmetric} Lie algebra generator\cpurple{s} {$\rT^{\ra}= - {\rT}^{a {\rm T}}$ relating to \cpurple{their} transpose}. 
\\
In contrast, in the
typical high-energy phenomenology literature, one uses in particular for the 
covariant derivative of the $su(n)$ gauge field
\bea  \label{eq:derivative-HEP-PH}
{D}_{\mu, A' } \equiv \nabla_{\mu} - \ii g \,  q_{{\bf R}} \,  A'_\mu - \ii q_{{X}} \cA'_{\mu},
\eea
such that the 1-form gauge field 
$A' \equiv A'_\mu \dd x^\mu = {A'}_\mu^{\ra} \rT'^{\ra} \dd x^\mu$
where ${A'}_\mu^{\ra} =\frac{1}{g} A_\mu^{\ra}$
and the {hermitian} (self-adjoint)
$\rT'^{\ra}=\ii \rT^{\ra}=\rT'^{\ra \dagger}$,
whose commutator satisfies $[\rT'^\rb, \rT'^\rc]= \ii f^{\rb \rc \rd} \rT'^\rd$
(note that this commutator does \emph{not} strictly satisfy the 
\emph{closed} binary operation 
$[\,\cdot \,,\cdot \,]:{\mathfrak {g}}\times {\mathfrak {g}}\to {\mathfrak {g}}$)
for the real-valued structure constant $f^{\rb \rc \rd}$.\\
The conversion between \eq{eq:derivative} and \eq{eq:derivative-HEP-PH},
say for a Dirac lagrangian of \eq{eq:SM-action}, is
\bea
A 
= A_\mu^{\ra} \rT^{\ra} \dd x^\mu
&=& 
- \ii g ({A'}_\mu^{\ra} \rT'^{\ra} \dd x^\mu)
= - \ii g A'.\cr
F = \dd A + A \wedge A &=& - \ii g( \dd A' - \ii g A' \wedge A') = - \ii g  F' .\cr
\bar\Psi \ii  (\nabla_\mu +  A_\mu ) \Psi \dd^4 x -
\frac{1}{g^2}\Tr(F \wedge * F) - \theta \frac{1}{8\pi^2} \Tr({F}\wedge {F})
&=&
\bar\Psi \ii  (\nabla_\mu - \ii g A'_\mu ) \Psi \dd^4 x +
\Tr(F' \wedge * F') + \theta \frac{g^2}{8\pi^2} \Tr({F'}\wedge {F'}).
\eea
}
that can be used to construct the \emph{connection 1-form},
where \cred{$q_{{\bf R}}$ and $q_{{X}}$ label the charge or the representation of the 
corresponding gauge fields}.

The $\nabla_{\mu}$ is a covariant derivative, 
which contains a spin connection for the Dirac spinor $\Psi$
with $\hat{S}^{ab} \equiv \frac{\ii}{4} [\hat \gamma^a, \hat \gamma^b]$
and $\omega_{\mu a b} \equiv \eta_{ac} \omega_{\mu}{}^c{}_b$,
\bea
\nabla_{\mu} \Psi \equiv
(\prt_{\mu} +\frac{1}{8} \eta_{ac} \omega_{\mu}{}^c{}_b [\hat \gamma^a, \hat \gamma^b] ) \Psi
 \equiv (\prt_{\mu} -\frac{\ii}{2}  \omega_{\mu a b}  \hat{S}^{ab} ) \Psi,
\eea
that can be projected by $P_L$, obtaining a  spin connection for the left-handed Weyl spinor $\psi_L = P_L \Psi
= \frac{1 - \hat \gamma^5}{2} \Psi$.
Suppose we choose Weyl representation of gamma matrices 
$\hat \gamma^{a}\equiv
\begin{pmatrix} 0 & {\hat \sigma}^{a}  \\
 \bar{\hat{\sigma}}^{a} & 0
\end{pmatrix}$ and 
$\hat \gamma^{5}\equiv
\begin{pmatrix}
-1 & 0 \\
0 & 1 
\end{pmatrix}$ 
for the flat orthonormal frame,
then ${\psi}^\dagger_L  (\ii \bar  \sigma^\mu \nabla_{\mu} ) \psi_L$ is obtained from the $P_L$ projection of Dirac lagrangian
\bea
{\Psi}^\dagger P_L \hat\gamma^0 (\ii   \gamma^\mu \nabla_{\mu} ) P_L {\Psi} 
&=& \ii  {\psi}^\dagger_L \hat\gamma^0 (\hat \gamma^{a'} e^\mu{}_{a'} \nabla_{\mu})\psi_L
= \ii  {\psi}^\dagger_L (
\begin{pmatrix} \bar{\hat \sigma}^{a'} & 0 \\
0 & \hat\sigma^{a'}
\end{pmatrix}
e^\mu{}_{a'} 
(\prt_{\mu} +\frac{1}{8} \omega_{\mu a b}
\begin{pmatrix} 
{\hat \sigma}^{a}  \bar{\hat \sigma}^{b} - {\hat \sigma}^{b}  \bar{\hat \sigma}^{a}  & 0 \\
0 & \bar{\hat \sigma}^{a}  {\hat \sigma}^{b} - \bar{\hat \sigma}^{b}  {\hat \sigma}^{a} 
\end{pmatrix}
)
)\psi_L\cr
\Rightarrow
{\psi}^\dagger_L  (\ii \bar  \sigma^\mu \nabla_{\mu} ) \psi_L
&=& \ii  {\psi}^\dagger_L  \bar{\hat \sigma}^{a'}  e^\mu{}_{a'}
(\prt_{\mu} +\frac{1}{8} \omega_{\mu a b}
({\hat \sigma}^{a}  \bar{\hat \sigma}^{b} - {\hat \sigma}^{b}  \bar{\hat \sigma}^{a} ) 
) \psi_L.
\label{eq:Weyl-lagrangian-curved}
\eea
In particular,
\bea
\nabla_{\mu} \psi_L \equiv
\prt_{\mu} +\frac{1}{8} \eta_{ac} \omega_{\mu}{}^c{}_b
({\hat \sigma}^{a}  \bar{\hat \sigma}^{b} - {\hat \sigma}^{b}  \bar{\hat \sigma}^{a} ) \psi_L.
\eea
The lagrangian
\eq{eq:Weyl-lagrangian-curved} is typically not hermitian, but call \eq{eq:Weyl-lagrangian-curved} as $\cL$,
then $\frac{1}{2}(\cL + \cL^\dagger)$ is hermtian thus real-valued.
The $\frac{1}{2}(\cL + \cL^\dagger)$ can be further simplified up to a total derivative term \cite{nakahara2018geometry},
by using
$\nabla_{\mu} \gamma_{\nu} = \nabla_{\mu} ( e_{\nu}{}^a \hat \gamma_a)=0$
based on $\nabla_{\mu}  \hat \gamma_a=0$ and
\eq{eq:nabla-e}'s $\nabla_{\mu} e_{\nu}{}^a=0$.

The $\cA_{\mu}$ corresponds to a ${\bf B-L}$'s $\U(1)_{{\bf Q}-N_c {\bf L}}$ gauge field
or a $\Z_{4,X}$ gauge field.
The $q_{{\bf R}} \,  A$ in  \eq{eq:SM-action} and \eq{eq:derivative}
contains the quantum number read from \Table{table:SMfermion}
\bea
  q_{{\bf R}}\,  A \equiv (q_{\tilde Y}  A_{{u(1)},\mu} 
+    \sum_{\ra=1}^3\frac{\varsigma^\ra}{2}   A_{{su(2)},\mu}^\ra
+  \sum_{\ra=1}^8\frac{\tau^\ra}{2} A_{{su(3)},\mu}^\ra )\dd x^\mu.
\eea

\item  In \eq{eq:SM-action},
Yang-Mills gauge theory has the action
$S_{\text{YM}}=\int  \frac{1}{g^2} \Tr(F_{} \wedge \star F_{})
=\int \cred{\hat{\cL}}_{\text{YM}} \cred{\sqrt{|{\rm g}|}}  \dd^4 x$ and Lagrangian $\cred{\hat{\cL}}_{\text{YM}}= -\frac{1}{4 {g^2}} F_{\mu\nu}^\ra F^{\ra{\mu\nu}} 
=-\frac{1}{4 {g^2}} 
{{\rm g}^{\mu \mu'}{\rm g}^{\nu \nu'}
F_{\mu\nu}^\ra F_{\mu'\nu'}^{\ra}}$.\footnote{Note that for a $p$-form $P$ in a total spacetime dimension $D$, the Hodge dual 
$(\star P)_{\mu_1 \dots \mu_{D-p}} 
\equiv \frac{1}{p!} \epsilon^{\nu_1 \dots \nu_p}{}_{\mu_1 \dots \mu_{D-p}} P_{\nu_1 \dots \nu_p}
 ={\rm g}^{\nu_1 \nu_1'}\dots {\rm g}^{\nu_p \nu_p'}
 \frac{1}{p!} {\sqrt{|{\rm g}|}} \tilde\epsilon_{\nu_1' \dots \nu_p' \mu_1 \dots \mu_{D-p}} P_{\nu_1 \dots \nu_p}$,
also $\star\star P=(-1)^{s+p(D-p)}P$ where $s$ is the number of negative eigenvalues in the metric.
So $\Tr [F \wedge *F]
=\Tr[ \frac{1}{2} F_{\mu\nu}(\dd x^\mu \wedge \dd x^\nu)
\wedge\frac{1}{2}(\frac{1}{2!}{\sqrt{|g|}}
\tilde\epsilon_{\mu_1\mu_2 {\nu_1  \nu_2}} F^{\nu_1 \nu_2})(\dd x^{\mu_1} \wedge \dd x^{\mu_2})]
=(-1)^{\rm{s}} \frac{1}{2}  \Tr[ F_{\mu\nu} F^{\mu\nu}] {\sqrt{|g|}} \dd^4 x$.}
The $F$ is the Lie algebra valued field strength curvature 2-form
\bea
F \coloneqq \dd A + A \wedge A,
\eea
with its Hodge dual {$\star F$}, all written \cpurple{as} differential forms.
For $u(1)$, $su(n)$, or $u(n)$ Lie algebra,
in the trace ``Tr'' we pick up an {anti-hermitian} (skew-self-adjoint)
Lie algebra \cred{fundamental} representation ${\bf R}$ whose {Lie algebra} generators {$\rT^a$} labeled by the gauge index ``$\ra$,''
such that {anti-hermitian} \cpurple{condition} requires
{$\rT^{a \dagger}= - {\rT^a}$}.
{We have also the subindex $I=1,2,3$ to specify the SM Lie algebra sectors $u(1)$, $su(2)$, or $su(3)$.}

Precisely 
$F=\frac{1}{2} F_{\mu\nu}( \dd x^\mu \wedge \dd x^\nu) =\frac{1}{2} F^\ra_{\mu\nu} \rT^\ra( \dd x^\mu \wedge \dd x^\nu)$,
and the commutator $[\rT^\rb, \rT^\rc]= f^{\rb \rc \rd} \rT^\rd$ with a \cred{real-valued} structure constant $f^{\rb \rc \rd}$, 
then 
$F^\ra_{\mu\nu} = \partial_\mu A^\ra_\nu -\partial_\nu A^\ra_\mu +  f^{\rb \rc \ra} A^\rb_{\mu} A^\rc_{\nu}$.
Lie algebra satisfies the Jacobi identity, 
$[\rT^\ra, [\rT^\rb, \rT^\rc]] + [\rT^\rb, [\rT^\rc, \rT^\ra]]
+ [\rT^\rc, [\rT^\ra, \rT^\rb]]=0$,
which implies 
$f^{\ra \rd \re} f^{\rb \rc \rd}
+f^{\rb \rd \re} f^{\rc \ra \rd}
+f^{\rc \rd \re} f^{\ra \rb \rd}
=0$ summed over repeated indices.

{Note} that $\Tr(\rT^\ra \rT^\rb) \equiv C({\bf R}) \delta^{\ra \rb}$ and 
$\sum_{a} \rT^\ra \rT^\ra \equiv C_2({\bf R}) \mathbb{I}_{d({\bf R}) \times d({\bf R})}$
for some constant of representation ${\bf R}$.
We have $\sum_{\ra}\Tr(\rT^\ra \rT^\ra)=C({\bf R}) d(G) =
C_2({\bf R}) d({\bf R})$ with 
the dimension of Lie group $d(G)$ 
and the rank of the representation matrix $d({\bf R})$.
Here for the $su(n)$ fundamental representation ${\bf R}$ in SM, we take 
\cred{$\Tr(\rT^\ra \rT^\rb)=-\frac{1}{2} \delta^{ab}$}.
Then we have $-\Tr(F \wedge * F)
=-(-1)^{\rm{s}}
\frac{1}{2}
 \Tr(F_{\mu\nu} F^{\mu\nu} ) {\sqrt{|{\rm g}|}} \dd^4x
= (-1)^{\rm{s}}
(\frac{1}{4} )
F^\ra_{\mu\nu} (F^\ra)^{\mu\nu} {\sqrt{|{\rm g}|}} \dd^4x$ 
with the $(-1)^{\rm{s}}$ as the sign of the determinant of the spacetime metric.
Here $(-1)^{\rm{s}}=-1$ in the Minkowski signature.
In \eq{eq:SM-action}, we normalize the $u(1)$ Yang-Mills theory slightly differently from the conventional $u(1)$ Maxwell theory by scaling a factor $\Tr(1)$ of the $u(1)$ by a 
\cred{$-\frac{1}{2}$}.

\item 
We write the wedge product of the field strength $F^n \equiv F \wedge F \wedge \dots \wedge F$ such as in \Sec{subsec:Summary}, similarly for other wedge products.
The first and second Chern classes of complex vector bundles $\cE$ are related to
the field strength curvatures via
\bea
c_1 &=& \frac{\Tr {F}}{2\pi}.\\
c_2 &=& \frac{1}{8\pi^2} (-\Tr({F}\wedge {F}) +  (\Tr {F}) \wedge(\Tr {F})).
\eea
The instanton number is
given by the integral $\int_{M^4} \frac{1}{8\pi^2} \Tr({F}\wedge {F})=\int_{M^4} - c_2 +\frac{1}{2} c_1^2$
over a 4d spacetime base manifold $M^4$.

\item  The commutator of two covariant derivatives
gives $[\nabla_\al, \nabla_\bt] V^\lambda=R^{\lambda}{}_{\nu \al \bt} V^\nu - T^{\lambda}{}_{ \al \bt} \nabla_{\lambda} V^\nu $
where the Riemann curvature tensor is
$$
R^{\mu}{}_{\nu \al \bt}
\coloneqq\prt_\al \Gamma^\mu{}_{\nu \bt}
-\prt_\bt \Gamma^\mu{}_{\nu \al}
+\Gamma^\mu{}_{\zeta \al}\Gamma^\zeta{}_{\nu \bt}
-\Gamma^\mu{}_{\zeta \bt}\Gamma^\zeta{}_{\nu \al}
$$
and the torsion tensor 
$T^{\lambda}{}_{ \al \bt} \coloneqq 
\Gamma^{\lambda}{}_{ \al \bt}
- \Gamma^{\lambda}{}_{\bt \al} \equiv 2 \Gamma^{\lambda}{}_{[ \al \bt]}$ is zero 
only by the torsion-free Levi-Civita connection.
Below we shall focus only on the torsion-free Levi-Civita connection. 

The curvature tensor 2-form $R^a{}_b$ is constructed out of the spin-connection 1-form 
$\omega^a{}_b \coloneqq \omega_{\mu}{}^a{}_b  \dd x^{\mu}$
as
\bea
R^a{}_b &\coloneqq &
\dd \omega^a{}_b 
+ \omega^a{}_c \wedge \omega^c{}_b 
= \frac{1}{2} R_{\mu \nu}{}^a{}_b \dd x^\mu \wedge \dd x^\nu
= \frac{1}{2} R^a{}_b{}_{\mu \nu} \dd x^\mu \wedge \dd x^\nu\cr
&=&\frac{1}{2}  
(\prt_\mu \omega_{\nu}{}^a{}_b -\prt_\nu \omega_{\mu}{}^a{}_b
+ \omega_{\mu}{}^a{}_c \omega_{\nu}{}^c{}_b - 
\omega_{\nu}{}^a{}_c \omega_{\mu}{}^c{}_b )\dd x^\mu \wedge \dd x^\nu 
.
\eea
\cred{Both $\omega^a{}_b$ and $R^a{}_b$ are
\emph{Lie-algebra valued differential forms}
with the $so(4)$ Lie algebra, 
namely the Euclidean version of the Lorentzian $so(1,3) \cong sl(2, \mathbb{C})$ Lie algebra.}
Here $R_{\mu \nu}{}^a{}_b=
R_{\mu \nu}{}^{\mu'}{}_{\nu'} e_{\mu'}{}^a e ^{\nu'} {}_b 
={\rm g}^{\mu' \mu''} R_{\mu \nu \mu'' \nu'} e_{\mu'} {}^a e^{\nu'} {}_b 
={\rm g}^{\mu' \mu''} R_{\mu'' \nu' \mu \nu} e_{\mu'}{}^a e^{\nu'} {}_b
= R^{\mu'}{}_{ \nu' \mu \nu} e_{\mu'}{}^a e^{\nu'} {}_b 
=R^a{}_b{}_{\mu \nu}$.
We raise or lower the indices of a tensor by the metric
$R_{\mu \nu}{}^{\mu'}{}_{\nu'}= {\rm g}_{\mu \al} {\rm g}^{\mu' \bt} R^{\al}{}_{\nu \bt \nu'}$.
Obviously $R^{\mu}{}_{\nu \al \bt}=-R^{\mu}{}_{\nu\bt  \al }$.
Moreover, the lower indiced Riemann curvature tensor
$R_{\mu \nu \al \bt}= {\rm g}_{\mu \mu'} R^{\mu'}{}_{\nu \al \bt}$ exhibits 
more algebraic (anti-)symmetric properties for exchanging indices':
$R_{\mu \nu \al \bt}=-R_{\nu \mu  \al \bt}$,
$R_{\mu \nu \al \bt}=-R_{\mu \nu \bt  \al}$,
$R_{\mu \nu \al \bt}=R_{\al \bt \mu \nu}$,
also the first Bianchi (algebraic) identity with antisymmetrized
$R_{\mu [\nu \al \bt]}=\frac{1}{3}(R_{\mu \nu \al \bt}+R_{\mu  \al \bt \nu}
+R_{\mu \bt \nu \al})=0$ thus similarly the antisymmetrized $R_{[\mu \nu \al \bt]}=0$.
There is also a differential identity called the second Bianchi (differential) identity,
the antisymmetrized $\nabla_{[\lambda }R_{\mu \nu] \al \bt}=0$, closely related to the Jacobi identity
$[\nabla_\lambda, [\nabla_\mu,\nabla_\nu]]+[\nabla_\mu, [\nabla_\nu,\nabla_\lambda]]+
[\nabla_\nu, [\nabla_\lambda,\nabla_\mu]]=0$.

In addition, other familiar types of curvatures or tensors include
the symmetric Ricci tensor 
$R_{\al \bt} = R_{\bt \al} \coloneqq R^{\lambda}{}_{\al \lambda \bt}$,
the Ricci scalar $R \coloneqq R^\al {}_\al=  {\rm g}^{\al \bt}R_{\al \bt}$. 

\item 
\cred{For some characteristic class or anomaly polynomial $I \equiv I_{d+2}$ in ${d+2}$ dimensions
defined such as 
in the de Rham cohomology (or Chern?Weil theory), the quantization of 
$\int I_{d+2}$ implies that the quantization of the \cpurple{overall coefficient (``level'') in  the corresponding $(d+1)$-dimensional action defined by the descend procedure, that is as} the integration of 
differential form $C \equiv C_{d+1}$ such that $I_{d+2} = \frac{1}{2\pi} \dd C_{d+1}$.}

\begin{enumerate}

\item
For one example, take $I$ to be the instanton number density $n_{\rm inst}$ in 4d
as a combination of \cpurple{certain Chern classes
of complex vector bundles $\cE$ of associated to a certain representation  of the gauge group $G$. For $SU(n)$ and $U(n)$ groups one ordinarily considers the defining representation. 
We have}:
\bea
n_{\rm inst}&\coloneqq&\frac{1}{8\pi^2} \Tr({F}\wedge {F})= - c_2 +\frac{1}{2} c_1^2 \equiv \frac{1}{2\pi} \dd \CS.\\
\CS &\coloneqq&  \frac{1}{4\pi}\Tr[A \wedge d A +
    \frac{2}{3}\,A \wedge A \wedge A]
\eea

\item
For another example, take $I$ to be the first Pontryagin class,
\bea
p_1 &\coloneqq&- \frac{1}{8 \pi^2}   \Tr[ R \wedge R]
= - \frac{1}{8 \pi^2}   R^a{}_b \wedge R^b{}_a
= - \frac{1}{8 \pi^2}
\frac{1}{2^2}
\tilde\epsilon^{\mu \nu \al \bt} R^a{}_b{}_{\mu \nu}  R^b{}_a{} {}_{\al \bt} \dd^4 x
\equiv -\frac{1}{2\pi} \dd \GCS.\\
\GCS &\coloneqq&  \frac{1}{4\pi}\Tr[\omega\wedge d\omega+
    \frac{2}{3}\,\omega\wedge \omega\wedge \omega]
=\frac{1}{4\pi}
\tilde\epsilon^{ \nu \al \bt} 
(\omega_{\nu}{}^a{}_b \prt_\al \omega_{\bt}{}^b{}_a 
+ 
\frac{2}{3}\, \omega_{\nu}{}^a{}_b \omega_{\al }{}^b{}_c \omega_{\bt}{}^c{}_a
) \dd^3 x.
\eea
Crucially \cpurple{$\Tr[A^4]=0$ and $\Tr[\omega^4]=0$ were used above.
This is} more generally for any even power of Lie algebra valued differential form \cpurple{of an odd degree},
due to the cyclicity of the trace and the anticommutativity of the differential form.

\item
\cred{For more examples, when we have a perturbative local anomaly of QFT in $d$d
classified by $k\in \Z$ and \cpurple{which is for $k=1$} captured by a $d+2$d anomaly polynomial 
$I_{d+2}$ (see Appendix
\ref{app:anomaly-polynomial-quantization}), we write the \emph{invertible} %(?) 
anomaly polynomial partition function
\bea
\exp(\ii k \int_{M^{d+2}} \theta  I_{d+2})
\eea
When $M^{d+2}$ is a closed ${d+2}$-manifold,  
$\int_{M^{d+2}} I_{d+2} \in \Z$
with  different $\theta \in [0, 2 \pi)$ specifying a different invertible field theory.}

When $M^{d+2}$ has a boundary $\prt M^{d+2} = M^{d+1}$, 
 we can consider this $M^{d+1}$ as a ${d+1}$d interface between two ${d+2}$d bulks with the lagrangian density
$\theta  I_{d+2}$ such that $\theta=0$ 
on one ${d+2}$d side and $\theta=2 \pi$ on the other ${d+2}$d side.
Define the relation $I_{d+2} = \dd I_{d+1} = \frac{1}{2\pi} \dd C_{d+1}$.
On the interface $M^{d+1} =\prt M^{d+2}$, we have an invertible topological\footnote{\cpurple{If the anomaly polynomial contains Pontryagin classes of the tangent bundle of the spacetime, and one chooses their differential form representatives using Levi-Civita connection, the theory is actually not topological in the usual sense, because the action will depend on metric.}} field theory (iTFT)
with the action
$$
S_{d+1}
= 2 \pi \int_{ M^{d+1}} I_{d+1}
=   \int_{ M^{d+1}} C_{d+1} 
= 2 \pi \int_{ M^{d+2}} \dd I_{d+1}
= 2 \pi \int_{ M^{d+2}}  I_{d+2}
\in 2 \pi \R
,$$ 
the $d+1$d iTFT partition function on the interface $M^{d+1}$
is 
\bea \label{eq:iTFT-S-U1}
\exp(\ii k   S_{d+1}) \in \U(1).
\eea 
Its value $S_{d+1}$ 
modulo $2 \pi $ is independent of the choice of the extension from the interface 
$M^{d+1}=\partial M^{d+2}=\partial M'^{d+2}$, because their difference is
$\exp(\ii k \cdot  2 \pi \int_{ M^{d+1} \sqcup (-M'^{d+2})} I_{d+1}) =1$,
where the gluing $M^{d+1} \sqcup (-M'^{d+2})$ is a closed manifold without boundary.
Since $S_{d+1}$ is metric-independent and extension-independent,
thus it is topological. \Eq{eq:iTFT-S-U1} is \cpurple{the partition function of an invertible TFT (coupled to background fields) for any $k\in\Z$}.

%{\bf Is it invertible or iTFT for a generic $k$ for CS vs GCS? for background fields? How about dynamical fields? How about $k \in \Q$ or $\R$?}

\item
\cred{When we have a nonperturbative global anomaly of QFT in $d$d
classified by $k\in \Z_n$, it is typically captured by a $d+1$d iTFT
on the $M^{d+1}$
\cpurple{for which the partition function} is only a $\Z_n$ subgroup of $\U(1)$:
\bea \label{eq:iTFT-S-Zn}
\exp(\ii k   S_{d+1})  \cpurple{\;\in \Z_n\subset \U(1)},
\eea 
where $S_{d+1} \in 2 \pi \frac{\Z_n}{n}$ valued.
}

\end{enumerate}

\end{enumerate}
    
\section{6d Anomaly Polynomial Generators for 4d Anomaly with U(1) Symmetry from Index Theorem}
\label{app:anomaly-polynomial-quantization}

Below we will characterize the 4d anomaly with $\U(1)$ symmetry
for fermionic (\cpurple{i.e. defined on manifolds with} $\Spin \times \U(1)$ or $\Spin \times_{\Z_2^\rF} \U(1) \equiv \Spin^c$ structure)
or bosonic (\cpurple{i.e. defined on manifolds with} $\SO \times \U(1)$ structure)
theories by writing down the 6d anomaly polynomial generators 
derived from the index theorem. See the comparison of these structures also in \cite{1405.7689, WanWang2018bns1812.11967}.

\begin{enumerate}

\item Compare the fermionic and bosonic cases:

\begin{enumerate}

\item Fermion: \cred{In the fermion case, we read the 4d anomaly and its associated 5d invertible theory
action
$S_5= 2 \pi \int_{ M^5} I_5 \in 2 \pi \R$
from the 6d anomaly polynomial 
$I_6=\dd I_5$ whose integration over a closed 6-manifold is valued in $\Z$,
from
the $\hat{\rA}$ genus and the Chern character $\ch(\mathcal{E})$,
\bea
\hat{\rA} \, \ch(\mathcal{E}),
\eea
where $\hat{\rA}$ and $\ch(\mathcal{E})$ are already given in \eq{eq:hatA} and \eq{eq:ch}:
\bea
\hat{\rA} &=&1-\frac{p_1}{24}+ \frac{7 p_1^2 - 4 p_2}{5760}+ \ldots, \\
\ch(\mathcal{E})&=&\mathrm{rank}\,\mathcal{E}+c_1(\mathcal{E})+
    \frac{1}{2}\left(c_1^2(\mathcal{E})-2c_2(\mathcal{E})\right)+
    \frac{1}{6}\left(
(c_1^3(\mathcal{E})-3c_1(\mathcal{E})c_2(\mathcal{E})+3c_3(\mathcal{E})
    \right)+\ldots
\eea
\cpurple{For a single left-handed Weyl fermion of charge $q$ we take $\mathcal{E}$ to be the complex line bundle associated with the corresponding representation of $\U(1)$}.
Hence, the fermionic 6d anomaly polynomial is
\bea \label{eq:I6f}
I_{6,f} =[\hat{\rA} \, \ch(\mathcal{E})]_6=
q^3 \frac{c_1^3}{6} -  
q\frac{c_1p_1}{24} \in \Z.
\eea
}

Consider a collection of \cred{left-handed} Weyl fermions in 4d with the global $\U(1)$ symmetry charges $q_i,i=1,\ldots, n_L$  respectively.\footnote{\cred{For a unit
charge 1 of an \emph{axial} $\U(1)$ symmetry in the Weyl fermion basis,
we choose the left-handed particle and right-handed anti-particle to have $q=1$.
We choose the right-handed particle and left-handed anti-particle to have $q=-1$.\\
For a unit
charge 1 of a \emph{vector} $\U(1)$ symmetry in the Weyl fermion basis,
we choose the left-handed particle and right-handed particle to have $q=1$.
We choose the left-handed anti-particle and right-handed anti-particle to have $q=-1$.}
} 

They have the following degree 6 anomaly polynomial which can be computed as the index of the 6d Dirac operator via Atiyah-Singer index theorem \cite{AlvarezGaume1983igWitten1984,Alvarez-Gaume:1984zlq}:
\begin{equation}
I_6= \left(\cred{\sum_{i=1}^{n_L}{q}_i^3}\right)\frac{c_1^3}{6} -  
\left(\cred{\sum_{i=1}^{n_L}{q}_i }\right)\frac{c_1p_1}{24} \in \Z.
\label{6d-anomaly-pol}
\end{equation}
Below we will consider two compatible 
$\Spin \times \U(1)$ and $\Spin \times_{\Z_2^\rF} \U(1) \equiv \Spin^c$ structures
for the fermion case with a U(1) symmetry.

\item Boson:
\cred{In contrast, in \cpurple{bosonic case we can produce a} 6d anomaly polynomial from
the $L$ genus and the Chern character $\ch(\mathcal{E})$.}

\cred{The Hirzbuch signature theorem has a formula expressing the signature $\sigma(M)\in\Z$ of manifold $M^d$,
if the dimension $d=4n$ thus $d = 0 \mod 4$,
\bea
\sigma(M^d)=\int_{M^d} L_n =\int_{M^d} L_{d/4} \equiv \< L_{d/4}, [M^d] \>
\eea
where the $L$ genus is given by
\bea
\label{eq:L}
L &=& L_0 + L_1 + L_2 + \dots=  1+ \frac{p_1}{3}+ \frac{-p_1^2+7 p_2}{45}+ \dots.
\eea
and the $[M^d]$ is the fundamental class of $M^d$. For example, $\sigma(M^4)=\< \frac{p_1}{3}, [M^4] \>$.}

\cpurple{For the signature operator twisted by a complex vector bundle $\mathcal{E}$ the index theorem gives the following modification of the formula above (see e.g. Theorem 3.1.5 in \cite{gilkey2018invariance})}
\bea
\cpurple{\Z\ni}\;\sigma_{\mathcal{E}}(M^d)=\sum_{4 i + 2 j=d}  2^j \int_{M^d} L_i \; \ch_j(\mathcal{E}).
\eea
Take $d=6$ and take $\mathcal{E}$ as the complex \cpurple{line} bundle associated with the $\U(1)$ representation $q=1$, % \cred{(of what? a single charge $q$ boson has no anomaly...)},
we have
\bea
\sigma_{\mathcal{E}}(M^6)= 
2^1 
\int_{M^6} L_1 \; \ch_1(\mathcal{E}) + 2^3 \int_{M^6}  \ch_3(\mathcal{E})
= 2 \frac{p_1}{3} c_1 + 8 \frac{1}{6} c_1^3.
\eea
Combin\cpurple{g} the above $\frac{4}{3} c_1^3 +\frac{2}{3} c_1 p_1$ 
with \cpurple{an obviously} integer class $-c_1 p_1$ (because
$c_1$ and $p_1$ are integer cohomology classes), 
we \cpurple{can obtain
the following} bosonic 6d anomaly polynomial for a bosonic theory:
\bea \label{eq:I6b}
I_{6,b} =
 \frac{4 c_1^3}{3} -  
\frac{c_1p_1}{3} \in \Z.
\eea
\cred{Note that $I_{6,b}=8 I_{6,f}$ for a single Weyl fermion with $q=1$.
To understand \cpurple{the relation} $I_{6,b}=8 I_{6,f}$ from a 4d anomaly perspective,
we can start with a fermionic theory with 8 Weyl fermions each with $q=1$ in 4d, 
and then we bosonize this theory by summing over the spin structure to a bosonic theory.
The $I_{6,b}=8 I_{6,f}$ captures the 4d anomaly of the bosonic and fermionic theories.}\\

\cred{\cpurple{Note that for $\CP^3$, 
although there exists a $\U(1)$ bundle such that $c_1p_1= 4$},\footnote{\cred{Let $h$ be a generator in $\H^2(\CP^n,\Z)=\Z$.
Note that  $T(\CP^n)$ $+$ a trivial complex line bundle $=$ a sum of $n+1$ tautological complex line bundles 
(the complex version of Theorem 4.5 of \cite{milnor1974characteristic}),
hence total Chern class $c(T\CP^n)=(1+h)^{n+1}$ and total Pontryagin class $p(T\CP^n)=(1+h^2)^{n+1}$.
For $n=3$, 
$T(\CP^3)$ $+$ a trivial complex line bundle $=$ a sum of 4 tautological complex line bundles,
hence $c(T\CP^3)=(1+h)^4$ and $p(T\CP^3)=(1+h^2)^4$.
So the minimum $c_1(\U(1))=h$ because the $h$ generates $\H^2(\CP^n,\Z)=\Z$
and $p_1(T\CP^3)=4h^2$, we have $c_1(\U(1))^3=h^3$ and $c_1(\U(1))p_1(T\CP^3)=4h^3$.
\label{ft:CPn-tautological}
}} 
so $c_1p_1/3=4/3$ is fractional, 
but $I_{6,b} = \frac{4 c_1^3}{3} -  \frac{c_1p_1}{3} =0$
or $\frac{1}{3}(c_1^3-c_1p_1) = 1/3 - 4/3 = -1$, which generates the \cpurple{integers. It follows that any integer value of $I_{6,b}$ can be realized. In particular it can be chosen as one of the two generators of $\Z^2$ group classifying the anomalies in the bosonic case}. }\\

\cred{There is also a 1+1d CFT interpretation for the ratio 8 between \eq{eq:I6b} and \eq{eq:I6f}.
For the bosonic CFT, we have the chiral central charge $c_-=8$ for the 1+1d E${}_8$ CFT 
that can implement U(1) symmetry on each of the eight $c_-=1$ compact chiral boson theory.
Although for the fermionic CFT, 
we have the minimal chiral central charge $c_-=1/2$ for the 1+1d free 
real-valued Majorana-Weyl fermion, it cannot implement a U(1) symmetry.
The minimal fermionic \cpurple{theory} that can implement a U(1) symmetry is a free complex-valued Weyl fermion, 
which has  a chiral central charge $c_-=1$.  So the ratio of $c_-$ for \cpurple{the bosonic CFT over $c_-$ for the fermionic one}
is 8. These two 2d CFTs correspond to the boundar\cpurple{ies} of two 3d gravitational Chern-Simons theories GCS with their level ratio 8, which further corresponds to the 4d first Pontryagin class
$p_1=-\dd \GCS/(2\pi)$ with their level ratio (namely $p_1/3$ over $p_1/24$) also 8.}\\

Below we will also consider a compatible 
$\SO \times \U(1)$ structure for the boson case with a U(1) symmetry.

\end{enumerate}

\item $\Spin \times \U(1)$ structure:

Assuming that $\Z_2^\rF$ is not included in the $\U(1)$, 
\cred{which means the spacetime-internal symmetry group structure is
$\Spin \times \U(1)$ structure},
the 6d anomaly polynomial $I_6$ above is in general a linear combination (over $\Z$, if all the charges 
${q}_j$ are integers) of the following two terms:
\begin{equation}
\label{eq:SpinU1-polynomial}
   I^{A}\coloneqq \frac{c_1^3}{6}-\frac{c_1p_1}{24} \in \Z,\qquad I^{B}\coloneqq{c_1^3} \in \Z.
\end{equation}

Those are the values of $I_6$ for the charge vectors \cred{$q=(1)$ and $q=(2,-1,-1)$} respectively. For general charges, we have:
\begin{equation}
    I_6=
    \left( \sum_{i=1}^{n_L}{q}_i\right)I^A
    +\left( \sum_{i=1}^{n_L}\frac{{q}_i^3-q_i}{6}\right)I^B \in \Z.
\end{equation}
Note that $(q^3-q)/6\in \Z$ for any $q\in \Z$. Of course, instead of $(I^A,I^B)$ as above, one can consider another pair related to it by a $\GL(2,\Z)$ transformation.

Moreover, $I^A$ and $I^B$ serve as the two generators of $\Hom(\Omega_6^{\Spin}(\B\U(1)),\Z)\cong \Z\times \Z$ by considering their integrals over 6-manifolds representing the elements in the bordism group. This can be argued as follows. First, the fact that $I^A$ and $I^B$ are integer-valued on any representative follows from Atiyah-Singer index theorem for the twisted Dirac operator. It is then enough to check that there exists a pair of representatives in the bordism group such that the values of $(I^A,I^B)$ on them form a basis in $\Z^2$. 

$\bullet$ For the first  such representative, $W_1$, let us take a spin 6-manifold $S^2\times S^2\times S^2$ with $c_1=a+b+c$, where $a,b,c$ are Poincar\'e dual to $[\pt\times S^2\times S^2],\,[S^2\times \pt\times S^2],\,[S^2\times S^2\times \pt]$ respectively. \cred{We have $c_1^3=(a+b+c)^3=6 a b c$, where $a,b,c$ all are degree 2 thus 
they commute in wedge product.}
So $(I^A,I^B)(W_1)=(1,6)$, which follows from the fact that signature of $S^2\times S^2$ is zero. 

$\bullet$ For the second representative, $W_2$, let us take a spin 6-manifold $\CP^3$ with $c_1=h$, the standard generator of $\H^2(\CP^3,\Z)=\Z$ (\cpurple{proportional to} the class of the K\"ahler form). Using the fact
 $p_1=4h^2$, \cred{in the $h=1$ case}, we obtain $(I^A,I^B)(W_2)=(0,1)$.

 \cpurple{Since (1,6) and (0,1) generate $\Z^2$, this already proves that $I^A$ and $I^B$ form a basis. As the dual basis in the bordism group, one can take $W_2$ and $W_1'=W_1 \# (-W_2)^{\# 6}$, constructed}
 \cred{using the connected sum $(\#)$ and the orientation reversal $(-)$}, 
\cpurple{so that $(I^A,I^B)(W_1')=(1,0)$}.

\item $\Spin \times_{\Z_2^\rF} \U(1) \equiv \Spin^c$ structure:

If instead we have $\Z_2^\rF\subset \U(1)$ (so that in particular $q_i$ are all necessarily odd), 
\cred{which means the spacetime-internal symmetry structure is
$\Spin \times_{\Z_2^\rF} \U(1) \equiv \Spin^c$ structure},
the general 6d anomaly polynomial is an integral  linear combination of the following two terms:
\begin{equation}
   \label{spinc-anomaly-basis} %\label{eq:Spinc-polynomial}
   I^C\coloneqq \frac{c_1^3}{6}-\frac{c_1p_1}{24}=\frac{(2c_1)^3-(2c_1)p_1}{48} \in \Z,\qquad 
   \cred{I^D\coloneqq {4c_1^3}=\frac{(2c_1)^3}{2}} \in \Z.
\end{equation}
Those are the values of $I_6$ for the charge vectors 
\cred{$q=(1)$ and $q=(3,-1,-1,-1)$} respectively.
 Note that in this case $c_1$ is in general not a well-defined integer cohomology class, only $2c_1$ is. This is because in general there is no globally well-defined $\U(1)$ bundle, only $\U(1)/\Z_2$ bundle, the first Chern class of which is \cred{$c_1'=2c_1\in\Z$}.\footnote{\cred{For $\Spin^c$, 
 the $\U(1) \supset {\Z_2^\rF}$ contains the fermion parity as a normal subgroup.\\
$\bullet$ For the original $\U(1)$ with $c_1(\U(1))$, 
the gauge bundle constraint is $w_2(TM)= 2 c_1 \mod 2$.
In the original $\U(1)$, fermions have odd charges under $\U(1)$,
while bosons have even charges under $\U(1)$.
Call the original U(1) gauge field $A$,
then $c_1=\frac{\dd A}{2 \pi} \in \frac{1}{2}\Z$.\\
$\bullet$ For the new $\U(1)'=\frac{\U(1)}{\Z_2^\rF}$ with $c_1(\U(1)')$,
the gauge bundle constraint is $w_2(TM)= c_1' = 2 c_1 \mod 2$.
Call the new $\U(1)'$ gauge field $A'$,
then $c_1'=\frac{\dd A'}{2 \pi}=\frac{\dd (2A)}{2 \pi} = 2 c_1\in 2\frac{1}{2}\Z = \Z$.\\
$\bullet$ To explain why $A' = 2 A$ or $ c_1' = 2 c_1$, we look at the Wilson line operator
$\text{$\exp(\ii q' \oint A')$ and $\exp(\ii q \oint A)$.}$
The original $\U(1)$ has charge transformation $\exp(\ii q \theta)$ with $\theta \in [0, 2 \pi)$,
while the new $\U(1)'$ has charge transformation $\exp(\ii q' \theta')$ with $\theta' \in [0, 2 \pi)$.
But the $\U(1)'=\frac{\U(1)}{\Z_2^\rF}$, so the $\theta=\pi$ in the old $\U(1)$ 
is identified as $\theta'=2\pi$ as a trivial zero
in the new $\U(1)'$.
In the original $\U(1)$, the $q \in \Z$ to be compatible with $\theta \in [0, 2 \pi)$.
In the new $\U(1)'$, the original $q$ is still allowed to have $2\Z$ to be compatible with $\theta \in [0, \pi)$;
but the new $q'=\frac{1}{2} q \in \Z$
and the new $\theta'= 2 \theta  \in [0, 2 \pi)$ are scaled accordingly.
Since the new $q'=\frac{1}{2} q \in \Z$, we show the new $A'=2 A$.}} 
It has to satisfy the condition $c_1'= 2c_1=w_2\mod 2$ where $w_2$ is the second Stiefel-Whitney class of the tangent bundle.

For general charges, we have:
\begin{equation}
    I_6=
    \left(\sum_{i=1}^{n_L}{q}_i\right)I^C
    +\left(\sum_{i=1}^{n_L}\frac{{q}_i^3-q_i}{24}\right)I^D \in \Z. 
\end{equation}
Note that $(q^3-q)/24\in \Z$ for any $q\in 2\Z+1$.

 Now $I^C$ and $I^D$ serve as the two generators of $\Hom(\Omega_6^{\Spin^c},\Z)\cong \Z\times \Z$. As before, the fact that they are \cpurple{integer-valued} follows from Atiyah-Singer index theorem. Finding a pair of representatives in the bordism group such that the values of $(I^C,I^D)$ on them form a basis in $\Z^2$ is a bit more involved. To construct it, we will first consider the following triple of representatives. 
 
$\bullet$ For the first, $V_1$, we take $W_1$ as above, but now considered as a Spin$^c$ 6-manifold. It has $(I^C,I^D)(V_1)=(1,24)$. 
 
$\bullet$ For the second, $V_2$, we take Spin$^c$ 6-manifold $S^2\times \CP^2$ with $2c_1=2a+h$ where $a$ is the Poincar\'e dual to $[\pt\times \CP^2]$ and $h$ is the standard generator of $\H^2(\CP^2,\Z) \cred{=\Z}$. Note that $w_2=h\mod 2$, so $2c_1$ indeed satisfies the necessary condition. Using the fact that $p_1=3h^2$ (footnote \ref{ft:CPn-tautological}), we get $(I^C,I^D)(V_2)=(0,3)$. 
 
$\bullet$ For the third, $V_3$, we take a quartic complex hypersurface in \cred{a non-spin} $\CP^4$ with $2c_1=h$, 
 the standard generator of 
 $\H^2(\CP^4,\Z) \cred{=\Z}$ induced on the cohomology of the quartic. It is consistent with the fact that $w_2=h\mod 2$ in this case. Using also the facts that $p_1=-11h^2$ and that the top-degree cohomolgy generator of the quartic is $4h^3$, we obtain $(I^C,I^D)(V_3)=(1,2)$. 
 
 Now let us take \cred{$U_1\coloneqq V_1\# (-V_2)^{\# 8}$ and $U_2 \coloneqq  V_1\# (-V_2)^{\# 7}\# (-V_3)$} where $\#$ denotes the connected sum and the minus sign denotes the orientation reversal. We have $(I^C,I^D)(U_1)=(1,0)$ and $(I^C,I^D)(U_2)=\cred{(0,1)}$ which do form a basis in $\Z^2$.  

 \item $\SO \times \U(1)$ structure:

If instead we consider a bosonic system without fermion parity symmetry $\Z_2^\rF$, 
which means the spacetime-internal symmetry structure is
$\SO \times \U(1)$ structure,
the general 6d anomaly polynomial is an integral  linear combination of the following two terms:
\begin{equation}
\label{eq:SOU1-polynomial}
   I^{C'}\coloneqq \frac{c_1^3}{3}-\frac{c_1p_1}{3} \in \Z,\qquad 
   I^{D'}\coloneqq {c_1^3} \in \Z.
\end{equation}
\cpurple{The anomaly polynomial in \eq{eq:I6b}'s is related to them as} $I_{6,b}=I^{C'} + I^{D'} =\frac{4c_1^3}{3}-\frac{c_1p_1}{3} \in \Z$.
As explained earlier, the ratio of the coefficients of $c_1p_1$ term in the \Eqn{eq:SOU1-polynomial}
\cpurple{and} \Eqn{eq:SpinU1-polynomial}
(but not the full generators) is 8. \cpurple{This is
because that is just the ratio of the coefficients in front of $p_1$ in 
the Hirzebruch signature theorem ($\frac{p_1}{3}$ from the L genus \eq{eq:L}) 
and the index theorem for Dirac operator ($-\frac{p_1}{24}$ from the $\hat{\rA}$ genus (\ref{eq:hatA}) respectively}.

\cred{To understand this $I^{C'}$ from a 4d anomaly perspective,
we can start with a fermionic theory with 8 Weyl fermions each with $q=1$ in 4d, 
and then we bosonize this theory to a bosonic theory with an anomaly 
$I_{6,b} =
 \frac{4 c_1^3}{3} -  
\frac{c_1p_1}{3}$.
To obtain an extra $-c_1^3$ anomaly in a 4d bosonic theory,
we can bosonize a \cpurple{triple of left-handed Weyl fermion with charge vector} $q=-(2,-1,-1)$ in 4d.
The bosonization is done by gauging $\Z_2^\rF$, namely summing over the spin structure, which changes
the $\Spin \times \U(1)$ structure to the $\SO \times \U(1)$ structure.}

 Now 6d anomaly polynomials 
 $I^{C'}$ and $I^{D'}$ serve as the two generators of $\Hom(\Omega_6^{\SO \times \U(1)},\Z)\cong \Z\times \Z$,
 \cpurple{and also correspond to the generators of the group of 5d iTFTs} $\TP_5(\B({\SO \times \U(1)}))=\Z^2$ \cite{WanWang2018bns1812.11967}. 
 As before, the fact that they are \cpurple{integer-valued} follows from a \cpurple{the index theorem for twisted signature operator considered above. Below we find a pair of representatives in the bordism group such that the values of $(I^{C'},I^{D'})$ on them form a basis in $\Z^2$.}

$\bullet$
For the first, $V_1'$, we take a U(1) bundle over the spin manifold $\CP^3$. 
\cpurple{For $h\in \H^2(\CP^3)$ as before we take $c_1(\U(1))=-h$. We then have $p_1(T\CP^3)=4h^2$, $c_1(\U(1))^3=-h^3$, 
$c_1(\U(1))p_1(T\CP^3)=-4h^3$ (footnote \ref{ft:CPn-tautological}). Paired with the fundamental class of the manifold $M=\CP^3$,
we have $\frac{1}{3}(c_1^3-c_1p_1) = -1/3 + 4/3 = +1$, $c_1^3=-1$. It follows that
$(I^{C'},I^{D'})(V_1')=(1,-1)$. 
}

\cred{$\bullet$
For the second, $V_2'$, we take a U(1) bundle over 
the base nonspin manifold  $M=S^2 \times \CP^2$,
\cpurple{such that $c_1(\U(1))=-h$, minus the standard generator of $\H^2(S^2,\Z)=\Z$,  Poincar\'e dual to [pt $\times \CP^2$]. We also have  $p_1(T\CP^2)=3h'^2$, from $p(T\CP^2)=(1+h'^2)^{3}$
where $h'$ is the generator of $\H^2(\CP^2,\Z)=\Z$.
Paired with the fundamental class of the base manifold $S^2\times \CP^2$,
we have $c_1^3([S^2\times \CP^2)=0$, $-c_1p_1([S^2\times \CP^2])=p_1([\CP^2])=3$.
Therefore $(I^{C'},I^{D'})(V_2')=(1,0)$.}}

\cpurple{Since (1,-1) and (1,0) generate $\Z^2$, this already proves that $I^{C'}$ and $I^{D'}$ form a basis. As the dual basis in the bordism group one can take $V_2'$ iteslf and $V_3'=({-V_1'}) \# {V_2'}$, so that $(I^{C'},I^{D'})(V_3')=(0,1)$.}

\item ${\Spin\times_{\Z_2^\rF}\Z_4}$ structure:

 Consider now the inclusion map $\Z_4\subset \U(1)$, in the case when $\Z_2^\rF\subset \Z_4$. The anomalies of $\Z_4$ symmetry have $\Z_{16}=\Hom(\Omega_5^{\Spin\times_{\Z_2^\rF}\Z_4},\U(1))$ classification. The inclusion induces the pullback map between the groups classifying the anomalies:
\begin{equation}
    \begin{array}{rcl}
    \Hom(\Omega_6^{\Spin^c},\Z) & \longrightarrow & \Hom(\Omega_5^{\Spin\times_{\Z_2^\rF}\Z_4},\U(1)),
    \\
    \rotatebox{90}{$\cong$}& &\rotatebox{90}{$\cong$}
    \\
        \Z^2 & \longrightarrow & \Z_{16}. 
    \end{array}
    \label{U1-Z4-anomaly-map-app} 
\end{equation}
To describe the map explicitly, we need to choose a basis for each group. For $\Z^2$ we choose the basis (\ref{spinc-anomaly-basis}). 
For $\Z_{16}$, 
we choose the basis element to be the anomaly of a single \cred{left-handed} Weyl fermion of charge $+1\mod 4$. By considering a \cred{left-handed} fermion with $\U(1)$ charge $+1$, we then immediately conclude that
\begin{equation}
    (k,\ell)=(1,0) \text{ thus } (\kappa_1,\kappa_2)=(1, - {1}/{24})\;\longmapsto\;\upnu=1\mod 16.
    \label{U1-Z4-map-1}
\end{equation}
Note that the \cred{left-handed} fermion of charge $3=-1\mod 4$ should necessarily have anomaly $-1\mod 16$, because the map (\ref{U1-Z4-anomaly-map-app}) is a homomorphism, and (\ref{6d-anomaly-pol}) changes the sign when all the fermion charges change signs. Therefore, by considering a \cred{left-handed} fermion with $\U(1)$ charge $+3$ we  conclude that
\begin{equation}
    \cred{(k,\ell)=(3,1) \text{ thus } (\kappa_1,\kappa_2)=(27, - {1}/{8})\;\longmapsto\; \upnu=-1\mod 16}.
    \label{U1-Z4-map-2}
\end{equation}
Combining (\ref{U1-Z4-map-1}) and (\ref{U1-Z4-map-2}) we get more generally:
\begin{equation}
     \cred{ (k,\ell) \text{ thus } (\kappa_1,\kappa_2)=(24 \ell +k, -k/24)  \longmapsto  \upnu=k-4\ell}
    \mod 16.
    \label{U1-Z4-map}
\end{equation}

\end{enumerate}

%\newpage
\bibliography{BSM-B+L-Categorical.bib}

\end{document}